\documentclass[12pt]{article}
\usepackage{amsmath}
\usepackage{amsfonts} 
\usepackage{graphicx,psfrag,epsf}
\usepackage{enumerate}
\usepackage{natbib}
\usepackage{url} 
\usepackage{hyperref}
\usepackage{graphicx}
\usepackage[english]{babel}
\usepackage{amsthm}
\usepackage{bbm}
\usepackage{soul}
\usepackage{paralist}
\usepackage{mathtools}
\usepackage{adjustbox}
\usepackage{booktabs}
\usepackage{xcolor}
\usepackage{amssymb}
\usepackage{multirow}  
\usepackage{array}     
\usepackage{dutchcal}
\usepackage{soul}
\usepackage{subcaption}

\allowdisplaybreaks

\newtheorem{theorem}{Theorem}

\newtheorem{corollary}{Corollary}
\newtheorem{assumption}{Assumption}
\newtheorem{proposition}{Proposition}

\theoremstyle{definition}
\newtheorem{definition}{Definition}

\theoremstyle{remark}

\newcommand{\indep}{\perp \!\!\! \perp}
\newcommand{\E}{\mathbb{E}}

\newcommand{\Var}{\mathbb{V}}
\newcommand{\Corr}{\mathbb{C}{\rm orr}}
\newcommand{\Cov}{\mathbb{C}{\rm ov}}

\newcommand\yo[1]{{\color{black} #1}}

\usepackage{tikz}
\usetikzlibrary{fit,positioning}
\usetikzlibrary{bayesnet}
\usepackage{color}

\usetikzlibrary{shapes,decorations,arrows,calc,arrows.meta,fit,positioning}
\tikzset{
    -Latex,auto,node distance =1 cm and 1 cm,semithick,
    state/.style ={ellipse, draw, minimum width = 0.7 cm},
    point/.style = {circle, draw, inner sep=0.04cm,fill,node contents={}},
    bidirected/.style={Latex-Latex,dashed},
    el/.style = {inner sep=2pt, align=left, sloped}
}

\usepackage{setspace}
\usepackage{bibspacing}
\newcommand{\blind}{0}

\begin{document}

\def\spacingset#1{\renewcommand{\baselinestretch}%
{#1}\small\normalsize} \spacingset{1}


\if0\blind
{
  \title{\bf 
  A Bayesian nonparametric approach to mediation and spillover effects with multiple mediators in cluster-randomized trials
  }
  \author{Yuki Ohnishi and Fan Li\\
    Department of Biostatistics \\ 
    Yale School of Public Health}
  \maketitle
} \fi

\if1\blind
{
  \bigskip
  \bigskip
  \bigskip
  \begin{center}
    {\LARGE\bf A Bayesian nonparametric approach to mediation and spillover effects with multiple mediators in cluster-randomized trials }
\end{center}
  \medskip
} \fi

\bigskip
\begin{abstract}
Cluster randomized trials (CRTs) with multiple unstructured mediators present significant methodological challenges for causal inference due to within-cluster correlation, interference among units, and the complexity introduced by multiple mediators. Existing causal mediation methods often fall short in simultaneously addressing these complexities, particularly in disentangling mediator-specific effects under interference that are central to studying complex mechanisms. To address this gap, we propose new causal estimands for spillover mediation effects that differentiate the roles of each individual's own mediator and the spillover effects resulting from interactions among individuals within the same cluster. We establish identification results for each estimand and, to flexibly model the complex data structures inherent in CRTs, we develop a new Bayesian nonparametric prior---the Nested Dependent Dirichlet Process Mixture---designed for flexibly capture the outcome and mediator surfaces at different levels. We conduct extensive simulations across various scenarios to evaluate the frequentist performance of our methods, compare them with a Bayesian parametric counterpart and illustrate our new methods in an analysis of a completed CRT. 
\end{abstract}

\noindent%
\vfill

\newpage
\spacingset{1.9} 

\section{Introduction}
\label{sec:intro}
In cluster-randomized trials (CRTs), entire clusters are randomly assigned to different treatment conditions, while post-treatment variables are typically measured for individual members within each cluster. 
Although the average causal effect is the conventional focus in CRTs, there is a growing interest in understanding mechanisms that explain the estimated causal effect. 
Causal mediation analysis has emerged as a valuable tool for this purpose, allowing researchers to decompose the total causal effect into natural indirect effects, mediated by intermediate variables, and natural direct effects that operate independently of the mediators.  
This decomposition can provide an evidence-based perspective for refining group-level interventions to maximize public health and social benefits \citep{Williams2016}. 

Several previous studies have developed mediation methods to address within-cluster correlation and interference in CRTs. For example, \citet{VanderWeele2009} and \citet{VanderWeele2013_CRT} provided a decomposition of the natural indirect effect into a spillover mediation effect and an individual mediation effect, and discussed nonparametric identification. 
\citet{cheng2024} developed the efficient influence function to motivate several doubly robust estimators for estimating the natural indirect effect and spillover mediation effect in CRTs. Using similar techniques as in causal mediation, others have addressed noncompliance in CRTs, where the treatment receipt is viewed as a special binary mediator \citep[e.g.,][]{Forastiere2016, Park2023, Ohnishi2024}. However, their primary interest lies in inferring the spillover effects among different compliance strata, addressing a different scientific question from the mediation context.  

A primary limitation of the aforementioned methods is that they have exclusively assumed a single mediator, whereas multiple mediators can be collected in a CRT and may jointly explain the total causal effect. 
The inclusion of multiple mediators poses unique challenges in CRTs, 
because it requires careful definitions of indirect and spillover effect estimands that represent various pathways. Although methods have been developed for studying multiple mediators under independent data \citep[e.g.][]{vanderweele2013_multiple, Daniel2015, Taguri2018, Kim2019, Xia2022}, they operated on the assumption of no interference and cannot be directly used to address spillover mediation effect in CRTs. 
To the best of our knowledge, no prior work has investigated the spillover mediation effects---where the treatment effect on one individual may be mediated through effects on other individuals within the same cluster---in the presence of multiple mediators. The lack of identification results and robust estimation strategies presents a barrier to offering a deeper understanding of the complex mechanisms through which cluster-level interventions exert their impact. 

\yo{This work provides a new treatment for causal mediation analysis with multiple mediators in CRTs. First, we develop mediation estimands by decomposing the natural indirect effect into the mediator-specific exit indirect effects and interaction effects, analogous to \citet{Xia2022} but within the context of CRTs. To address within-cluster interference, we further consider a novel and practically useful decomposition of the exit indirect effects into individual and spillover components to investigate finer mechanisms. We then establish the requisite structural assumptions and characterize the nonparametric point identification formula of each estimand. Second, given that parametric modeling for complex estimands is susceptible to misspecification bias, we propose a new Bayesian nonparametric (BNP) prior---the nested dependent Dirichlet process mixture (nDDPM)---specifically designed to flexibly model components of the derived identification formulas and to ensure robust analysis of clustered data. While BNP methods have been studied for causal mediation with independent data \citep[e.g.,][]{Kim2019, Roy2022}, none are designed to model clustered data. 
We complement the nDDPM prior development with a theoretical analysis of its induced correlation structure and weak support property. Finally, we conduct extensive simulations to evaluate the frequentist performance of our proposed methods. The results demonstrate that our methods outperform existing ones in terms of accuracy and robustness under different data generating processes.} 

\yo{
\subsection{A motivating empirical application}
\label{sec:motivating_example}

Child undernutrition remains a significant public health challenge, contributing to increased morbidity and mortality in young children and leading to long-term adverse effects on cognitive development and education. While global rates of undernutrition have declined, substantial disparities persist across regions, highlighting the need for targeted interventions. Conditional cash transfer (CCT) programs emerged as a promising policy tool to improve child nutritional outcomes by addressing key determinants such as poverty, food insecurity, and healthcare access. 
Although it has been suggested that the effectiveness of CCTs depends on their ability to enforce health service utilization \citep{Manley2013}, the specific causal mechanisms through which CCTs influence child nutrition remain unclear. 

We reanalyze data from the Nicaraguan Red de Protecci\'on Social (RPS) CRT, which evaluated the effect of CCTs among households living in poverty across 42 \textit{comarcas} (administrative regions and unit of randomization) in Nicaragua \citep{Charters2023}. Specifically, we investigate how two intermediate variables with an unknown causal structure---child health check-ups and household dietary diversity---might explain the program's impact on child nutritional outcomes (Figure \ref{fig:DAG_mediators_sub1}). 
To accurately assess causal mechanisms due to these two intermediate variables with an unknown causal structure in a clustered data context, it is imperative to refine estimands that can separate direct effects from mediator-specific indirect effects and further understand to what extent the mediator-specific indirect effect is explained by mediators from other members of the same cluster. Such methodological rigor would yield insights for understanding causal pathways and is crucial for optimizing the design and implementation of future policy programs to maximize their benefits for child health and development.

}

\section{Assumptions, Estimands, and Identification}
\label{sec:setup}

\subsection{Notation and data structure}
\label{sec:notation}

We consider a CRT with $ I $ clusters. For cluster $ i \in \{1, \ldots, I\} $, we denote $ N_i $ as the number of individuals in cluster $ i $ (cluster size), $ A_i \in \{0, 1\} $ as the cluster-level treatment assignment, with $A_i = 1$ if it is assigned treatment and $A_i = 0$ otherwise, and $ \mathbf{V}_i \in \mathcal{V}=\mathbb{R}^{d_V \times 1} $ as a vector of cluster-level baseline covariates. 
The total number of individuals in the study is denoted by $ N = \sum_{i=1}^{I} N_i$.
For individual $ j \in \{1, \ldots, N_i\} $ in cluster $ i $, we observe a vector of individual-level baseline covariates $ \mathbf{X}_{ij} \in \mathcal{X}=\mathbb{R}^{d_X \times 1} $, and write $ \mathbf{X}_i = [\mathbf{X}_{i1}, \ldots, \mathbf{X}_{iN_i}]^\top \in \mathbb{R}^{N_i \times d_X} $. 
Let $ \mathbf{C}_i = \{\mathbf{V}_i, \mathbf{X}_i\} $ represent all baseline covariates in cluster $ i $ and $ \mathbf{C}_{ij} = \{\mathbf{V}_i, \mathbf{X}_{ij}\} $ represent baseline covariates of individual $j$ in cluster $ i $. 
We observe the individual-level outcome $ Y_{ij} \in \mathbb{R} $, and consider multiple individual-level mediators measured in the treatment-outcome pathway. 
For ease of presentation, we focus on the scenario with two mediators, $M_{ij}^{(k)} \in \mathbb{R}$ for $k = 1, 2$, but our methods can be extended to accommodate more mediators (Supplementary Material Section \ref{sec:K_mediator_identification}). 
Additionally, we do not assume any causal ordering between $ M_{ij}^{(1)} $ and $ M_{ij}^{(2)} $, as when multiple mediators are considered in the study, there is typically a lack of knowledge about their causal structures \citep{Taguri2018, Xia2022}. 
We let $ \mathbf{Y}_i = [Y_{i1}, \ldots, Y_{iN_i}]^\top \in \mathbb{R}^{N_i \times 1} $, $ \mathbf{M}_i^{(k)} = [M_{i1}^{(k)}, \ldots, M_{iN_i}^{(k)}]^\top \in \mathbb{R}^{N_i \times 1} $, and $ \mathbf{M}_{i(-j)}^{(k)} \in \mathbb{R}^{(N_i - 1) \times 1} $ as the vector of mediators from cluster $ i $ excluding individual $ j $. 
Finally, we let $ \mathbf{A} $, $ \mathbf{M}^{(1)} $, $ \mathbf{M}^{(2)} $, and $ \mathbf{Y} $ be the ($ I \times 1 $)-dimensional vector of treatment assignments and the ($ N \times 1 $)-dimensional vectors of the first mediators, the second mediators and outcomes, respectively. 
\yo{In the RPS CRT, $ \mathbf{A} $, $ \mathbf{M}^{(1)} $, $ \mathbf{M}^{(2)} $, and $ \mathbf{Y} $ represent the cluster-level CCT program intervention, child health check-up conditions, dietary diversity score, and child nutritional status.}
Figure \ref{fig:DAG_mediators_sub1} provides a graphical representation of the causal structure between the observed variables.

\begin{figure}[htbp]
    \centering
    \begin{subfigure}[t]{0.49\textwidth}
    \centering
    \scalebox{0.7}{
    \begin{tikzpicture}
        \node[state] (a) at (0,0) {$A_i$};
        \node[state] (m1) at (1.5,1.5) {$M_{ij}^{(1)}$};
        \node[state] (m2) at (4, 1.5)  {$\mathbf{M}_{i(-j)}^{(1)}$};
        \node[state] (y) at (6,0) {$Y_{ij}$};
        \node[state] (c) at (-2,0) {$\mathbf{X}_i$};
        \node[state] (n) at (-4,0) {$\left\{N_i,\mathbf{V}_i\right\}$};
        \node[state] (m3) at (1.5,-1.5) {$M_{ij}^{(2)}$};
        \node[state] (m4) at (4, -1.5) {$\mathbf{M}_{i(-j)}^{(2)}$};

        \path (a) edge (y);
        \path (a) edge (m1);
        \path (a) edge (m2);
        \path (a) edge (m3);
        \path (a) edge (m4);
        
        \path (m1) edge (y);
        \path (m2) edge (y);
        \path (m3) edge (y);
        \path (m4) edge (y);
        
        \path (c) edge[bend left=80] (y);
        \path (c) edge (m1);
        \path (c) edge (m3);
        \path (c) edge[bend left=40] (m2);
        \path (c) edge[bend left=-40] (m4);
        
        \path (n) edge[bend left=-70] (y);
        \path (n) edge[bend left=-20] (m3);
        \path (n) edge[bend left=-40] (m4);
        \path (n) edge[bend left=20] (m1);
        \path (n) edge[bend left=40] (m2);
        \path (n) edge (c);
        
        \draw [dashed,-] (m1) edge (m2);
        \draw [dashed,-] (m1) edge (m3);
        \draw [dashed,-] (m1) edge (m4);
        \draw [dashed,-] (m2) edge (m3);
        \draw [dashed,-] (m2) edge (m4);
        \draw [dashed,-] (m3) edge (m4);
    \end{tikzpicture}}
    \caption{Mediation DAG}
    \label{fig:DAG_mediators_sub1}
    \end{subfigure}
    \begin{subfigure}[t]{0.49\textwidth}
    \centering
    \scalebox{0.7}{
    \begin{tikzpicture}[>=stealth, font=\small]
        \node[draw, circle, thick] (Ai) at (0,0) {$A_i$};
        \node[draw, circle, thick] (Yij) at (8,0) {$Y_{ij}$};

        \node[draw, thick, rectangle, rounded corners,
              minimum width=5cm, minimum height=4.0cm] (macro) at (4,0) {};

        \node (M1ij) at (3, 0.9)   {$M_{ij}^{(1)}$};
        \node (M1iMinusJ) at (5, 0.9)   {$\mathbf{M}_{i(-j)}^{(1)}$};
        \node (M2ij) at (3,-0.9)   {$M_{ij}^{(2)}$};
        \node (M2iMinusJ) at (5,-0.9)   {$\mathbf{M}_{i(-j)}^{(2)}$};

        \draw[dashed] (M1ij) ellipse (0.7 and 0.4);
        \draw[dotted] (M1iMinusJ) ellipse (0.7 and 0.4);
        \draw[dashed] (M2ij) ellipse (0.7 and 0.4);
        \draw[dotted] (M2iMinusJ) ellipse (0.7 and 0.4);

        \node[draw, dash dot dot, thick, rounded corners,
              minimum width=4cm, minimum height=1.5cm] 
              (topBox) at (4,0.8) [fit=(M1ij)(M1iMinusJ)] {};

        \node[draw, dash dot dot, thick, rounded corners,
              minimum width=4cm, minimum height=1.5cm] 
              (bottomBox) at (4,-0.8) [fit=(M2ij)(M2iMinusJ)] {};

        \draw[->] (Ai) -- (macro.west);
        \draw[->, bend right=85] (Ai) to (Yij);

        \draw[->, dashed, bend left=34] (M1ij) to (Yij);
        \draw[->, dotted, bend left=14] (M1iMinusJ) to (Yij);
        \draw[->, dashed, bend right=34] (M2ij) to (Yij);
        \draw[->, dotted, bend right=14] (M2iMinusJ) to (Yij);

        \draw[->, dash dot dot, thick] (topBox) -- (Yij);
        \draw[->, dash dot dot, thick] (bottomBox) -- (Yij);
    \end{tikzpicture}}
    \caption{Proposed estimands DAG}
    \label{fig:DAG_mediators_sub2}
    \end{subfigure}

    \caption{(a) Mediation directed acyclic graph for a CRT with two unordered mediators. Here, $\mathbf{X}_i$ and $N_i$ are baseline covariates and sample size of cluster $i$, $A_i$ is treatment assignment, $M_{ij}^{(k)}$ and $Y_{ij}$ are mediator and outcome of individual $j$ in cluster $i$, and $\mathbf{M}_{i(-j)}^{(k)}$ is the vector of mediators excluding individual $j$. 
    \yo{(b) Graphical representation of proposed estimands with an unknown causal structure. The dash-dot, dashed, and dotted lines represent the EIE, EIME, and ESME, respectively.}}
    \label{fig:DAG_mediators}
\end{figure}

We adopt the potential outcomes framework, and define $M_{ij}^{(1)}(\mathbf{A})$, $M_{ij}^{(2)}(\mathbf{A})$ as the potential mediator variables under assignment vector $ \mathbf{A} $, and $Y_{ij}(\mathbf{A}, \mathbf{M}^{(1)}, \mathbf{M}^{(2)})$ as the potential outcomes for unit $ j $ in cluster $ i $ when $ \mathbf{A}, \mathbf{M}^{(1)}, \mathbf{M}^{(2)} $ were the vectors of assignments and mediators in the whole study population.
\begin{assumption}[Cluster-level SUTVA]
\label{asmp:sutva}
    Cluster-level stable unit treatment value assumption (SUTVA) consists of two parts: (1) \textit{(No interference between clusters)}: $ M_{ij}^{(1)}(\mathbf{A}) = M_{ij}^{(1)}(A_i)$, $ M_{ij}^{(2)}(\mathbf{A}) = M_{ij}^{(2)}(A_i)$, and $Y_{ij}(\mathbf{A}, \mathbf{M}^{(1)}, \mathbf{M}^{(2)}) = Y_{ij}(A_i, \mathbf{M}^{(1)}_{i}, \mathbf{M}^{(2)}_{i})$. (2) \textit{(No multiple versions of treatment)} if $ A_i = A_i' $ then $ M_{ij}^{(k)}(A_i) = M_{ij}^{(k)}(A_i')$ and if $ A_i = A_i' $ $\mathbf{M}^{(k)}_{i}= \mathbf{M}^{(k)'}_{i} $ then $ Y_{ij}(A_i, \mathbf{M}^{(1)}_{i}, \mathbf{M}^{(2)}_{i}) = Y_{ij}(A_i', \mathbf{M}^{(1)'}_{i}, \mathbf{M}^{(2)'}_{i})$, for $k=1,2$.
\end{assumption}

\noindent
\yo{The first part of Assumption \ref{asmp:sutva} allows for mediator interference within each cluster but rules out interference across clusters, which is reasonable in our context due to the physical separation across clusters that minimizes cross-cluster interactions. The second part of Assumption \ref{asmp:sutva} ensures that the mediators and outcomes respond similarly to the treatment across all clusters without variations in treatment execution. This is a standard assumption in causal inference and its validity is typically guaranteed by the experimental design.} 
We define $ M_{i j}^{(k)}(a) $ as the potential mediator variable under condition $ a \in \{0, 1\} $, $ \mathbf{M}_{i}^{(k)}(a) = [M_{i 1}^{(k)}(a), \ldots, M_{i N_i}^{(k)}(a)]^\top $ as the vector of potential mediator variables for all individuals in cluster $ i $, and $ \mathbf{M}_{i (-j)}^{(k)}(a) $ as the vector excluding the $ j $th element in $ \mathbf{M}_{i}^{(k)}(a) $. 
We define $ Y_{i j}(a, \mathbf{m}_i^{(1)}, \mathbf{m}_i^{(2)}) $ as the potential outcome if cluster $ i $ had been randomized to condition $ a $ and the two mediators of all individuals in cluster $ i $, $ \mathbf{M}_{i}^{(1)} $ and $ \mathbf{M}_{i}^{(2)} $, were set to $ \mathbf{m}_i^{(1)} $ and $ \mathbf{m}_i^{(2)} $ respectively.
The subscript of the vector $ \mathbf{m}_i^{(k)} $ indicates the dependence of the mediator vector on the cluster size $ N_i $.
Also, notice that one can equivalently represent $ Y_{i j}(a, \mathbf{m}_i^{(1)}, \mathbf{m}_i^{(2)}) = Y_{i j}(a, m_{i j}^{(1)}, \mathbf{m}_{i(-j)}^{(1)}, m_{i j}^{(2)}, \mathbf{m}_{i(-j)}^{(2)}) $; this notation explicitly distinguishes an individual’s own mediator from the mediators of the remaining cluster members. 

The only possibly observable potential outcome is the one where, if $ A_i $ were set to $ a $, the mediators of all the units in cluster $i$  were set to the value they would have taken under condition $a$. Throughout we use the following notation for potential outcomes of this type: 
$ Y_{i j}(a) = Y_{i j}(a, \mathbf{M}_{i}^{(1)}(a), \mathbf{M}_{i}^{(2)}(a)) = Y_{i j}(a, M_{i j}^{(1)}(a), \mathbf{M}_{i(-j)}^{(1)}(a), M_{i j}^{(2)}(a), \mathbf{M}_{i(-j)}^{(2)}(a) ) $. 
We define the collection of all random variables in cluster $ i $ as $\mathbf{W}_i = \{ \mathbf{C}_i, \mathbf{M}_i^{(1)}(0), \mathbf{M}_i^{(1)}(1), \mathbf{M}_i^{(2)}(0), \allowbreak \mathbf{M}_i^{(2)}(1), \mathbf{Y}_i(0, \mathbf{m}_i^{(1)}, \mathbf{m}_i^{(2)}), \mathbf{Y}_i(1, \mathbf{m}_i^{(1)}, \mathbf{m}_i^{(2)})\}$ 
for all $ \mathbf{m}_i^{(1)}, \mathbf{m}_i^{(2)} \in \mathbb{R}^{N_i \times 1}$.
Next, we introduce the following assumptions on the complete data $ \{ (\mathbf{W}_1, A_1, N_1), (\mathbf{W}_2, A_2, N_2), \ldots, (\mathbf{W}_I, A_I, N_I)\} $.

\begin{assumption}[Cluster randomization]
    \label{asmp:randomization}
    The treatment assignment for each cluster is an independent realization from a Bernoulli distribution with $ p(A_i =1 ) = \pi \in (0,1) $.
\end{assumption}

\begin{assumption}[Super-population framework]
\label{asmp:superpopulation}
    (a) The cluster size $N_i$ follows an unknown distribution $ \mathcal{P}^{N}$ over a finite  support on $ \mathbb{N}^{+} $. (b) Conditional on $ N_i $, the joint  distribution $ \mathcal{P}^{W, A \mid N} $ can be decomposed into $ \mathcal{P}^{W \mid N} \times \mathcal{P}^{ A } = \mathcal{P}^{Y \mid M^{(1)}, M^{(2)}, C, N} \times \mathcal{P}^{M^{(1)}, M^{(2)} \mid C, N} \times\mathcal{P}^{ C \mid N} \times\mathcal{P}^{ A } $. Furthermore, positivity holds such that the conditional density $f_{M^{(1)}, M^{(2)} \mid \mathbf{C}, A, N} (\mathbf{m}^{(1)},\mathbf{m}^{(2)} \mid \mathbf{c}, a, n) >0$ for any $\{ \mathbf{m}^{(1)},\mathbf{m}^{(2)}, \mathbf{c}, a, n \}$ over their valid support.
\end{assumption}

\noindent
Assumption \ref{asmp:randomization} eliminates unmeasured confounding for both the treatment-mediator and the treatment-outcome relationships, and is guaranteed by the cluster-randomization study design. Assumption \ref{asmp:superpopulation} extends \citet{Wang2024} and conceptualizes a super-population of clusters with a finite size of individuals within each cluster. 

\subsection{Causal mediation estimands}
\label{sec:estimands}

We focus on the cluster-average treatment effect defined as $\mathrm{TE_C} = \E \left[ \frac{1}{N_i} \sum_{j=1}^{N_i} \left\{ Y_{ij}(1) - Y_{ij}(0) \right\}   \right]$ \citep{kahan2024demystifying}.
The cluster-average treatment effect can be decomposed into two parts: the natural direct effect (NDE) and the natural indirect effect (NIE), i.e., $ \mathrm{TE_C} = \mathrm{NIE_C} + \mathrm{NDE_C} $, where $\mathrm{NIE_C} = \E \left[ \frac{1}{N_i} \sum_{j=1}^{N_i} \left\{ Y_{ij}(1, \mathbf{M}_i^{(1)}(1), \mathbf{M}_i^{(2)}(1)) - Y_{ij}(1, \mathbf{M}_i^{(1)}(0), \mathbf{M}_i^{(2)}(0)) \right\}   \right]$ and $\mathrm{NDE_C} = \E \left[ \frac{1}{N_i} \sum_{j=1}^{N_i} \left\{ Y_{ij}(1, \mathbf{M}_i^{(1)}(0), \mathbf{M}_i^{(2)}(0)) - Y_{ij}(0, \mathbf{M}_i^{(1)}(0), \mathbf{M}_i^{(2)}(0)) \right\}   \right]$. 
In the presence of multiple mediators where the mediators have an unknown causal structure, \citet{Xia2022} have shown that the NIE can be decomposed into the mediator-specific exit indirect effect (EIE) and inter-mediator interaction effects (INT) as  $ \mathrm{NIE}_{\mathrm{C}} = \mathrm{EIE}_{\mathrm{C}}^{(1)} + \mathrm{EIE}_{\mathrm{C}}^{(2)} - \mathrm{INT}_{\mathrm{C}}^{(1,2)} $, where for $ k = 1,2 $,
\begin{align*}
    \mathrm{EIE}_{\mathrm{C}}^{(k)} 
    &= \E \left[ \frac{1}{N_i}\sum_{j=1}^{N_i} \left \{ Y_{ij}(1, \mathbf{M}_i^{(k)}(1), \mathbf{M}_i^{(3-k)}(1)) - Y_{ij}(1, \mathbf{M}_i^{(k)}(0), \mathbf{M}_i^{(3-k)}(1)) \right \} \right] \\
    \mathrm{INT}_{\mathrm{C}}^{(1,2)} 
    &=  \E \left[ \frac{1}{N_i}\sum_{j=1}^{N_i} \left \{ Y_{ij}(1, \mathbf{M}_i^{(k)}(1), \mathbf{M}_i^{(3-k)}(1)) - Y_{ij}(1, \mathbf{M}_i^{(k)}(1), \mathbf{M}_i^{(3-k)}(0))  \right \} \right] \nonumber\\
    &- \E \left[ \frac{1}{N_i}\sum_{j=1}^{N_i} \left \{Y_{ij}(1, \mathbf{M}_i^{(k)}(0), \mathbf{M}_i^{(3-k)}(1)) - Y_{ij}(1, \mathbf{M}_i^{(k)}(0), \mathbf{M}_i^{(3-k)}(0)) \right \} \right] ,
\end{align*}
Although the EIE is not the finest possible estimand in CRTs, it is relevant because it picks up all indirect effects of the intervention that exit the mediator set through one specific mediator of all individuals in a cluster and moves towards $Y_{ij}$ immediately after. That is, $\mathrm{EIE}_{\mathrm{C}}^{(k)}$ includes all indirect effects from all individuals in a cluster making up the $\mathrm{NIE}_{\mathrm{C}}$ whose last stop before $Y_{ij}$ is $\mathbf{M}_i^{(k)}$. \yo{The dash-dot lines in Figure \ref{fig:DAG_mediators_sub2} represent the EIE with an unknown causal structure.} The EIE estimands reduce to familiar path-specific estimands when the causal structure between mediators is known (Supplementary Material Section \ref{sec:DAG_estimands}). The $\mathrm{INT}_{\mathrm{C}}^{(1,2)}$ is the difference between two indirect effects through a mediator with the other mediator fixed at different counterfactual values, so it is the indirect effect through one mediator modified by levels of the other mediator within the same cluster and is the overlapping component measured by both $\mathrm{EIE}_{\mathrm{C}}^{(1)}$ and $\mathrm{EIE}_{\mathrm{C}}^{(2)}$. 
\yo{Since the definition of $\mathrm{INT}_{\mathrm{C}}^{(1,2)}$ is invariant to the choice of $k$, it suffices to fix $k = 1$ (or $2$) to define the effect. As a result, this decomposition is also invariant to the ordering and labeling of $\mathbf{M}_i^{(1)}$ and $\mathbf{M}_i^{(2)}$, and can be attractive when the mediators have an unknown causal structure.} 


Extending \citet{VanderWeele2009} and \citet{cheng2024} with a single mediator, we further decompose $ \mathrm{EIE}_{\mathrm{C}}^{(k)} $ into the exit spillover mediation effect (ESME) and the exit individual mediation effect (EIME) as $ \mathrm{EIE}_{\mathrm{C}}^{(k)} = \mathrm{ESME}_{\mathrm{C}}^{(k)} + \mathrm{EIME}_{\mathrm{C}}^{(k)}$, where 
\begin{align*}
    \mathrm{ESME}_{\mathrm{C}}^{(k)} &= \E \left[ \frac{1}{N_i} \sum_{j=1}^{N_i} \left\{ Y_{ij}(1, M_{ij}^{(k)}(1), \mathbf{M}_{i(-j)}^{(k)}(1), \mathbf{M}_i^{(3-k)}(1)) - Y_{ij}(1, M_{ij}^{(k)}(1), \mathbf{M}_{i(-j)}^{(k)}(0), \mathbf{M}_i^{(3-k)}(1)) \right\} \right], \\
    \mathrm{EIME}_{\mathrm{C}}^{(k)} &= \E \left[ \frac{1}{N_i} \sum_{j=1}^{N_i} \left\{ Y_{ij}(1, M_{ij}^{(k)}(1), \mathbf{M}_{i(-j)}^{(k)}(0), \mathbf{M}_i^{(3-k)}(1)) - Y_{ij}(1, M_{ij}^{(k)}(0), \mathbf{M}_{i(-j)}^{(k)}(0), \mathbf{M}_i^{(3-k)}(1)) \right\} \right].
\end{align*}
The $\mathrm{ESME}_{\mathrm{C}}^{(k)}$ captures the indirect effect of an intervention that occurs through the $k$th mediators of other individuals within the same cluster, rather than through the individual's own mediator. It accounts for spillover effects that arise due to the unmeasured interaction of individuals,  
\yo{but should be distinguished from the traditional spillover effects in causal analyses under network interference \citep[e.g., ][]{Aronow2017}. Specifically, the ESME compares two mean potential outcomes that differ in the mediator values for other individuals under a fixed cluster-level treatment. In contrast, the traditional spillover effect is typically defined as the mean difference between two potential outcomes for a unit with a fixed treatment assignment for that unit, but with different treatment configurations for other units within the network.}
On the other hand, $\mathrm{EIME}_{\mathrm{C}}^{(k)}$ measures the indirect effect of an intervention on an outcome through an individual's $k$th mediator. It captures the part of the mediation effect that operates specifically through changes in the individual's mediator, holding constant the mediators of others in the same cluster. 
\yo{The dash lines and dot lines in Figure \ref{fig:DAG_mediators_sub2} represent the EIME and ESME with an unknown causal structure, respectively.}
Although the above decomposition focuses on two mediators, it remains informative when the mediators are independent or causally ordered. These connections are discussed and graphically illustrated in Figures \ref{fig:DAG_esimtands_causallyordered} and \ref{fig:DAG_esimtands_causallyindenpendent} in the supplementary material. 

\yo{In the RPS CRT, EIE$^{(1)}$ represents the effect of the CCT program that influences child nutritional levels within a household through the child health check-up conditions of all households in a comarcas, capturing the cluster-level indirect effects of the CCT program on a household’s child nutritional status mediated by child health check-up conditions. The EIE$^{(2)}$ is similarly interpreted but focuses on comarcas-level dietary diversity instead of child health check-up conditions. The INT estimand represents the difference between the effect of the CCT program operating through child health check-up conditions, assuming those conditions are fixed at the level they would have taken had the household received the CCTs, and the effect of the CCT program operating through dietary diversity, assuming child health check-up conditions are fixed at the level they would have taken had the household not received the CCTs. Furthermore, the EIME$^{(1)}$ refers to the effect of the CCT program operating through a specific household’s child health check-up conditions, directly influencing that household’s child nutritional status, whereas the ESME$^{(1)}$ describes the effect of the CCT program acting through the child health check-up conditions of all other households within the same comarcas, directly influencing a single household’s child nutritional status possibly through unmeasured interactions between households. Collectively, quantifying these estimands helps elucidate the causal mechanisms in a clustered data setting, thereby identifying critical intervention targets that could be leveraged to optimize desired changes in child nutritional outcomes.}


\subsection{Nonparametric identification}
\label{sec:identification}
To identify the proposed causal mediation estimands, we introduce a set of additional identification assumptions and provide the nonparametric identification results. 
\begin{assumption}[Sequential ignorability]
\label{asmp:si}
    (i) $ Y_{ij}(a,\mathbf{m}_i^{(1)},\mathbf{m}_i^{(2)}) \indep \{ \mathbf{M}_i^{(1)}(a),  \mathbf{M}_i^{(2)}(a) \} \mid  A_i=a, \mathbf{C}_i, N_i  $ and (ii) $ Y_{ij}(a,\mathbf{m}_i^{(1)},\mathbf{m}_i^{(2)}) \indep \{ \mathbf{M}_i^{(1)}(a'),  \mathbf{M}_i^{(2)}(a') \} \mid   \mathbf{C}_i, N_i  $ for all $ i, j$, $ a,a' \in \{0,1\}$, and $\mathbf{m}_i^{(1)}$, $ \mathbf{m}_i^{(2)} $ over their valid support.
\end{assumption}
\begin{assumption}[Conditional homogeneity]\label{asmp:cond_homogeneity} For $a, a' \in \{0,1\}$, and $\mathbf{m}_i^{(3-k)}$ over its valid support, we assume $\E [ \frac{1}{N_i} \sum_{j=1}^{N_i} \{ Y_{ij}(1, M_{ij}^{(k)}(1), \mathbf{M}_{i(-j)}^{(k)}(1), \mathbf{m}_i^{(3-k)})
         - Y_{ij}(1, M_{ij}^{(k)}(a), \mathbf{M}_{i(-j)}^{(k)}(0), \mathbf{m}_i^{(3-k)} \} \mid \mathbf{M}_i^{(3-k)}(a')=\mathbf{m}_i^{(3-k)}, \mathbf{C}_i, N_i ] 
        = \E [ \frac{1}{N_i} \sum_{j=1}^{N_i} \{ Y_{ij}(1, M_{ij}^{(k)}(1), \mathbf{M}_{i(-j)}^{(k)}(1), \mathbf{m}_i^{(3-k)}) -\\ Y_{ij}(1, M_{ij}^{(k)}(a), \mathbf{M}_{i(-j)}^{(k)}(0), \mathbf{m}_i^{(3-k)}) \} \mid \mathbf{C}_i, N_i ]$, for $ k \in \{ 1, 2 \}$.
\end{assumption}
Assumption \ref{asmp:si} extends the standard sequential ignorability assumption in \citet{Imai2013} to the context of clustered data with multiple mediators, ruling out unmeasured mediator-outcome confounding. 
Assumption \ref{asmp:cond_homogeneity} extends the identification assumption of \citet{Xia2022} from independent data to CRTs. 
\yo{In the context of the RPS CRT, when $k=1$, this assumption implies that the treatment effect of the CCT program exiting through child health check-ups on nutritional outcomes does not vary within levels of the household dietary diversity after adjusting for all baseline covariates. An analogous interpretation can be made for $k=2$, which concerns the effect that exits through household dietary diversity. 
In technical terms, this assumption only requires that the cluster-average treatment effect mediated through $\mathbf{M}^{(1)}_{i}$ remains unchanged by $\mathbf{M}^{(2)}_{i}$ (and vice versa) after adjusting for baseline covariates. It does not require conditional independence between $\mathbf{M}^{(1)}_{i}$ and $\mathbf{M}^{(2)}_{i}$, nor independence between the potential mediators and potential outcomes. The implication of Assumption \ref{asmp:cond_homogeneity} is further explored under a multivariate normal model in Supplementary Material \ref{sec:implication_assumption5}.}
Under Assumptions \ref{asmp:sutva}--\ref{asmp:cond_homogeneity}, the $\mathrm{EIE}_{\mathrm{C}}^{(k)}$ is point identified.
\begin{theorem}
\label{thm:eie_identification}
        Under Assumption \ref{asmp:sutva}--\ref{asmp:cond_homogeneity}, $ \mathrm{EIE}_{\mathrm{C}}^{(k)} $ are nonparametrically identified as follows: 
        \begin{align*}
             \E_{\mathbf{C}, N} \left[ \frac{1}{N} \sum_{j=1}^{N} \left \{\int_{\mathcal{M}^{(3-k)}} \int_{\mathcal{M}^{(k)}} \mu_{\mathbf{C}, N}(1, \mathbf{m}^{(k)}, \mathbf{m}^{(3-k)}) \right. dF_{\mathbf{M}^{(k)} \mid A=1, \mathbf{C}, N} (\mathbf{m}^{(k)}) dF_{\mathbf{M}^{(3-k)} \mid A=1, \mathbf{C}, N} (\mathbf{m}^{(3-k)}) \right. \\
             \left. - \int_{\mathcal{M}^{(3-k)}} \int_{\mathcal{M}^{(k)}} \mu_{\mathbf{C}, N}(1, \mathbf{m}^{(k)}, \mathbf{m}^{(3-k)})  \left.  dF_{\mathbf{M}^{(k)} \mid A=0, \mathbf{C}, N} (\mathbf{m}^{(k)}) dF_{\mathbf{M}^{(3-k)} \mid A=1, \mathbf{C}, N} (\mathbf{m}^{(3-k)}) \right\}\right],
        \end{align*}
        where $\mu_{\mathbf{C}, N}(a, \mathbf{m}^{(k)}, \mathbf{m}^{(3-k)})=\E \left[ Y_{\cdot j} \mid A=a, \mathbf{M}^{(k)}=\mathbf{m}^{(k)}, \mathbf{M}^{(3-k)}=\mathbf{m}^{(3-k)}, \mathbf{C}, N \right]$.
\end{theorem}

\noindent Furthermore, $\mathrm{INT}_{\mathrm{C}}^{(1,2)}$ is identified as the difference between the identified $\mathrm{NIE}_C$ and the exit effects (Supplementary Material \ref{sec:identification_INT}).
Theorem \ref{thm:eie_identification} gives the g-computation formula to estimate $ \mathrm{EIE}_{\mathrm{C}}^{(k)} $ by specifying the mediator and outcome models for $Y$, $\mathbf{M}^{(1)}$, and $\mathbf{M}^{(2)}$. For example, multilevel parametric regression models that appropriately account for the within-cluster correlations between the observed mediators and outcomes in the same cluster may be applied to derive a plug-in estimator for $ \mathrm{EIE}_{\mathrm{C}}^{(k)}$. 

While Assumptions \ref{asmp:sutva}--\ref{asmp:cond_homogeneity} are sufficient to identify the EIE, an additional assumption is required for the identification of $\mathrm{ESME}_{\mathrm{C}}^{(k)}$ to address the spillover mediation effects.
\begin{assumption}[Cross-world inter-individual mediator independence]
\label{asmp:no_cross_intra_corr}
For all $i$, $j \neq j'$, $a \neq a'$, and $k \in \{1,2\}$, we assume $M_{ij}^{(k)}(a) \indep M_{ij'}^{(k)}(a') \mid \{\mathbf{C}_{i}, N_i \}.$
\end{assumption}

\noindent
\yo{In the RPS CRT, Assumption \ref{asmp:no_cross_intra_corr} implies that, for two different households within the same comarcas, the potential child health check-up conditions and dietary diversity (under two different counterfactuals) are conditionally independent once we account for baseline covariates. This could be plausible because even households that reside in the same area can differ in critical ways---such as socioeconomic status, parents’ educational backgrounds, health-seeking behaviors, cultural norms, and food preferences---resulting in unique child health check-up histories and dietary patterns. Once these baseline differences are properly controlled for, it is reasonable to treat the cross-world health conditions and dietary choices of separate households as independent.}
In technical terms, this assumption assumes away the residual correlation for any two cross-world potential mediators measured from two different individuals in the same cluster after adjusting for the measured within-cluster information, but allows for arbitrary residual correlations between single-world potential values, $M_{ij}^{(k)}(a) $ and $ M_{ij'}^{(k)}(a)$, as well as within-individual cross-world potential values, $M_{ij}^{(k)}(a) $ and $ M_{ij}^{(k)}(a')$. A graphical representation of Assumption \ref{asmp:no_cross_intra_corr} is provided in Supplementary Material Section \ref{sec:discussion_assumptions_estimands_supp}. 
Given Assumptions \ref{asmp:sutva}--\ref{asmp:no_cross_intra_corr},  $\mathrm{ESME}_{\mathrm{C}}^{(k)}$ is point identified as follows.


        
        
        
        
    

\begin{theorem} 
\label{thm:esme_identification}
Under Assumption \ref{asmp:sutva}--\ref{asmp:no_cross_intra_corr}, $\mathrm{ESME}_{\mathrm{C}}^{(k)}$ are nonparametrically identified as follows:
    \begin{align*}
        &\E\left[ \frac{1}{N}\sum_{j=1}^{N} \left \{  \int_{\mathcal{M}^{(3-k)}} \int_{\mathcal{M}^{(k)}} \kappa_{\mathbf{C}, N}(a, m_{\cdot j}, \mathbf{m}^{(k)}_{\cdot (-j)}, \mathbf{m}^{(3-k)}) \right.\right. \\
        & \hspace{10.em} dF_{M^{(k)}_{\cdot j} \mid A=1,\mathbf{C}, N} (m_{\cdot j})dF_{\mathbf{M}^{(k)}_{\cdot, (-j)} \mid A=1,\mathbf{C}, N} ( \mathbf{m}^{(k)}_{\cdot (-j)}) dF_{\mathbf{M}^{(3-k)} \mid A=1,\mathbf{C}, N} (\mathbf{m}^{(3-k)})\\
        & \left.\left. \hspace{3.5em}  -\int_{\mathcal{M}^{(3-k)}} \int_{\mathcal{M}^{(k)}} \kappa_{\mathbf{C}, N}(a, m_{\cdot j}, \mathbf{m}^{(k)}_{\cdot (-j)}, \mathbf{m}^{(3-k)}) \right.\right. \\
        & \left.\left. \hspace{9.5em} dF_{M^{(k)}_{\cdot j} \mid A=1,\mathbf{C}, N} (m_{\cdot j})dF_{\mathbf{M}^{(k)}_{\cdot, (-j)} \mid A=0,\mathbf{C}, N} ( \mathbf{m}^{(k)}_{\cdot (-j)}) dF_{\mathbf{M}^{(3-k)} \mid A=1,\mathbf{C}, N} (\mathbf{m}^{(3-k)}) \right\}\right]
    \end{align*}
    where $\kappa_{\mathbf{C}, N}(a, m_{\cdot j}, \mathbf{m}^{(k)}_{\cdot (-j)}, \mathbf{m}^{(3-k)})=\E\left[  Y_{\cdot  j} \middle| A=1, M_{\cdot j}^{(k)}=m_{\cdot j}, \mathbf{M}_{\cdot(-j)}^{(k)}=\mathbf{m}^{(k)}_{\cdot (-j)},  \mathbf{M}^{(3-k)}=\mathbf{m}^{(3-k)}, \mathbf{C}, N  \right] $.
\end{theorem}


\section{Causal Mediation via Bayesian Nonparametrics}
\label{sec:methodologies}

Although all potential mediators and outcomes are never jointly observed, the identification results established in Section \ref{sec:identification} imply that we can estimate the effects based on functions of the observed data. Consequently, we adopt the Bayesian g-computation approach to obtain the posterior distribution of the causal estimands of interest, and the generic algorithm proceeds as follows: (1) specify models for all observed mediators conditional on covariates and cluster size, the outcomes model conditional on all mediators, covariates, and cluster size, and prior distributions for model parameters, (2) derive the posterior distribution of model parameters and get a draw from their respective posterior distributions, (3) draw a sample from the posterior predictive distributions of the mediators $\mathbf{M}^{(1)}$ and $\mathbf{M}^{(2)}$ given the posterior draws of model parameters, and (4) draw a sample from the posterior predictive distributions of the outcome given the posterior draws of model parameters and mediators. 



\subsection{The Nested Dependent Dirichlet Process Mixtures (nDDPM)}
To mitigate potential bias due to model misspecification, we propose a more flexible Bayesian nonparametric model designed for CRTs, termed as the Nested Dependent Dirichlet Process Mixture (nDDPM) model. The nDDPM model builds upon the nested Dirichlet process (nDP) \citep{Rodriguez2008} by incorporating an additional dependence structure \citep{Maceachern1999}, which flexibly accounts for covariates at both the cluster and individual levels. 
While the nDP offers flexibility in modeling a collection of dependent distributions, it assumes that the distributions are simply exchangeable at both cluster and individual levels. 
To address this limitation, we propose the nested dependent Dirichlet process (nDDP) and its mixture model (nDDPM) to incorporate covariate dependence into the nDP framework to effectively capture heterogeneous distributions that vary with respect to the covariates. The key idea behind the nDDP is to define a set of random measures that are marginally nDP-distributed for every possible combination of covariates $\mathbf{c} = (\mathbf{v}, \mathbf{x}) \in \mathcal{V} \times \mathcal{X} = \mathcal{C} $. The nDDPM uses the nDDP as a prior for the mixing distribution. 
\yo{To formally characterize the nDDP prior, we first present a corollary, as a result from Kolmogorov’s consistency theorem and Sklar's theorem \citep{Sklar1959}, as an intermediate step to introduce a copula-based definition of nDDP. 
\begin{corollary}[Generalization of \citet{Barrientos2012}]
\label{cor:Komogorov_Sklar}
    Let $\Psi_{\mathcal{X}} = \{\psi_{x_1,\dots,x_d} : x_1,\dots,x_d \in \mathcal{X}, d > 1\}$ be a collection of copula functions and $\mathcal{D}_{\mathcal{X}} = \{F_x : x \in \mathcal{X}\}$ a collection of one-dimensional probability distributions defined on a common measurable space $(S,\mathcal{B})$, where $S \subseteq \mathbb{R}$. Assume that for every integer $d > 1$, $x_1,\dots,x_d \in \mathcal{X}$, $u_i \in [0,1]$, $i = 1,\dots,d$, $k \in \{1,\dots,d\}$, and permutation $\pi = (\pi_1,\dots,\pi_d)$ of $\{1,\dots,d\}$, the elements in $\Psi_{\mathcal{X}}$ satisfy the following consistency conditions: (i) $\psi_{x_1,\dots,x_d}(u_1,\dots,u_d) = \psi_{x_{\pi_1},\dots,x_{\pi_d}}(u_{\pi_1},\dots,u_{\pi_d})$, and
    (ii) $\psi_{x_1,\dots,x_d}(u_1,\dots,u_{k-1},1,u_{k+1},\dots,u_d) = \psi_{x_1,\dots,x_{k-1},x_{k+1},\dots,x_d}(u_1,\dots,u_{k-1},u_{k+1},\dots,u_d)$. 
Then there exists a probability space $(\Omega,\mathcal{F},P)$ and a stochastic process $Y : \mathcal{X} \times \Omega \to S,$
such that $P\{\omega \in \Omega : Y(x_1,\omega) \le t_1,\dots,Y(x_d,\omega) \le t_d\} = \psi_{x_1,\dots,x_d}(F_{x_1}(t_1),\dots,F_{x_d}(t_d)),$
for any $t_1,\dots,t_d \in \mathbb{R}$.
\end{corollary}

Let $\Psi_{\mathcal{C}}^{\theta}$ be a set of copulas satisfying the consistency conditions of Corollary \ref{cor:Komogorov_Sklar}.
Let $G_{\mathcal{C}}^0 = \{G_{\mathbf{c}}^0 : \mathbf{c} \in \mathcal{C} \}$ be a set of probability measures defined on a common measurable space $(S, \mathcal{B})$, where $S \subseteq \mathbb{R}^{q}$, $q \in \mathbb{N}$, and $\mathcal{B}$ is the Borel $\sigma$-algebra of $S$. Let $\mathcal{P}(S)$  be the set of all Borel probability measures defined on $(S, \mathcal{B})$. The nDDP model is formally defined.
\begin{definition}\label{def:1}
For any $k \in \mathbb{N}$, let $\{F^{*}_{\mathbf{c},k}: \mathbf{c} \in \mathcal{C}\}$ be a $\mathcal{P}(S)$-valued stochastic process on an appropriate probability space $(\Omega, \mathcal{F}, P)$ such that: (i) $s^{*}_{1}, s^{*}_{2}, \ldots$ are independent random variables of the form $s^{*}_{k}:\Omega \to [0,1]$ for all $k$, with the common Beta distribution with parameter $(1,\alpha)$; (ii) $u^{*}_{1k}, u^{*}_{2k}, \ldots$ are independent random variables of the form $u^{*}_{lk}:\Omega \to [0,1]$ for all $l$, with the common Beta distribution with parameter $(1,\beta)$; (iii) $\boldsymbol{\theta}^{*}_{1k}, \boldsymbol{\theta}^{*}_{2k}, \ldots$ are independent stochastic processes of the form $\boldsymbol{\theta}^{*}_{lk}: \mathcal{C} \times \Omega \to S$ for all $l$, with common finite dimensional distributions determined by the set of copulas $\Psi_{\mathcal{C}}^{\theta}$ and the set of a marginal distribution $G_{\mathcal{C}}^0$; (iv) For every $\mathbf{c} \in \mathcal{C}$, $B \in \mathcal{B}$ and almost every $\omega \in \Omega$,
    \begin{equation*}
        F_k^{*}(\mathbf{c}, \omega)(B) = \sum_{l=1}^{\infty} w^{*}_{lk}(\omega)\delta_{\boldsymbol{\theta}^{*}_{lk}(\mathbf{c}, \omega)(B)},~~~ F(\mathbf{c}, \omega)(B) = \sum_{k=1}^{\infty} \pi^{*}_k(\omega) F_k^{*}(\mathbf{c}, \omega)(B),
    \end{equation*}
where $w^{*}_{lk}(\omega) =  u^{*}_{lk}(\omega)\prod_{i=1}^{l-1}(1-u^{*}_{ik}(\omega))$ and $\pi^{*}_k(\omega) =  s^{*}_{k}(\omega)\prod_{i=1}^{k-1}(1-s^{*}_{i}(\omega))$.
A process $\mathcal{H} = \{ F(\mathbf{c}, \cdot) : \mathbf{c} \in \mathcal{C} \} $ is referred to as the \emph{Atom-Dependent Nested Dependent Dirichlet Process (AD-nDDP)}. 
\end{definition}
}
\noindent
Alternatively, we say that a collection of distributions $\{F_{\mathbf{c},1},\ldots, F_{\mathbf{c},I}\}$ follows an AD-nDDP if $F_{\mathbf{c},i}(\cdot) \sim Q_{\mathbf{c}} \equiv \sum_{k=1}^{\infty} \pi^{*}_k(\omega) \, \delta_{F^{*}_{\mathbf{c},k}(\cdot)},$
where $F_{\mathbf{c}, k}^{*} \doteq F_k^{*}(\mathbf{c}, \cdot)$ and $F_{\mathbf{c}, k} \doteq F_k(\mathbf{c}, \cdot)$.
The key aspect of AD-nDDP is that each element of the collection $\{F^{*}_{\mathbf{c},k}\}_{k=1}^{\infty}$ follows a single-weights DDP \citep{Maceachern1999,Barrientos2012}.
In Definition \ref{def:1}, we introduce covariate dependence only through the atoms of $F^{*}_{\mathbf{c},k}$, while the weights $w^{*}_{lk}$ remain free of the covariates. 
We denote $\{F_{\mathbf{c},1}, \ldots, F_{\mathbf{c},I}\} \sim \text{AD-nDDP}(\alpha, \beta, \Psi_{\mathcal{C}}^{\theta}, G^0_{\mathbf{c}})$ to indicate that the collection $\{F_{\mathbf{c},1}, \ldots, F_{\mathbf{c},I}\}$ marginally follows the AD-nDDP for every possible value of covariates $\mathbf{c} \in \mathcal{C}$.  
\yo{In the nDDPM model, common nested stick-breaking representations are used across all clusters and individuals. Specifically, each cluster is allocated to a top-level distribution from a global stick-breaking process, then has its own second-level stick-breaking structure for individual mixtures, while the atom parameters vary with covariates. This setup yields a unified global partitioning scheme while still allowing ample heterogeneity among clusters. 
The global mixture weights remain shared across clusters, ensuring partial exchangeability and stable borrowing of information at the cluster level. The top-level partitioning scheme in the nDDPM creates a global ``menu'' of distributions from which each cluster’s distribution is drawn. 
This mechanism determines how clusters group or separate in their overall distributional shapes, before further variation is introduced at the individual level.
The top-level partitioning is assumed to be exchangeable, leaving any covariate-driven heterogeneity at the cluster level is therefore handled elsewhere, typically through the second-level atoms or lower-level modeling structure (e.g., the AD-nDDPM’s covariate-dependent atoms).}

Beyond Definition \ref{def:1}, one can further incorporate cluster-level covariates into the weights $\pi^{*}_k$, which govern the allocation of probability measures to clusters---that is, defining $\pi^{*}_k = \pi^{*}_k(\mathbf{v})$ as functions of the features through the generalized stick-breaking processes (e.g., \citet{Dunson_2008}). 
While keeping $F^{*}_{\mathbf{c},k}$ as a single-weights DDP, this approach allows the weights $\pi^{*}_k$ to depend on cluster-level covariates, capturing heterogeneity that depends on cluster characteristics. 
By allowing both the weights $\pi^{*}_k$ and the atoms $\boldsymbol{\theta}^{*}_{lk}(\mathbf{c})$ to depend on covariates, we obtain a more flexible modeling framework that can capture complex data structures at both cluster and individual levels. 
We refer to this model as the \textit{Fully-Dependent Nested Dependent Dirichlet Process (FD-nDDP)}.  \yo{The detailed definition of the FD-nDDPM is provided in the supplementary material Section \ref{sec:FD_nDDPM}, along with its statistical properties. Figure \ref{fig:plate_diagrams_nDDPs} illustrates the stick-breaking representation for the nDP and the proposed nDDP models. The FD-nDDP model is expected to gain additional inferential efficiency compared to the AD-nDDP model when the clusters exhibit substantial heterogeneity, and the cluster-level covariates effectively capture that heterogeneity. }

\begin{figure}[!ht]
    \centering
\resizebox{0.9\textwidth}{!}{ 
\begin{tikzpicture}[
  >=stealth,                 
  node distance=1.2cm,  
  latent/.style={circle, draw=black, minimum size=8mm},
  param/.style={circle, draw=black, fill=gray!10, minimum size=8mm},
  observed/.style={circle, draw=black, fill=gray!30, minimum size=8mm},
  plate/.style={rectangle, draw=black, rounded corners, inner sep=6pt, label={[anchor=south east]south east:#1}}
]

\newcommand{\drawnDP}{
  \node[param] (aalpha) {$a_\alpha$};
  \node[param, right=0.15cm of aalpha] (balpha) {$b_\alpha$};
  \node[latent, below=0.4cm of aalpha] (alpha) {$\alpha$};

  \draw[->] (aalpha) -- (alpha);
  \draw[->] (balpha) -- (alpha);

  \node[param, right=1.2cm of aalpha] (Abeta) {$a_\beta$};
  \node[param, right=0.15cm of Abeta] (Bbeta) {$b_\beta$};
  \node[latent, below=0.4cm of Abeta] (beta) {$\beta$};

  \draw[->] (Abeta) -- (beta);
  \draw[->] (Bbeta) -- (beta);

  \node[param, right=1.35cm of Abeta] (G0) {$G^0$};

  \node[latent, below=1.4cm of beta] (Wk) {$w_{lk}^{*}$};  
  \node[latent, right=1.2cm of Wk] (thetastar) {$\theta_{lk}^*$};
  \node[latent, below=1.6cm of alpha] (pi) {$\pi_k$};

  \draw[->] (alpha) -- (pi);
  \draw[->] (beta) -- (Wk);
  \draw[->] (G0) -- (thetastar);

  \node[plate={$l = 1,\dots,\infty$}, minimum width=3cm, minimum height=2.5cm] 
    (plateL) [fit=(Wk)(thetastar)] {};

  \node[plate={$k = 1,\dots,\infty$}, minimum width=6.2cm, minimum height=3.8cm] 
    (plateK) [fit=(pi)(plateL)] {}; 

  \node[latent, below=5.2cm of pi] (zetai) {$\zeta_i$};
  \node[latent, below=3.5cm of Wk] (xii) {$\xi_{ij}$};  
  \node[latent, right=1.2cm of xii] (thetaITi) {$\theta_{\zeta\xi}$};
  \node[observed, below=0.4cm of thetaITi] (Yij) {$Y_{ij}$};

  \draw[->] (pi) -- (zetai);
  \draw[->] (Wk) -- (xii);
  \draw[->] (thetastar) -- (thetaITi);
  \draw[->] (zetai) -- (thetaITi);
  \draw[->] (xii) -- (thetaITi);
  \draw[->] (thetaITi) -- (Yij);

  \node[plate={$j = 1,\dots,N_i$}, minimum width=3.cm, minimum height=4.0cm] 
    (plateJ) [fit=(thetaITi)(xii)(Yij)] {};

  \node[plate={$i = 1,\dots,I$}, minimum width=6.2cm, minimum height=5.5cm] 
    (plateI) [fit=(zetai)(plateJ)] {};
}

\newcommand{\drawADnDDP}{
  \node[param] (aalpha) {$a_\alpha$};
  \node[param, right=0.15cm of aalpha] (balpha) {$b_\alpha$};
  \node[latent, below=0.4cm of aalpha] (alpha) {$\alpha$};

  \draw[->] (aalpha) -- (alpha);
  \draw[->] (balpha) -- (alpha);

  \node[param, right=1.2cm of aalpha] (Abeta) {$a_\beta$};
  \node[param, right=0.15cm of Abeta] (Bbeta) {$b_\beta$};
  \node[latent, below=0.4cm of Abeta] (beta) {$\beta$};

  \draw[->] (Abeta) -- (beta);
  \draw[->] (Bbeta) -- (beta);

  \node[latent, below=1.4cm of beta] (Wk) {$w_{lk}^{*}$};  
  \node[latent, right=1.2cm of Wk] (thetastar) {$\theta_{lk}^*(c)$};
  \node[latent, below=1.6cm of alpha] (pi) {$\pi_k$};

  \node[param, above=2.35cm of thetastar] (G0) {$G^0_c$};

  \draw[->] (alpha) -- (pi);
  \draw[->] (beta) -- (Wk);
  \draw[->] (G0) -- (thetastar);

  \node[plate={$l = 1,\dots,\infty$}, minimum width=3cm, minimum height=2.5cm] 
    (plateL) [fit=(Wk)(thetastar)] {};

  \node[plate={$k = 1,\dots,\infty$}, minimum width=6.2cm, minimum height=3.8cm] 
    (plateK) [fit=(pi)(plateL)] {}; 

  \node[latent, below=5.0cm of pi] (zetai) {$\zeta_i$};
  \node[latent, below=3.3cm of Wk] (xii) {$\xi_{ij}$};  
  \node[latent, right=1.2cm of xii] (thetaITi) {$\theta_{\zeta\xi}(c)$};
  \node[observed, below=0.4cm of thetaITi] (Yij) {$Y_{ij}$};
  \node[observed, left=0.5cm of Yij] (Xij) {$X_{ij}$};
  \node[observed, below=0.2cm of zetai] (Vi) {$V_{i}$};

  \draw[->] (pi) -- (zetai);
  \draw[->] (Wk) -- (xii);
  \draw[->] (thetastar) -- (thetaITi);
  \draw[->] (zetai) -- (thetaITi);
  \draw[->] (xii) -- (thetaITi);
  \draw[->] (thetaITi) -- (Yij);
  \draw[->] (Xij) -- (thetaITi);
  \draw[->] (Vi) -- (thetaITi);

  \node[plate={$j = 1,\dots,N_i$}, minimum width=3.cm, minimum height=4.cm] 
    (plateJ) [fit=(thetaITi)(xii)(Yij)] {};

  \node[plate={$i = 1,\dots,I$}, minimum width=6.2cm, minimum height=5.5cm] 
    (plateI) [fit=(zetai)(plateJ)(Vi)] {};
}

\newcommand{\drawFDnDDP}{
  \node[param] (aalpha) {$a_\alpha$};
  \node[param, right=0.15cm of aalpha] (balpha) {$b_\alpha$};
  \node[latent, below=0.4cm of aalpha] (alpha) {$\alpha_v$};

  \draw[->] (aalpha) -- (alpha);
  \draw[->] (balpha) -- (alpha);

  \node[param, right=1.2cm of aalpha] (Abeta) {$a_\beta$};
  \node[param, right=0.15cm of Abeta] (Bbeta) {$b_\beta$};
  \node[latent, below=0.4cm of Abeta] (beta) {$\beta$};

  \draw[->] (Abeta) -- (beta);
  \draw[->] (Bbeta) -- (beta);

  \node[latent, below=1.4cm of beta] (Wk) {$w_{lk}^{*}$};  
  \node[latent, right=1.2cm of Wk] (thetastar) {$\theta_{lk}^*(c)$};
  \node[latent, left=0.9cm of Wk] (pi) {$\pi_k(v)$};
  \node[latent, left=1.1cm of pi] (Gamma) {$\Gamma_k$};

  \node[param, above=2.4cm of thetastar] (G0) {$G^0_c$};

  \node[param, above=2.75cm of Gamma] (mu_gamma) {$\mu_{\Gamma}$};
  \node[param, right=.2cm of mu_gamma] (sigma_gamma) {$\Sigma_{\Gamma}$};

  \draw[->] (alpha) -- (pi);
  \draw[->] (beta) -- (Wk);
  \draw[->] (G0) -- (thetastar);
  \draw[->] (Gamma) -- (pi);
  \draw[->] (mu_gamma) -- (Gamma);
  \draw[->] (sigma_gamma) -- (Gamma);

  \node[plate={$l = 1,\dots,\infty$}, minimum width=3cm, minimum height=2.5cm] 
    (plateL) [fit=(Wk)(thetastar)] {};

  \node[plate={$k = 1,\dots,\infty$}, minimum width=8.4cm, minimum height=3.8cm] 
    (plateK) [fit=(pi)(plateL)(Gamma)] {}; 

  \node[latent, below=4.2cm of pi] (zetai) {$\zeta_i(v)$};
  \node[latent, below=3.3cm of Wk] (xii) {$\xi_{ij}$};  
  \node[latent, right=1.2cm of xii] (thetaITi) {$\theta_{\zeta\xi}(c)$};
  \node[observed, below=0.4cm of thetaITi] (Yij) {$Y_{ij}$};
  \node[observed, left=0.5cm of Yij] (Xij) {$X_{ij}$};
  \node[observed, below=0.25cm of zetai] (Vi) {$V_i$};

  \draw[->] (pi) -- (zetai);
  \draw[->] (Wk) -- (xii);
  \draw[->] (thetastar) -- (thetaITi);
  \draw[->] (zetai) -- (thetaITi);
  \draw[->] (xii) -- (thetaITi);
  \draw[->] (thetaITi) -- (Yij);
  \draw[->] (Xij) -- (thetaITi);
  \draw[->] (Vi) -- (zetai);
  \draw[->] (Vi) -- (thetaITi);

  \node[plate={$j = 1,\dots,N_i$}, minimum width=3.cm, minimum height=4.0cm] 
    (plateJ) [fit=(thetaITi)(xii)(Yij)] {};

  \node[plate={$i = 1,\dots,I$}, minimum width=6.2cm, minimum height=5.5cm] 
    (plateI) [fit=(zetai)(plateJ)(Vi)] {};
}

\begin{scope} 
\node[anchor=south east, font=\bfseries] at (1.0, 0.8) {(a) nDP};
  \drawnDP
\end{scope}


\begin{scope}[shift={(6.6,0)}]  
\node[anchor=south east, font=\bfseries] at (2.3, 0.8) {(b) AD-nDDP};
  \drawADnDDP
\end{scope}


\begin{scope}[shift={(15.8,0)}]  
\node[anchor=south east, font=\bfseries] at (-0.2, 0.8) {(c) FD-nDDP};
  \drawFDnDDP
\end{scope}

\end{tikzpicture}
}
\caption{\yo{Plate diagrams for the stick-breaking representations of the nDP, AD-nDDP, and FD-nDDP models. The upper panels depict the model structure, while the lower panels represent the observational process. The outer panels correspond to cluster-level components, whereas the inner panels illustrate individual-level models and observations. The subscripts of the atom $ \theta_{\zeta\xi} $ indicate that it is indexed by the latent cluster-level and individual-level class indicators, $ \zeta_i $ and $ \xi_{ij} $. In the proposed nDDP models, the atoms in the upper panels, $ \theta_{\zeta\xi}(c) $, are stochastic processes indexed by any value of $ c $ and are generated from the base measure $ G_c^0 $. At the observation level, the outcomes are assumed to be generated from the corresponding processes indexed by the covariates observed for each cluster and individual.}}
\label{fig:plate_diagrams_nDDPs}
\end{figure}

\yo{
\subsubsection{Basic properties}
\label{sec:properties_ADnDDPM}
We here discuss the statistical properties of the AD-nDDP models. As the AD-nDDP is defined as a set of random measures that are marginally nDP-distributed for every possible combination of covariates $\mathbf{c} \in \mathcal{C}$, many properties of the nDP model hold for a given $\mathbf{c}$. First, the following properties hold for any $i$, $\mathbf{c}$, and measurable set $A \in \mathcal{B}$, 
\begin{equation}
\label{eq:expectation_variance}
    \E[F_{\mathbf{c}, i}(A)] = G_{\mathbf{c}}^0(A) \text{ and } \Var[F_{\mathbf{c}, i}(A)] = \frac{G_{\mathbf{c}}^0(A)(1-G_{\mathbf{c}}^0(A))}{\beta + 1}.
\end{equation}
The proof of \eqref{eq:expectation_variance} is provided in the Supplementary Material. 
For a given set $A \in \mathcal{B}$, the nDDP prior implies the following correlation structure among a collection of random variables $\{F_{\mathbf{c}, 1}(A), \ldots, F_{\mathbf{c}, I}(A) : \mathbf{c} \in \mathcal{C}\}$.
\begin{proposition} [Correlation analysis]
\label{prop:correlation_measures}
For any $\mathbf{c}, \mathbf{c}'  \in \mathcal{C}$, cluster $i,j \in \{1,\ldots,I\}$, any measurable sets $A,B \in \mathcal{B}$, with $\rho_{\mathbf{c},\mathbf{c}'}(A,B) = P\left\{ \boldsymbol{\theta}_{lk}^{*}(\mathbf{c}) \in A, \boldsymbol{\theta}_{lk}^{*}(\mathbf{c}') \in B  \right\}$, we have
    \begin{equation*}
        \Corr(F_{\mathbf{c}, i}(A), F_{\mathbf{c}', j}(B)) = \begin{cases}
            \displaystyle\frac{\rho_{\mathbf{c},\mathbf{c}'}(A,B) - G_{\mathbf{c}}^0(A)G_{\mathbf{c}'}^0(B)}{ \sqrt{G_{\mathbf{c}}^0(A)(1-G_{\mathbf{c}}^0(A))G_{\mathbf{c}'}^0(B)(1-G_{\mathbf{c}'}^0(B))}},  \hfill \text{ if } i=j\\
            \displaystyle\frac{\rho_{\mathbf{c},\mathbf{c}'}(A,B) - G_{\mathbf{c}}^0(A)G_{\mathbf{c}'}^0(B)}{(\alpha+1) \sqrt{G_{\mathbf{c}}^0(A)(1-G_{\mathbf{c}}^0(A))G_{\mathbf{c}'}^0(B)(1-G_{\mathbf{c}'}^0(B))}},  \hfill \text{ if } i \neq j.
        \end{cases}
    \end{equation*}
\end{proposition}
\noindent
The proof is provided in the Supplementary Material. The joint distribution $\rho_{\mathbf{c},\mathbf{c}'}(A,B)$ is characterized by the set of copulas $\Psi_{\mathcal{C}}^{\theta}$ and the set of marginals $G_{\mathcal{C}}^0$.
The correlation is useful for investigating the dependence among random probability measures. In particular, the correlation for the distributions of units with $\mathbf{c}$ and $\mathbf{c}'$ in the \textit{same} cluster is greater (by a factor of $\frac{1}{\alpha+1}$) than for the units in \textit{different} clusters. Furthermore, if we ignore the covariate dependence, i.e, setting $\mathbf{c} = \mathbf{c}'$, we can recover the correlation of the nDP given in \citet{Rodriguez2008}, as a special case. Within the same cluster, the correlation approaches $1$ for a common measurable set $A \in \mathcal{B}$ as $\mathbf{c} \to \mathbf{c}'$.

Let $\mathcal{P}(S)^{\mathcal{C}}$ be the set of all $\mathcal{P}(S)$-valued functions defined on $\mathcal{C}$ and $\mathcal{B}(\mathcal{P}(S)^{\mathcal{C}})$ be the Borel $\sigma$-algebra generated by the product topology of weak convergence. The support of the nDDP is the smallest closed set in $\mathcal{B}(\mathcal{P}(S)^{\mathcal{C}})$  with $P \circ \mathcal{H}^{-1}$-measure one. Assume that $\Theta \subseteq S$ is the common support of $G_{\mathbf{c}}^0$ for every $\mathbf{c} \in \mathcal{C}$. We give sufficient conditions for the full weak support of the AD-nDDP, where $\mathcal{P}(\Theta)^{\mathcal{C}}$ is the set of all $\mathcal{P}(\Theta)$-valued functions defined on $\Theta$, with $\mathcal{P}(\Theta)$ being the set of all probability measures defined on $(\Theta, \mathcal{B}(\Theta))$.
\begin{proposition}[Weak support property]
    \label{prop:weak_support}
    Let $\{ F_{\mathbf{c},i} : \mathbf{c} \in \mathcal{C}\}$ be an AD-nDDP for each $i$. If $\Psi_{\mathcal{C}}^{\theta}$ is a collection of copulas with positive density with respect to Lebesgue measure, on the appropriate Euclidean space, then $\mathcal{P}(\Theta)^{\mathcal{C}}$ is the weak support of the process.
\end{proposition}
\noindent
The proof is provided in the Supplementary Material. This proposition indicates that the AD-nDDP prior assigns positive probability to every weak neighborhood of any collection of distributions, implying its theoretical robustness for approximating any true collection of distributions arbitrarily well. Consequently, even if the true data‐generating process is complex, the posterior can still concentrate around the true distributions as more data become available. Hence, in practice, letting only the atoms depend on covariates (AD-nDDP) already provides sufficient theoretical flexibility. Similar properties of the FD-nDDP, including its correlation structure and weak support property, have also been established and are provided in Supplementary Material Section \ref{sec:FD_nDDPM}. 
}

\subsubsection{Model specifications}
\label{sec:model_specification}
We propose to model the observed mediators and outcomes using the AD-nDDPM. For continuous outcomes and mediators, we posit the model for unit $j$ in cluster $i$ as follows:
\begin{equation}
\label{eq:nDDPMmodel}
    \begin{split}
        Y_{ij} \mid \mathbf{C}^{y}_{ij}=\mathbf{c}_y &\sim \int_{\Theta} \mathrm{N}\left( \cdot \mid \boldsymbol{\theta}^{\top}\mathbf{c}_y, \sigma^2 \right) dF_{\mathbf{c}_y,i}(\boldsymbol{\theta},\sigma)\\ 
        M_{ij}^{(1)},M_{ij}^{(2)} \mid \mathbf{C}^{m}_{ij}=\mathbf{c}_m &\sim \int \mathrm{MVN}\left( \cdot \mid (\boldsymbol{\gamma}_1^\top \mathbf{c}_m,  \boldsymbol{\gamma}_2^\top \mathbf{c}_m)^\top, \boldsymbol{\Sigma} \right) dF_{\mathbf{c}_m,i}(\boldsymbol{\gamma}_{1}, \boldsymbol{\gamma}_{2}, \boldsymbol{\Sigma} )\\
        F_{\mathbf{c}_y,i} \sim \text{AD-nDDP}&(\alpha_{y}, \beta_{y}, \Psi_{\mathcal{C}}^{\theta}, G_{\mathbf{c}_{y}}^0),~~~
        F_{\mathbf{c}_m,i} \sim \text{AD-nDDP}(\alpha_{m}, \beta_{m}, \Psi_{\mathcal{C}}^{\theta}, G_{\mathbf{c}_{m}}^0),     
        \end{split}
\end{equation}
where $\alpha_{y}, \alpha_{m} \sim \mathrm{Ga}(a_{\alpha},b_{\alpha})$, 
$\beta_{y}, \beta_{m}  \sim \mathrm{Ga}(a_{\beta},b_{\beta})$, 
$G_{\mathbf{c}_{y}}^0 = \mathrm{MVN}(\boldsymbol{\mu}_0,\boldsymbol{\Sigma}_0)\mathrm{IG}(a_0,b_0)$, 
$G_{\mathbf{c}_{m}}^0 = \mathrm{MVN}(\mathbf{m}_0,\mathbf{S}_0)\mathrm{MVN}(\mathbf{m}_0,\mathbf{S}_0)\mathrm{IW}(\nu_0, \boldsymbol{\Psi}_0)$, 
$\mathbf{C}^{y}_{ij}=[A_i, g_{ij}^{m}(\mathbf{M}_{i}), g_{ij}^{x}(\mathbf{X}_{i}), \mathbf{V}_i^\top, N_i]^\top$,
$\mathbf{C}^{m}_{ij}=[A_i, g_{ij}^{x}(\mathbf{X}_{i}), \mathbf{V}_i^\top, N_i]^\top$ with $g_{ij}^{m}(\cdot)$ and $g_{ij}^{x}(\cdot)$ being fixed dimensional summary functions of $\mathbf{M}_{i}$ and $\mathbf{X}_{i}$, and 
$\Psi_{\mathcal{C}}^{\theta}$ is the independence copula.
Since the dimensions of $ \mathbf{M}_{i} $ and $ \mathbf{X}_{i} $ can vary across clusters in CRTs, we consider adjusting for summary functions with fixed dimensions in the models. This is a practical approach to model between-unit interference in regression models  \citep{VanderWeele2013_CRT,Ogburn2024,cheng2024}.
In this article, we consider a bivariate summary function
$g_{ij}^{m}(\mathbf{M}_{i})=\left\{ M_{ij}, \frac{1}{N_i - 1} \sum_{\substack{k=1 \\ k \ne j}}^{N_i} M_{ik} \right\}$
of $ \mathbf{M}_{i} $, such that $ Y_{ij} $ is assumed to be affected by $ \mathbf{M}_{i} $ via one's own mediator and the average mediator values of other same-cluster members. Similarly, $g_{ij}^{x}(\mathbf{X}_{i})=\left\{ X_{ij}, \frac{1}{N_i - 1} \sum_{\substack{k=1 \\ k \ne j}}^{N_i} X_{ik} \right\}$ can be used as a summary for $ \mathbf{X}_{i} $. We can enrich the functions by adding more terms.

There are several considerations for specifying the AD-nDDPM.
Firstly, we must decide between employing a single common AD-nDDPM prior and using independent AD-nDDPM priors for the mediator and outcome models. While a single common AD-nDDPM prior can be applied to both models by taking the product of $ G_{\mathbf{c}_{y}}^0 $ and $ G_{\mathbf{c}_{m}}^0 $ as the base measure, we opt for independent priors instead. This approach provides greater flexibility by allowing each model to have a different number of distributional clusters. 

Secondly, we need to choose appropriate mixture kernels for the outcome and mediator models. In the simplest case, where both the outcome and mediators are continuous variables, it is reasonable to select Gaussian kernels. Generally, when multiple mediators are measured for each individual, they are likely to be correlated, and ignoring this correlation can lead to invalid inferences. To account for the correlation structure between $ M_{ij}^{(1)} $ and $ M_{ij}^{(2)} $, we use a multivariate Gaussian kernel for the mediators. When the mediators (or, outcome) are dichotomous or polychotomous---for example, $ M_{ij}^{(1)} \in \{0,1\} $---we introduce a Gaussian latent variable $ Z^{(1)}_{ij} \sim \mathrm{N}(\mu_{z}, \sigma^2_{z}) $ for the discrete mediator. We then model the correlation among the latent variables (representing the discrete mediators) and the observed continuous mediators using a multivariate kernel \citep{Albert1993,Chib1998}. 
We detail this modeling strategy in Supplementary Material \ref{sec:extension_discrete_gibbs}.

\yo{Finally, the specification of processes $ \{\boldsymbol{\theta}_{lk}(\mathbf{c}_y) : \mathbf{c}_y \in \mathcal{C}_y\}$  provides analysts with a degree of flexibility in modeling. As discussed in  Section \ref{sec:properties_ADnDDPM}, stochastic processes can be defined with given marginal distributions via copula functions, specifying the dependence structures between the marginals with $\mathbf{c}_y$ and $\mathbf{c}_y'$ by choosing an appropriate copula function $\Psi_{\mathcal{C}}^{\theta}$.
In this article, we assume the atoms are indexed by a finite-dimensional parameter vector, i.e., linear models, and assume $\Psi_{\mathcal{C}}^{\theta}$ to be the independence copula. Specifically, we consider $ \boldsymbol{\theta}_{lk}(\mathbf{c}_y) = (\boldsymbol{\theta}_{lk}^\top \mathbf{c}_y, \sigma_{lk}^2) $, i.e., the mean function of the Gaussian kernel to be a linear combination of the covariates, while the variance of the kernel remains independent of the covariates. The atoms for mediators $ \{\boldsymbol{\gamma}_{lk}(\mathbf{c}_m) : \mathbf{c}_m \in \mathcal{C}_m\}$ are specified in the same manner. Additional flexibility may be obtained by incorporating covariates into the variance function or by specifying different dependence structures. 
}

It is worth mentioning that \citet{Ho2013} applied the nDP to infer the average treatment effects in CRTs without any intermediate variables. However, they only placed an nDP prior on the random effects of a linear mixed model, allowing for potential interactions between clusters and individuals within clusters. While their model relaxes the distributional assumption on the random effects in the linear mixed model, it still makes strong structural assumptions about how the parametric fixed effects are correlated with the outcome (i.e., linearity assumption). In contrast, our model is intrinsically functional, placing nDDPM priors on the functional space of the outcome and mediator models.
\yo{In particular, our nDDPM model is expected to capture non-linearity by partitioning the input space into regions, with each region explained by a local linear model. The flexibility of the nDDPM arises from its ability to allow the mixture components to vary with covariates or other inputs, enabling the model to dynamically adapt to different regions of the input space. This is similar to how splines approximate non-linear functions, but in nDDPM, the partitioning and linear models are driven by a probabilistic framework that incorporates dependence structures, allowing for smooth transitions between mixture components. 
Additionally, Proposition \ref{prop:weak_support} provides a theoretical justification that the AD-nDDP can approximate any distributions arbitrarily well. This ensures the model can flexibly adapt to both global and local patterns in the data,} 
corresponding to a broader set of data-generating processes and providing greater flexibility in modeling clustered observations in CRTs.

\subsection{Posterior inference}

We employ an approximated blocked Gibbs sampler \citep{Ishwaran2000} based on a two-level truncation of the stick-breaking representation of the DP proposed by \citet{Rodriguez2008}.
This algorithm proceeds by first selecting conservative upper bounds on the number of latent classes at both cluster-level $K_C$ and individual-level within each cluster $K_I$. 
Let $\zeta_i \in \{1,\ldots,K_C\}$ and $\xi_{ij} \in \{1,\ldots,K_I\}$ denote the latent class indicators for the cluster $i$ and individual $j$ therein; that is, the cluster-level indicator and the individual-level indicator, respectively.
We specify Multinomial distributions $\zeta_i \sim \mathrm{MN}(\boldsymbol{\pi}^{*})$ on $\zeta_i$ and $\xi_{ij} \sim \mathrm{MN}(\mathbf{w}^{*}_{\zeta_i})$ on $\xi_{ij}$,  where $\boldsymbol{\pi}^{*} = (\pi^{*}_{1},\ldots,\pi^{*}_{K_C})^\top$  and  $\mathbf{w}^{*}_{\zeta_i} = (w^{*}_{1\zeta_i},\ldots,w^{*}_{K_{I}\zeta_i})^\top$ contain the weights from the AD-nDDPM. 
\citet{Rodriguez2008} demonstrated that an accurate approximation to the exact DP is obtained as long as the truncation bound is sufficiently large. 
To ensure this, we ran several MCMC iterations with different values of $K_C$ and $K_I$ and increased them after an iteration if all clusters were occupied. We terminated this process when the number of occupied clusters was less than $K_C$ and $K_I$. 

\yo{In the absence of strong prior knowledge or external contextual information, we consider proper, weakly informative conjugate priors for all model parameters. For the AD-nDDPM, the posterior updates are analytically tractable, admitting fully Gibbs-sampled inference. In contrast, the FD-nDDPM involves covariate-dependent mixture weights, which preclude closed-form updates for certain parameters. We hence employ Metropolis–Hastings steps within the Gibbs sampler for posterior inference, incurring additional computational cost due to the evaluation of acceptance ratios at each iteration. Full details of the prior specifications and sampling algorithms are provided in Supplementary Section \ref{sec:gibbs_details}.}

\section{Simulation Studies}
\label{sec:simulations}

\label{sec:simulation_designs}
We examine the performance of the proposed methods in estimating mediation effects in CRTs.
Specifically, we evaluate the frequentist properties of the proposed Bayesian methods for estimating the NIE, EIE and ESME, \yo{with comparisons to their variations when the mediator and outcome are represented via frequestist linear mixed models (LMMs)---the most popular approach for analyzing CRTs in health sciences \citep{wang2025mixed}---and via the nDP model proposed in \citet{Ho2013}.
In particular, the model of \citet{Ho2013} can be viewed as a linear mixed model with random effects represented by nDP, allowing for potential interactions between the cluster and the individuals within the cluster and relaxing the distributional assumption on the random effects. We estimate each mediation effect estimand using our derived nonparametric identification results, applying them to summarize the model fit using either the LMM or the nDP approach. Notably, each comparison method can be incorporated into our identification formulas which are derived under a common set of causal assumptions to target common causal estimands. However, these methods differ in their underlying modeling assumptions, which in turn affect their flexibility and capacity to capture complex data-generating mechanisms.}

For our proposed Bayesian methods, we use the model \eqref{eq:nDDPMmodel} described in Section \ref{sec:model_specification} with proper, weakly informative conjugate priors. \yo{Specifically, $\boldsymbol{\theta}_{lk} \sim \mathrm{MVN}(\mathbf{0}, 10^2 \times I_{d_y})$, $\sigma^2_{lk} \sim \mathrm{IG}(2.0, 1.0)$, $\boldsymbol{\gamma}_{1,lk}, \boldsymbol{\gamma}_{2,lk} \sim \mathrm{MVN}(\mathbf{0}, 10^2 \times I_{d_m})$, $\boldsymbol{\Sigma} \sim \mathrm{IW}(2.0, I_2)$, $\alpha_y \sim \mathrm{Ga}(1.0,1.0)$, $\alpha_m \sim \mathrm{Ga}(1.0,1.0)$, $\beta_y \sim \mathrm{Ga}(1.0,1.0)$, and $\beta_m \sim \mathrm{Ga}(1.0,1.0)$, where $I_{d}$ is the identity matrix of dimension $d$.} The initial parameter values were randomly drawn from the prior distributions, and the posterior samples were obtained by running a chain for $2000$ MCMC iterations after an initial $1000$ burn-in iteration. 
Convergence was monitored by the trace plots of the sampled parameters, confirming that the chains had reached stationarity and exhibited good mixing without apparent trends or autocorrelation.

We simulate $100$ datasets and evaluate the bias and root mean square error (RMSE) of a point estimator, as well as frequentist coverage and interval length of the confidence/credible interval estimator. \yo{The evaluation of the interval estimators is provided in the supplementary material \ref{sec:results_tables}.} For the LMM, we used the cluster bootstrap to construct the confidence intervals. \yo{Although the posterior mode is a commonly used alternative, we report the posterior mean as the point estimator due to its computational tractability under complex mixture models and its Bayes optimality under the squared error loss.}

\yo{For the data-generating process, we consider three scenarios, each with increasing complexity, ranging from Scenario 1 to Scenario 3. Key features of simulation scenarios and complete descriptions of the data-generating processes for each scenario are provided in the supplementary material \ref{sec:simulation_details}.
We started with Scenario 1, a baseline linear mixed model with normally distributed mediators and outcomes, moderate intra-cluster correlation, and a random intercept for each cluster. This setup provides a relatively straightforward case with linear relationships and medium effect sizes ($\approx 1.0$), and moderate outcome noise $\sigma = 1.0$.
Scenario 1 features a single linear mixed model without latent classes, whereas Scenarios 2 and 3 incorporate mixtures of normal components and multiple latent classes at both cluster and individual levels. The key difference between Scenarios 2 and 3 is that class membership probabilities in Scenario 3 depend on the cluster size, inducing varying effect sizes and noise levels across components. In all scenarios, we capture partial interference by including cluster-level averages of mediators in the outcome. Overall, these designs yield moderate signal-noise ratios, reflecting effect sizes and error variances akin to real-world CRTs. We generate $I=100$ clusters.
Finally, we approximate the true causal estimands using a Monte Carlo simulation approach by generating and averaging the potential outcomes for a super-population of $3,000,000$ clusters.}

\subsection{Simulation results}
\label{sec:baseline_simulation}

\begin{figure*}
    \centering
    \includegraphics[width=\textwidth]{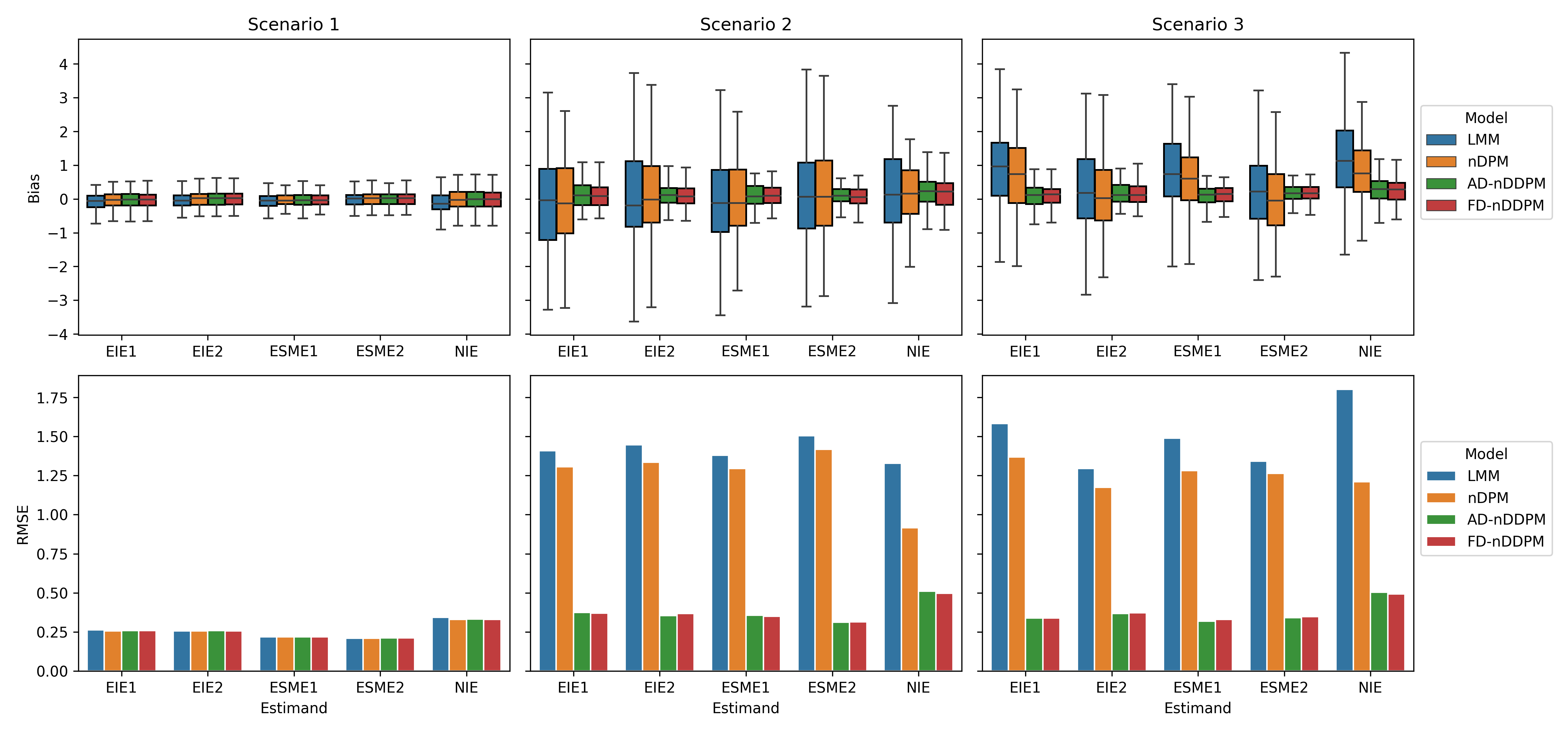}
    \caption{Boxplots of bias and bar charts of root mean squared error (RMSE) for point estimates of the key estimands across three baseline scenarios in Simulation 1. The bias boxplots summarize the distribution over 100 simulation replicates.}
    \label{fig:simulation1_bias_mse_boxplot}
\end{figure*}

Figure \ref{fig:simulation1_bias_mse_boxplot} reports the results for all scenarios. Overall, the results consistently show that our nDDPM methods have the smallest bias and MSE across scenarios. In particular, under Scenario 1, the LMM and the nDP are expected to perform well because they are correctly specified. Even in this simplest scenario, our nDDPM model exhibits similar performance to them in estimating the EIEs, ESMEs, and NIE without notable efficiency loss. 
\yo{If many effects are near-linear or additive, there may be some loss of statistical power compared to a well-specified parametric model. This potential power loss is a common concern for nonparametric models. However, it is also important to note that parametric models can be difficult to specify correctly, especially when there are many additive terms and non-linear effects. Our primary motivation is to address these potential model misspecifications.} 
Under Scenarios 2 and 3, our methods significantly outperform the other methods in terms of both bias and RMSE for the EIEs, ESMEs, and NIE, and the nDP shows superior performance over the simple LMM in all metrics. 
Additionally, as shown in Supplementary Material Section \ref{sec:additonal_simulation}, our methods yield much shorter interval lengths compared to competing methods while still achieving well-calibrated frequentist coverage probabilities for all estimands. 

\yo{
We also observe that the FD-nDDPM performs similarly to, or sometimes slightly underperforms relative to, the AD-nDDPM. This observation aligns with the robustness result of the AD-nDDP prior presented in Proposition~\ref{prop:weak_support}. In the FD-nDDPM, cluster-level covariates are explicitly used to inform how clusters are partitioned at the top level, which increases flexibility but also adds complexity. When the number of clusters is small, this complexity may not be well-supported by the data, leading to high-variance estimates of how covariates affect the top-level partition. In such cases, a simpler model such as the AD-nDDPM is often more stable and computationally efficient, while it captures cluster-level heterogeneity through individual-level atoms. To balance flexibility and complexity in practice, we recommend evaluating the predictive performance of these models using a predictive criterion such as the log pseudo marginal likelihood, as illustrated in Section \ref{sec:analysis}.
}

\yo{
Additional simulation scenarios are provided in Supplementary Material \ref{sec:additonal_simulation}. These scenarios examine the robustness of our methods under various conditions, including misspecified error terms, nonlinear fixed effects, dichotomous mediators, and a smaller number of clusters ($I=30$). Overall, the results demonstrate even more pronounced improvements in accurately estimating the EIEs and ESMEs, particularly in terms of precision (lower bias and MSE) and uncertainty (shorter interval length and coverage probabilities closer to $95\%$). These findings underscore the practical utility of our proposed methods for causal mediation analysis in CRTs, especially in real-world settings where the true data-generating process may be unknown and potentially complex due to intermediate outcomes. 
}

\section{Empirical Analysis of the RPS CRT}
\label{sec:analysis}

We apply the proposed Bayesian mediation methods to analyze the RPS CRT illustrated in Section \ref{sec:motivating_example}.
We conducted analyses using both the AD-nDDPM and FD-nDDPM models, employing proper, weakly informative conjugate priors. As default choices, we consider $\boldsymbol{\theta}_{lk} \sim \mathrm{MVN}(\mathbf{0}, 10^2 \times I_{d_y})$, $\sigma^2_{lk} \sim \mathrm{IG}(2.0, 1.0)$, $\boldsymbol{\gamma}_{1,lk}, \boldsymbol{\gamma}_{2,lk} \sim \mathrm{MVN}(\mathbf{0}, 10^2 \times I_{d_m})$, $\boldsymbol{\Sigma} \sim \mathrm{IW}(2.0, I_2)$, $\alpha_y \sim \mathrm{Ga}(1.0,1.0)$, $\alpha_m \sim \mathrm{Ga}(1.0,1.0)$, $\beta_y \sim \mathrm{Ga}(1.0,1.0)$, and $\beta_m \sim \mathrm{Ga}(1.0,1.0)$. \yo{We evaluate the predictive performance of the models using the log pseudo marginal likelihood (LPML; \citealp{Geisser1979}), and find the LPML for nDPM, AD-nDDPM, and FD-nDDPM models, three Bayesian models under consideration, are $30.72$, $33.81$, and $31.81$, respectively. Since a higher LPML indicates a better predictive fit in terms of LOO predictive densities, the AD-nDDPM emerges as the top model for predictive accuracy. We therefore mainly focus on the AD-nDDPM method in the subsequent discussion. We also provide sensitivity analyses to examine the robustness of our analysis results with respect to different prior specifications in Supplementary Material Section \ref{sec:sensitivity_analysis}.} 



\begin{figure*}
    \centering
    \includegraphics[width=0.9\textwidth]{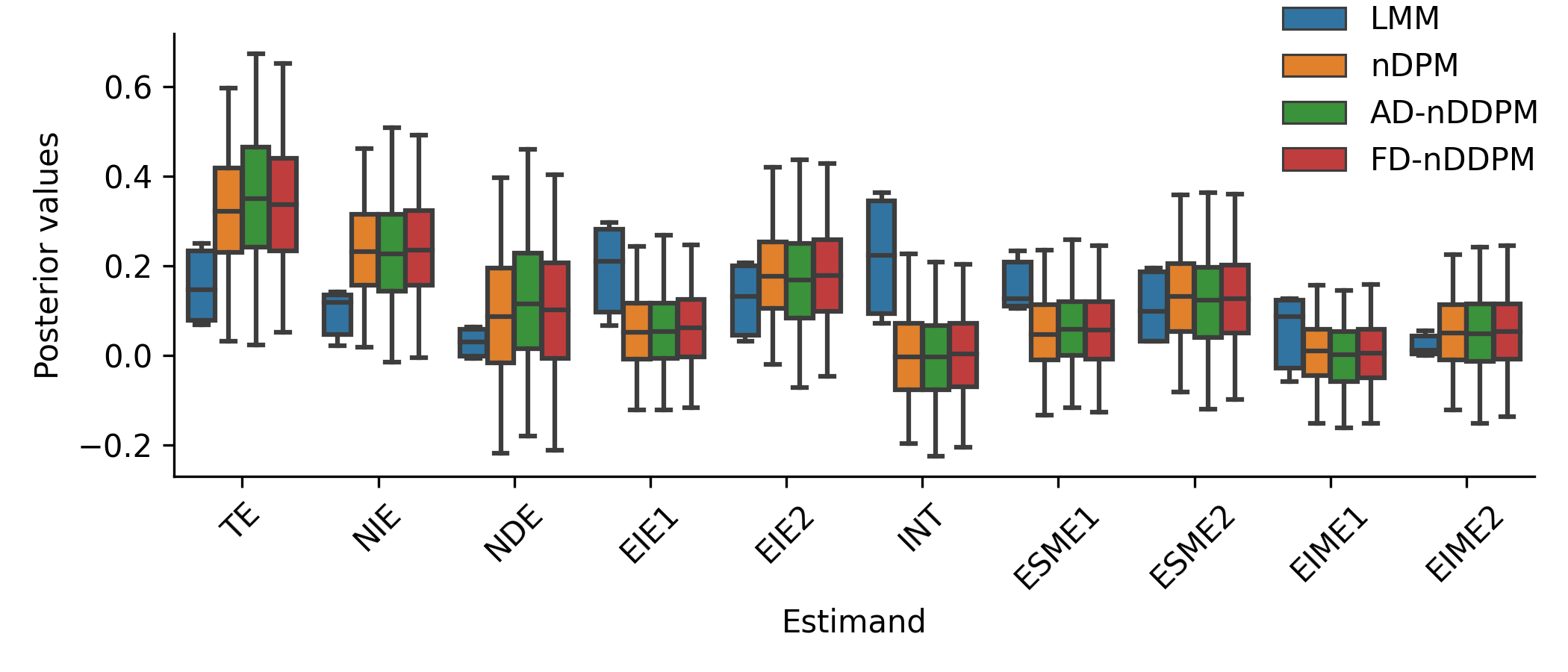}
    \caption{Boxplots of posterior samples of all estimands, under LMM, nDPM, AD-nDDPM and FD-nDPMM for the empirical analysis of the RPS cluster randomized trial.}
    \label{fig:analysis_result}
\end{figure*}

The analysis results are summarized in Figure \ref{fig:analysis_result} and provide insights into how the CCT influences child nutritional status through potential pathways. More detailed numerical results are also provided in the supplementary material \ref{sec:results_tables}. Overall, the TE of the intervention is positive, with a posterior mean of $0.353$ and a $95\%$ credible interval of $(0.023, 0.673)$. This indicates a notable improvement in child height-for-age $z$-scores due to the program. Decomposing the TE into the NIE and the NDE reveals that the NIE (posterior mean $= 0.232$; $95\%$ credible interval $=(-0.015, 0.509)$) accounts for a larger portion of the TE than the NDE (posterior mean $= 0.121$; $95\%$ credible interval $=(-0.180, 0.460)$). This suggests the a non-trivial proportion of positive impact of the CCT may be mediated through the specified mediators—child health check-ups and household dietary diversity—rather than through the direct effects of the intervention.

Further examination of the NIE through mediator-specific EIE shows that the EIE for household dietary diversity ($\mathrm{EIE}^{(2)}$, posterior mean $= 0.171$; $95\%$ credible interval $=(-0.072, 0.437)$) is larger than that for child health check-ups ($\mathrm{EIE}^{(1)}$, posterior mean $= 0.058$; $95\%$ credible interval $=(-0.122, 0.268)$). This indicates that improvements in household dietary diversity play a more substantial role in explaining the treatment effect on child nutritional status. In contrast, the mediation through health check-ups appears somewhat limited. Interestingly, the interaction effect (INT) between the two mediators is estimated to be close to null, with a posterior mean of $-0.004$ and a $95\%$ credible interval of $(-0.224, 0.209)$.
This suggests that, although the causal structure between the two mediators is unknown \emph{a priori}, there is no synergistic or antagonistic interaction between child health check-ups and household dietary diversity in influencing the outcome and the total effect seems to be mediated independently through the two mediators.

\yo{Digging deeper into the EIE for household dietary diversity, we observe that the $\mathrm{ESME}^{(2)}$ (posterior mean $= 0.119$; $95\%$ credible interval $=(-0.119, 0.363)$) exceeds the $\mathrm{EIME}^{(2)}$ (posterior mean $= 0.052$; $95\%$ credible interval $=(-0.152, 0.242)$). Additionally, the posterior probability that  $\mathrm{EIME}^{(2)}$  is greater than zero is $83.8 \%$, compared to $70.9\%$ for $\mathrm{EIME}^{(2)}$. These findings suggest that spillover effects, mediated through household dietary diversity within the same comarcas, contribute significantly more to improvements in individual outcome than individual household changes alone.
For the mediator related to health check-ups, both the $\mathrm{ESME}^{(1)}$ (posterior mean = 0.060; $95\%$ credible interval = $(-0.117, 0.257)$) and the $\mathrm{EIME}^{(1)}$ (posterior mean = $-0.002$; $95\%$ credible interval = $(-0.161, 0.145)$) are relatively small. The posterior probability that $\mathrm{ESME}^{(1)}$ is greater than zero is $75.4\%$, whereas for $\mathrm{EIME}^{(1)}$ it is only $50.8\%$, indicating essentially no effect. These results suggest that while the health check-up conditions of other households within the same comarcas may have some spillover influence on individual household outcomes (with a $75.4\%$ posterior probability), a household’s own health check-up status has minimal impact on its own outcome. Finally, Figure \ref{fig:analysis_result} also presents results from the LMM and nDPM approaches, using the same model specifications as in Section \ref{sec:simulations}. The nDPM yields point estimates comparable to those from the AD-nDDPM, with slightly narrower credible intervals that may have overstated the uncertainty. In contrast, the LMM produces notably smaller estimates for the NIE and EIE through household dietary diversity, but substantially larger INT effect and EIE through child health check-ups. As a result, the estimated ESME and ISME mediated by child health check-ups are larger in magnitude under the LMM compared to the AD-nDDPM. These discrepancies are likely attributable to parametric model misspecification in the LMM, as demonstrated by our simulation results.}

\yo{Under a Bayesian inference framework, interpretability need not hinge solely on whether zero lies within a credible interval. Instead, Bayesian methods provide the additional measure of posterior probability, which quantifies the likelihood that an effect is positive (or negative). For example, although the $95\%$ central credible interval for the natural indirect effect (NIE) includes zero, its posterior probability of being greater than zero is approximately $96\%$, suggesting a substantial chance that the true effect is in fact positive. This nuanced interpretation moves beyond a binary ``significant'' vs. ``not significant'' decision based on the frequentist p-value and highlights that while uncertainty remains, the data still lean toward a positive effect. In this way, Bayesian posterior probabilities provide a direct assessment of how strongly the evidence supports a given effect direction, even when the credible interval may encompass values near zero. Table \ref{tab:posterior_estimates} in Supplementary Material Section \ref{sec:remarks_analysis} displays the posterior probabilities for each estimand.}
In summary, our analyses revealed that the CCT's effectiveness in enhancing child nutritional status is largely mediated through improvements in household dietary diversity, particularly via spillover effects within each distinct administrative region. These results underscore the importance of targeting household-level factors in interventions aimed at improving child health outcomes in low-income settings. By revealing these finer causal mechanisms and clarifying the roles of individual mediator-specific effects and the spillover effects, our analysis could potentially offer insights into development of future interventions that further enhance the household dietary diversity and leverage connections within the \emph{comarcas} to maximize its impact on population health outcomes.

\section{Concluding Remarks}
\label{sec:conclusion}

In this paper, we have addressed the challenges of causal mediation analysis in CRTs involving multiple mediators and interference. Our first contribution is the introduction of new causal estimands that decompose the NIE into mediator-specific EIEs and INT and further dissect the EIEs into EIMEs and ESMEs. These decompositions provide a detailed understanding of how each mediator and their interactions contribute to an intervention's overall effect. We also present the identification formula for each estimand.
Our second methodological contribution is the development of the nDDPM, specifically designed for modeling complex data in CRTs. The nDDPM effectively captures distributional heterogeneity and complex clustering structures inherent in CRTs and avoids restrictive parametric assumptions. Extensive simulations demonstrate the strong performance of our method regarding accuracy and robustness under data-generating processes with varying levels of complexity. 

We have introduced two types of nDDPM models: AD-nDDPM and FD-nDDPM. In our simulation studies comparing the FD-nDDPM with other methods, we found that it performs similarly to the AD-nDDPM, indicating that additionally incorporating covariate dependence into the cluster-level weights offers limited enhancement. \yo{This observation is also supported by Proposition \ref{prop:weak_support}, ensuring the robustness of the AD-nDDP prior.} This suggests that the individual-level atom-based dependence structure of the AD-nDDPM sufficiently captures the necessary dependence for cluster-level distribution assignments. While the FD-nDDPM expands the modeling framework for CRTs by providing greater flexibility to capture distributional heterogeneity driven by covariates, it requires extra computational complexity due to the absence of closed-form posterior updates for certain parameters, necessitating Metropolis-Hastings steps within the MCMC algorithm. On the other hand, although the AD-nDDPM emerges as a practical and effective choice---suggesting that simpler models may suffice in certain contexts without compromising performance and while reducing computational demands---the FD-nDDPM model could offer valuable insights into cluster heterogeneity varying with cluster-level covariates and might gain efficiency in specific contexts where clusters exhibit extreme heterogeneity. 

We have primarily focused on unstructured mediators, and for ease of exposition, specifically addressed the case of two mediators. Extensions to scenarios involving $K$ mediators are provided in the supplementary material. When the causal structure among mediators is known, we can define alternative estimands and potentially enhance inferential efficiency by leveraging this knowledge. This is particularly advantageous with temporally ordered mediators; however, how to precisely define spillover mediation effects under the knowledge of temporal ordering remains an area for future research.
Moreover, we have invoked structural assumptions for the identification of estimands, such as the cross-world independence assumption. 
Developing sensitivity analysis methods for structural assumptions remains an important direction for future research.
Finally, in Supplementary Material \ref{sec:Interpretation_IE}, we provide a reinterpretation of our results as interventional mediation effects \citep{Vansteelandt2017}, which remains valid even if this structural assumption does not hold.

\section*{Acknowledgement}
Research in this article was supported by the Patient-Centered Outcomes Research Institute\textsuperscript{\textregistered} (PCORI\textsuperscript{\textregistered} Award ME-2023C1-31350). The statements presented in this article are solely the responsibility of the authors and do not necessarily represent the views of PCORI\textsuperscript{\textregistered}, its Board of Governors or Methodology Committee.

\begin{singlespace}
\bibliographystyle{Chicago}
\bibliography{literature}

\begin{thebibliography}{}

\bibitem[\protect\citeauthoryear{Albert and Chib}{Albert and
  Chib}{1993}]{Albert1993}
Albert, J.~H. and S.~Chib (1993).
\newblock Bayesian analysis of binary and polychotomous response data.
\newblock {\em Journal of the American Statistical Association\/}~{\em
  88\/}(422), 669--679.

\bibitem[\protect\citeauthoryear{Albert and Nelson}{Albert and
  Nelson}{2011}]{Albert2011}
Albert, J.~M. and S.~Nelson (2011).
\newblock Generalized causal mediation analysis.
\newblock {\em Biometrics\/}~{\em 67}, 1028--1038.

\bibitem[\protect\citeauthoryear{Antoniak}{Antoniak}{1974}]{Antoniak1974}
Antoniak, C.~E. (1974).
\newblock {Mixtures of Dirichlet Processes with Applications to Bayesian
  Nonparametric Problems}.
\newblock {\em The Annals of Statistics\/}~{\em 2\/}(6), 1152 -- 1174.

\bibitem[\protect\citeauthoryear{Aronow and Samii}{Aronow and
  Samii}{2017}]{Aronow2017}
Aronow, P.~M. and C.~Samii (2017).
\newblock {Estimating average causal effects under general interference, with
  application to a social network experiment}.
\newblock {\em The Annals of Applied Statistics\/}~{\em 11\/}(4), 1912 -- 1947.

\bibitem[\protect\citeauthoryear{Barrientos, Jara, and Quintana}{Barrientos
  et~al.}{2012}]{Barrientos2012}
Barrientos, A.~F., A.~Jara, and F.~A. Quintana (2012).
\newblock {On the Support of MacEachern’s Dependent Dirichlet Processes and
  Extensions}.
\newblock {\em Bayesian Analysis\/}~{\em 7\/}(2), 277 -- 310.

\bibitem[\protect\citeauthoryear{Benkeser and Ran}{Benkeser and
  Ran}{2021}]{BenkeserRan2021}
Benkeser, D. and J.~Ran (2021).
\newblock Nonparametric inference for interventional effects with multiple
  mediators.
\newblock {\em Journal of Causal Inference\/}~{\em 9\/}(1), 172--189.

\bibitem[\protect\citeauthoryear{Charters, Kaufman, and Nandi}{Charters
  et~al.}{2023}]{Charters2023}
Charters, T.~J., J.~S. Kaufman, and A.~Nandi (2023, 01).
\newblock {A Causal Mediation Analysis for Investigating the Effect of a
  Randomized Cash-Transfer Program in Nicaragua}.
\newblock {\em American Journal of Epidemiology\/}~{\em 192\/}(1), 111--121.

\bibitem[\protect\citeauthoryear{Cheng and Li}{Cheng and Li}{2024}]{cheng2024}
Cheng, C. and F.~Li (2024).
\newblock Semiparametric causal mediation analysis in cluster-randomized
  experiments.

\bibitem[\protect\citeauthoryear{Chib and Greenberg}{Chib and
  Greenberg}{1998}]{Chib1998}
Chib, S. and E.~Greenberg (1998, 06).
\newblock {Analysis of multivariate probit models}.
\newblock {\em Biometrika\/}~{\em 85\/}(2), 347--361.

\bibitem[\protect\citeauthoryear{Daniel, De~Stavola, Cousens, and
  Vansteelandt}{Daniel et~al.}{2015}]{Daniel2015}
Daniel, R.~M., B.~L. De~Stavola, S.~N. Cousens, and S.~Vansteelandt (2015).
\newblock Causal mediation analysis with multiple mediators.
\newblock {\em Biometrics\/}~{\em 71\/}(1), 1--14.

\bibitem[\protect\citeauthoryear{Diana, Matechou, Griffin, and Johnston}{Diana
  et~al.}{2020}]{Diana2020}
Diana, A., E.~Matechou, J.~Griffin, and A.~Johnston (2020).
\newblock {A hierarchical dependent Dirichlet process prior for modelling bird
  migration patterns in the UK}.
\newblock {\em The Annals of Applied Statistics\/}~{\em 14\/}(1), 473 -- 493.

\bibitem[\protect\citeauthoryear{Dunson and Park}{Dunson and
  Park}{2008}]{Dunson_2008}
Dunson, D.~B. and J.-H. Park (2008, 04).
\newblock {Kernel stick-breaking processes}.
\newblock {\em Biometrika\/}~{\em 95\/}(2), 307--323.

\bibitem[\protect\citeauthoryear{Escobar and West}{Escobar and
  West}{1995}]{Escobar1995}
Escobar, M.~D. and M.~West (1995).
\newblock Bayesian density estimation and inference using mixtures.
\newblock {\em Journal of the American Statistical Association\/}~{\em
  90\/}(430), 577--588.

\bibitem[\protect\citeauthoryear{Ferguson}{Ferguson}{1974}]{Ferguson1974}
Ferguson, T.~S. (1974).
\newblock {Prior Distributions on Spaces of Probability Measures}.
\newblock {\em The Annals of Statistics\/}~{\em 2\/}(4), 615--629.

\bibitem[\protect\citeauthoryear{Forastiere, Mealli, and
  VanderWeele}{Forastiere et~al.}{2016}]{Forastiere2016}
Forastiere, L., F.~Mealli, and T.~J. VanderWeele (2016, 4).
\newblock Identification and estimation of causal mechanisms in clustered
  encouragement designs: Disentangling bed nets using bayesian principal
  stratification.
\newblock {\em Journal of the American Statistical Association\/}~{\em 111},
  510--525.

\bibitem[\protect\citeauthoryear{Geisser and Eddy}{Geisser and
  Eddy}{1979}]{Geisser1979}
Geisser, S. and W.~F. Eddy (1979).
\newblock A predictive approach to model selection.
\newblock {\em Journal of the American Statistical Association\/}~{\em
  74\/}(365), 153--160.

\bibitem[\protect\citeauthoryear{Gelfand and Dey}{Gelfand and
  Dey}{1994}]{Gelfand_Dey1994}
Gelfand, A.~E. and D.~K. Dey (1994).
\newblock Bayesian model choice: Asymptotics and exact calculations.
\newblock {\em Journal of the Royal Statistical Society. Series B
  (Methodological)\/}~{\em 56\/}(3), 501--514.

\bibitem[\protect\citeauthoryear{Ho, Tu, Ghosh, and Tiwari}{Ho
  et~al.}{2013}]{Ho2013}
Ho, M.-W., W.~Tu, P.~Ghosh, and R.~C. Tiwari (2013).
\newblock A nested dirichlet process analysis of cluster randomized trial data
  with application in geriatric care assessment.
\newblock {\em Journal of the American Statistical Association\/}~{\em
  108\/}(501), 48--68.

\bibitem[\protect\citeauthoryear{Imai and Yamamoto}{Imai and
  Yamamoto}{2013}]{Imai2013}
Imai, K. and T.~Yamamoto (2013).
\newblock Identification and sensitivity analysis for multiple causal
  mechanisms: Revisiting evidence from framing experiments.
\newblock {\em Political Analysis\/}~{\em 21}, 141--171.

\bibitem[\protect\citeauthoryear{Ishwaran and Zarepour}{Ishwaran and
  Zarepour}{2000}]{Ishwaran2000}
Ishwaran, H. and M.~Zarepour (2000).
\newblock Markov chain monte carlo in approximate dirichlet and beta
  two-parameter process hierarchical models.
\newblock {\em Biometrika\/}~{\em 87\/}(2), 371--390.

\bibitem[\protect\citeauthoryear{Kahan, Blette, Harhay, Halpern, Jairath,
  Copas, and Li}{Kahan et~al.}{2024}]{kahan2024demystifying}
Kahan, B.~C., B.~S. Blette, M.~O. Harhay, S.~D. Halpern, V.~Jairath, A.~Copas,
  and F.~Li (2024).
\newblock Demystifying estimands in cluster-randomised trials.
\newblock {\em Statistical Methods in Medical Research\/}~{\em 33\/}(7),
  1211--1232.

\bibitem[\protect\citeauthoryear{Kim, Daniels, Hogan, Choirat, and Zigler}{Kim
  et~al.}{2019}]{Kim2019}
Kim, C., M.~J. Daniels, J.~W. Hogan, C.~Choirat, and C.~M. Zigler (2019, 9).
\newblock Bayesian methods for multiple mediators: Relating principal
  stratification and causal mediation in the analysis of power plant emission
  controls.
\newblock {\em Annals of Applied Statistics\/}~{\em 13}, 1927--1956.

\bibitem[\protect\citeauthoryear{MacEachern}{MacEachern}{1999}]{Maceachern1999}
MacEachern, S. (1999, 01).
\newblock Dependent nonparametric processes.
\newblock {\em Proceedings of the Section on Bayesian Statistical Science,
  American Statistical Association\/}, 50--55.

\bibitem[\protect\citeauthoryear{MacEachern}{MacEachern}{2000}]{Maceachern2000}
MacEachern, S. (2000).
\newblock Dependent dirichlet processes.
\newblock {\em Technical Report\/}.
\newblock
  \url{http://www.gatsby.ucl.ac.uk/$\sim$porbanz/talks/MacEachern2000.pdf}.

\bibitem[\protect\citeauthoryear{Manley, Gitter, and Slavchevska}{Manley
  et~al.}{2013}]{Manley2013}
Manley, J., S.~Gitter, and V.~Slavchevska (2013).
\newblock How effective are cash transfers at improving nutritional status?
\newblock {\em World Development\/}~{\em 48}, 133--155.

\bibitem[\protect\citeauthoryear{Ogburn, Sofrygin, Díaz, and van~der
  Laan}{Ogburn et~al.}{2024}]{Ogburn2024}
Ogburn, E.~L., O.~Sofrygin, I.~Díaz, and M.~J. van~der Laan (2024).
\newblock Causal inference for social network data.
\newblock {\em Journal of the American Statistical Association\/}~{\em
  119\/}(545), 597--611.

\bibitem[\protect\citeauthoryear{Ohnishi and Sabbaghi}{Ohnishi and
  Sabbaghi}{2024}]{Ohnishi2024}
Ohnishi, Y. and A.~Sabbaghi (2024).
\newblock {A Bayesian Analysis of Two-Stage Randomized Experiments in the
  Presence of Interference, Treatment Nonadherence, and Missing Outcomes}.
\newblock {\em Bayesian Analysis\/}~{\em 19\/}(1), 205 -- 234.

\bibitem[\protect\citeauthoryear{Park and Kang}{Park and Kang}{2023}]{Park2023}
Park, C. and H.~Kang (2023).
\newblock Assumption-lean analysis of cluster randomized trials in infectious
  diseases for intent-to-treat effects and network effects.
\newblock {\em Journal of the American Statistical Association\/}~{\em 118},
  1195--1206.

\bibitem[\protect\citeauthoryear{Reich and Fuentes}{Reich and
  Fuentes}{2007}]{Reich2007}
Reich, B.~J. and M.~Fuentes (2007).
\newblock {A multivariate semiparametric Bayesian spatial modeling framework
  for hurricane surface wind fields}.
\newblock {\em The Annals of Applied Statistics\/}~{\em 1\/}(1), 249 -- 264.

\bibitem[\protect\citeauthoryear{Rodríguez, Dunson, and Gelfand}{Rodríguez
  et~al.}{2008}]{Rodriguez2008}
Rodríguez, A., D.~B. Dunson, and A.~E. Gelfand (2008).
\newblock The nested dirichlet process.
\newblock {\em Journal of the American Statistical Association\/}~{\em
  103\/}(483), 1131--1154.

\bibitem[\protect\citeauthoryear{Roy, Daniels, Kelly, and Roy}{Roy
  et~al.}{2022}]{Roy2022}
Roy, S., M.~J. Daniels, B.~J. Kelly, and J.~Roy (2022, 8).
\newblock A bayesian nonparametric approach for causal inference with multiple
  mediators.

\bibitem[\protect\citeauthoryear{Sethuraman}{Sethuraman}{1994}]{Sethuraman1994}
Sethuraman, J. (1994).
\newblock A constructive definition of dirichlet priors.
\newblock {\em Statistica Sinica\/}~{\em 4\/}(2), 639--650.

\bibitem[\protect\citeauthoryear{Sklar}{Sklar}{1959}]{Sklar1959}
Sklar, M. (1959).
\newblock Fonctions de repartition an dimensions et leurs marges.
\newblock {\em Publ. inst. statist. univ. Paris\/}~{\em 8}, 229--231.

\bibitem[\protect\citeauthoryear{Taguri, Featherstone, and Cheng}{Taguri
  et~al.}{2018}]{Taguri2018}
Taguri, M., J.~Featherstone, and J.~Cheng (2018, 1).
\newblock Causal mediation analysis with multiple causally non-ordered
  mediators.
\newblock {\em Statistical Methods in Medical Research\/}~{\em 27}, 3--19.

\bibitem[\protect\citeauthoryear{Teh, Jordan, Beal, and Blei}{Teh
  et~al.}{2006}]{Teh2006}
Teh, Y.~W., M.~I. Jordan, M.~J. Beal, and D.~M. Blei (2006).
\newblock Hierarchical dirichlet processes.
\newblock {\em Journal of the American Statistical Association\/}~{\em
  101\/}(476), 1566--1581.

\bibitem[\protect\citeauthoryear{VanderWeele}{VanderWeele}{2009}]{VanderWeele2009}
VanderWeele, T.~J. (2009).
\newblock Direct and indirect effects for neighborhood-based clustered and
  longitudinal data.
\newblock {\em Sociological Methods and Research\/}~{\em 38}, 515--544.

\bibitem[\protect\citeauthoryear{VanderWeele, Hong, Jones, and
  Brown}{VanderWeele et~al.}{2013}]{VanderWeele2013_CRT}
VanderWeele, T.~J., G.~Hong, S.~M. Jones, and J.~L. Brown (2013).
\newblock Mediation and spillover effects in group-randomized trials: A case
  study of the 4rs educational intervention.
\newblock {\em Journal of the American Statistical Association\/}~{\em 108},
  469--482.

\bibitem[\protect\citeauthoryear{VanderWeele and Vansteelandt}{VanderWeele and
  Vansteelandt}{2013}]{vanderweele2013_multiple}
VanderWeele, T.~J. and S.~Vansteelandt (2013, 12).
\newblock Mediation analysis with multiple mediators.
\newblock {\em Epidemiologic Methods\/}~{\em 2}, 95--115.

\bibitem[\protect\citeauthoryear{VanderWeele, Vansteelandt, and
  Robins}{VanderWeele et~al.}{2014}]{VanderWeele2014_effect_decomposition}
VanderWeele, T.~J., S.~Vansteelandt, and J.~M. Robins (2014).
\newblock Effect decomposition in the presence of an exposure-induced
  mediator-outcome confounder.
\newblock {\em Epidemiology (Cambridge, Mass.)\/}~{\em 25\/}(2), 300--306.

\bibitem[\protect\citeauthoryear{Vansteelandt and Daniel}{Vansteelandt and
  Daniel}{2017}]{Vansteelandt2017}
Vansteelandt, S. and R.~M. Daniel (2017).
\newblock Interventional effects for mediation analysis with multiple
  mediators.
\newblock {\em Epidemiology\/}~{\em 28\/}(2).

\bibitem[\protect\citeauthoryear{Wang, Harhay, Small, Morris, and Li}{Wang
  et~al.}{2025}]{wang2025mixed}
Wang, B., M.~O. Harhay, D.~S. Small, T.~P. Morris, and F.~Li (2025).
\newblock On the mixed-model analysis of covariance in cluster-randomized
  trials.
\newblock {\em Statistical Science\/}.

\bibitem[\protect\citeauthoryear{Wang, Park, Small, and Li}{Wang
  et~al.}{2024}]{Wang2024}
Wang, B., C.~Park, D.~S. Small, and F.~Li (2024).
\newblock Model-robust and efficient covariate adjustment for
  cluster-randomized experiments.
\newblock {\em Journal of the American Statistical Association\/}~{\em 0\/}(0),
  1--13.

\bibitem[\protect\citeauthoryear{Williams}{Williams}{2016}]{Williams2016}
Williams, N.~J. (2016, 9).
\newblock Multilevel mechanisms of implementation strategies in mental health:
  Integrating theory, research, and practice.
\newblock {\em Administration and Policy in Mental Health and Mental Health
  Services Research\/}~{\em 43\/}(5), 783--798.

\bibitem[\protect\citeauthoryear{Xia and Chan}{Xia and Chan}{2022}]{Xia2022}
Xia, F. and K.~C.~G. Chan (2022, 12).
\newblock Decomposition, identification and multiply robust estimation of
  natural mediation effects with multiple mediators.
\newblock {\em Biometrika\/}~{\em 109}, 1085--1100.

\bibitem[\protect\citeauthoryear{Zhang, Wade, and Bochkina}{Zhang
  et~al.}{2024}]{zhang2024}
Zhang, H., S.~Wade, and N.~Bochkina (2024).
\newblock Covariate-dependent hierarchical dirichlet process.

\end{thebibliography}
\end{singlespace}

\clearpage

\appendix

\section{Discussion on assumptions and estimands}
\label{sec:discussion_assumptions_estimands_supp}
\yo{
\subsection{Implication of identification assumptions}
\label{sec:implication_assumption5}
As discussed in the main manuscript, Assumption \ref{asmp:cond_homogeneity} is a weaker condition than the conditional independence assumption of mediators (e.g., $\mathbf{M}_{i}^{(1)}(a) \indep \mathbf{M}_{i}^{(2)}(a') \mid \mathbf{C}_i, N_i$ for all $a, a'$), which has been previously invoked in the literature on mediation analysis with multiple mediators (for independent rather than clustered data) \citep{Taguri2018}. However, the dependence structure implied by Assumption \ref{asmp:cond_homogeneity} (i.e., mean independence) may not be entirely straightforward to interpret. We here investigate the implication of this assumption with a simple example.
Suppose the potential outcome follows a linear structure model:
\begin{equation*}
    Y_{ij}(a,m_{ij}^{(1)},\mathbf{m}_{i(-j)}^{(1)},m_{ij}^{(2)},\mathbf{m}_{i(-j)}^{(2)}) 
    = \beta_0^a + \beta_1^a m_{ij}^{(1)} + \beta_2^a \overline{\mathbf{m}}_{i(-j)}^{(1)} + \beta_3^a m_{ij}^{(2)} + \beta_4^a \overline{\mathbf{m}}_{i(-j)}^{(2)} + g^a(\mathbf{C}_{ij},N_i) + \epsilon_{ij}^a,
\end{equation*}
where $\epsilon_{ij}^a \sim \mathrm{N}(0,\sigma_y^2)$. Under this structure, each side of Assumption \ref{asmp:cond_homogeneity} is written as:
\begin{align*}
    \text{LHS} &= \E \Big[ \beta_1^a \big( M_{ij}^{(1)}(1) - M_{ij}^{(1)}(a) \big) \;+\; \beta_3^a \big( \overline{\mathbf{M}}_{i(-j)}^{(1)}(1) - \overline{\mathbf{M}}_{i(-j)}^{(1)}(0) \big) \;\Big|\; \mathbf{M}_{i}^{(2)}(1), \mathbf{C}_i, N_i \Big], \\
    \text{RHS} &= \E \Big[ \beta_1^a \big( M_{ij}^{(1)}(1) - M_{ij}^{(1)}(a) \big) \;+\; \beta_3^a \big( \overline{\mathbf{M}}_{i(-j)}^{(1)}(1) - \overline{\mathbf{M}}_{i(-j)}^{(1)}(0) \big) \;\Big|\; \mathbf{C}_i, N_i \Big].
\end{align*}
For simplicity, consider $a = a' = 1$, although the following argument holds for any $a, a' \in \{0,1\}$. Then, by identity, the assumption implies that, for $j = 1, \ldots, N_i$, 
\begin{equation}
\label{eq:cond_effect_equivalence}
    \E \Big[ M_{ij}^{(1)}(1) - M_{ij}^{(1)}(0) \;\Big|\; \mathbf{M}_{i}^{(2)}(1), \mathbf{C}_i, N_i \Big]
    \;=\; 
    \E \Big[ M_{ij}^{(1)}(1) - M_{ij}^{(1)}(0) \;\Big|\; \mathbf{C}_i, N_i \Big].
\end{equation}

Next, suppose the joint distribution of the two mediators, under different treatment assignments for different individuals, follows a multivariate normal distribution. That is, for two individuals $j$ and $k$,
\begin{align*}
    \begin{pmatrix}
      M_{ij}^{(1)}(1) \\
      M_{ij}^{(1)}(0) \\
      M_{ij}^{(2)}(1) \\
      M_{ij}^{(2)}(0) \\
      M_{ik}^{(1)}(1) \\
      M_{ik}^{(1)}(0) \\
      M_{ik}^{(2)}(1) \\
      M_{ik}^{(2)}(0) \\
    \end{pmatrix}
    \sim \mathrm{MVN}\!\left( 
    \begin{pmatrix}
      \boldsymbol{\mu}(\mathbf{C}_{ij}, N_i) \\
      \boldsymbol{\mu}(\mathbf{C}_{ik}, N_i) 
    \end{pmatrix},  
    \sigma^2 \begin{pmatrix}
      R & M  \\
      M & R
    \end{pmatrix} \right),
\end{align*}
where 
\[
\boldsymbol{\mu}(\mathbf{C}_{ij}, N_i) 
= 
\bigl(\mu_1^{(1)}(\mathbf{C}_{ij}, N_i),
\;\mu_0^{(1)}(\mathbf{C}_{ij}, N_i),
\;\mu_1^{(2)}(\mathbf{C}_{ij}, N_i),
\;\mu_0^{(2)}(\mathbf{C}_{ij}, N_i)\bigr)^\top
\]
is the mean function of the mediators, with the superscript and subscript respectively denoting mediator types and treatment conditions. The matrices
\[
    R \;=\; \begin{pmatrix}
      1 & \alpha_1 & \alpha_0 & \alpha_2 \\
      \alpha_1 & 1 & \alpha_2 & \alpha_0 \\
      \alpha_0 & \alpha_2 & 1 & \alpha_1 \\
      \alpha_2 & \alpha_0 & \alpha_1 & 1
    \end{pmatrix}, 
    \quad
    M \;=\; \begin{pmatrix}
      \rho_0 & 0 & \rho_1 & 0 \\
      0 & \rho_0 & 0 & \rho_1 \\
      \rho_1 & 0 & \rho_0 & 0 \\
      0 & \rho_1 & 0 & \rho_0 \\
    \end{pmatrix}
\]
define the correlation structure across mediators, treatment conditions, and individuals. Table \ref{tab:correlation_parameters} provides definitions of these correlation parameters. Note that the zero entries in $M$ reflect the cross-world inter-individual mediator independence from Assumption \ref{asmp:no_cross_intra_corr}. Figure \ref{fig:DAG_asmp_crossworld_between_ind_indep} illustrates Assumption \ref{asmp:no_cross_intra_corr} using the graphical representation and explains that it is only the conceptually weakest type of correlation that is assumed away. 

\begin{table*}
    \centering
    \caption{\yo{Correlation parameters that define the correlation structure between mediator types, counterfactuals, and individuals. The correlation of mediators between different individuals under different treatment assignments is assumed to be zero by Assumption \ref{asmp:no_cross_intra_corr}.}}
    \begin{adjustbox}{width=12cm}
        \begin{tabular}{r ccc}
        \toprule
            \textbf{Parameters} &  Individuals &  Mediator variables & Treatment condition\\
            \midrule 
             $\alpha_0$ & same & different & 
             same \\
             $\alpha_1$ & same & same & different \\
             $\alpha_2$ & same & different & different \\
             $\rho_0$ & different & same & same \\
             $\rho_1$ & different & different & same \\
        \bottomrule
        \end{tabular}
    \end{adjustbox}
    \label{tab:correlation_parameters}
\end{table*}

From the conditional expectation formula for the multivariate normal distribution, the above dependence structure implies:
\begin{align}\label{eq:condition_corr}
    &\E \Big[\;\mathbf{M}_{i}^{(1)}(1) - \mathbf{M}_{i}^{(1)}(0)\;\Big|\;\mathbf{M}_{i}^{(2)}(1), \mathbf{C}_i, N_i \Big] \nonumber\\
    &= \boldsymbol{\mu}_1^{(1)}(\mathbf{C}_{i}, N_i) 
    \;+\;  
    \begin{pmatrix}
    \alpha_0 & \rho_1 & \cdots & \rho_1 \\
    \rho_1 & \alpha_0 & \cdots & \rho_1 \\
    \vdots & \vdots & \ddots & \vdots \\
    \rho_1 & \rho_1 & \cdots & \alpha_0
    \end{pmatrix}
    \begin{pmatrix}
    1 & \rho_0 & \cdots & \rho_0 \\
    \rho_0 & 1 & \cdots & \rho_0 \\
    \vdots & \vdots & \ddots & \vdots \\
    \rho_0 & \rho_0 & \cdots & 1
    \end{pmatrix}^{-1}
    \Big\{\mathbf{M}_{i}^{(2)}(1) \;-\; \boldsymbol{\mu}_1^{(2)}(\mathbf{C}_{i}, N_i)\Big\} \nonumber\\
    &\quad\;-\; \boldsymbol{\mu}_0^{(1)}(\mathbf{C}_{i}, N_i) 
    \;-\;
    \begin{pmatrix}
    \alpha_2 & 0 & \cdots & 0 \\
    0 & \alpha_2 & \cdots & 0 \\
    \vdots & \vdots & \ddots & \vdots \\
    0 & 0 & \cdots & \alpha_2
    \end{pmatrix}
    \begin{pmatrix}
    1 & \rho_0 & \cdots & \rho_0 \\
    \rho_0 & 1 & \cdots & \rho_0 \\
    \vdots & \vdots & \ddots & \vdots \\
    \rho_0 & \rho_0 & \cdots & 1
    \end{pmatrix}^{-1}
    \Big\{\mathbf{M}_{i}^{(2)}(1) \;-\; \boldsymbol{\mu}_1^{(2)}(\mathbf{C}_{i}, N_i)\Big\}.
\end{align}

Hence, for \eqref{eq:cond_effect_equivalence} to hold, we need $\alpha_0 = \alpha_2$ and $\rho_1=0$. In other words, Assumption \ref{asmp:cond_homogeneity}, which comes down to Equation \eqref{eq:condition_corr}, only requires the between-mediator independence for different individuals (i.e.,  $\rho_1=0$), but it still allows the between-mediator dependence for the same individual in both same and cross-world ($\alpha_0 = \alpha_2 \neq 0$). This is a substantially weaker condition than the common requirement that all different types of mediators be independent in both same and cross-world settings. Figure \ref{fig:dependence_structure} illustrates the cross-world, between-mediator, and inter-individual dependence structures among potential mediators. Under Assumption \ref{asmp:cond_homogeneity}, we account for the dependence of potential mediators within the same individual. This is advantageous because there is no compelling reason to assume that potential mediators within one person are independent, given their shared biological and sociological characteristics.

\begin{figure}
    \centering
    \begin{tikzpicture}

    \draw[thick] (-1.5, -1) rectangle (5.5, 3);
    
    \node[state] (m00) at (0, 2) {$M_{ij}^{(1)}(0)$};
    \node[state] (m01) at (4, 2) {$M_{ij}^{(2)}(0)$};
    \node[state] (m10) at (0, 0) {$M_{ik}^{(1)}(0)$};
    \node[state] (m11) at (4, 0) {$M_{ik}^{(2)}(0)$};
    
    \node at (2, -2.0) {$A_i=0$};
    \node at (10, -2.0) {$A_i=1$};

    \path[<->] (m00) edge node[left] {$\rho_0$} (m10);
    \path[<->] (m00) edge node[above] {$\rho_1$} (m11);
    \path[<->] (m00) edge node[above] {$\alpha_0$} (m01);
    
    \draw[thick] (6.5, -1) rectangle (13.5, 3);
    
    \node[state] (m00b) at (8, 2) {$M_{ij}^{(1)}(1)$};
    \node[state] (m01b) at (12, 2) {$M_{ij}^{(2)}(1)$};
    \node[state] (m10b) at (8, 0) {$M_{ik}^{(1)}(1)$};
    \node[state] (m11b) at (12, 0) {$M_{ik}^{(2)}(1)$};

    \draw[<->] (m00) .. controls (2.5, 3.5) and (6, 3.5) .. node[above] {$\alpha_1$} (m00b);
    \draw[<->] (m00) .. controls (2.5, 4.5) and (6, 4.5) .. node[above] {$\alpha_2$} (m01b);
    \draw[dashed,-] (m00) .. controls (2.5, -1) and (6, -2.0) .. (m10b);
    \draw[dashed,-] (m00) .. controls (2.5, -1.5) and (6, -3.5) .. (m11b);

\end{tikzpicture}

    \caption{\yo{Graphical representation of the cross-world, between-mediator, and inter-individual dependence structures among mediators. The outer boxes represent cross-world scenarios under different treatment assignments. The dashed lines depict the cross-world inter-individual independence specified in Assumption \ref{asmp:no_cross_intra_corr}.}}
    \label{fig:dependence_structure}
\end{figure}

}

\begin{figure}
    \centering
    \begin{tikzpicture}

        \node[state] (m00) at (0, 2) {$M_{ij}^{(k)}(0)$};
        \node[state] (m01) at (4, 2) {$M_{ij'}^{(k)}(0)$};
        \node[state] (m10) at (0, 0) {$M_{ij}^{(k)}(1)$};
        \node[state] (m11) at (4, 0) {$M_{ij'}^{(k)}(1)$};
        
        \node at (-2, 2) {$A_i=0$};
        \node at (-2, 0) {$A_i=1$};
        
        \node at (0, 3) {$j$};
        \node at (4, 3) {$j'$};
        
        \path[<->] (m00) edge (m10);
        \path[<->] (m01) edge (m11);
        \path[<->] (m00) edge (m01);
        \path[<->] (m10) edge (m11);
        
        \draw[dashed,-] (m00) -- (m11);
        \draw[dashed,-] (m01) -- (m10);
    
    \end{tikzpicture}
    \caption{Graphical representation of Assumption \ref{asmp:no_cross_intra_corr}. Mediators are allowed to be correlated between individuals in a single world, as well as between cross-world and single-world mediators within individuals (solid arrows). Only mediators of different individuals in cross-worlds are assumed to be conditionally independent (dashed lines).}
    \label{fig:DAG_asmp_crossworld_between_ind_indep}
\end{figure}

\subsection{Connections of the mediator-specific EIE estimands to path-specific effects}
\label{sec:DAG_estimands}


In this section, we discuss the connection of $\mathrm{EIE}_{\mathrm{C}}^{(k)}$ estimand with well-studied mediation estimands assuming a known causal structure between mediators. Specifically, we explain that, with two mediators, $\mathrm{EIE}_{\mathrm{C}}^{(1)}$ and $\mathrm{EIE}_{\mathrm{C}}^{(2)}$ can reduce to path specific effects under two specific scenarios with a known causal structure between the two mediators: one scenario where $\mathbf{M}_i^{(1)}$ causes $\mathbf{M}_i^{(2)}$ as depicted in Figure \ref{fig:DAG_esimtands_causallyordered} (A) and (B), and the other one where $\mathbf{M}_i^{(1)}$ and $\mathbf{M}_i^{(2)}$ are causally independent as depicted in Figure \ref{fig:DAG_esimtands_causallyindenpendent}  (A) and (B). 

\subsubsection{Causally ordered mediators}
When there is prior knowledge that $\mathbf{M}_i^{(1)}$ causes $\mathbf{M}_i^{(2)}$, extending the framework of path-specific effects developed in \citet{Albert2011} and \citet{Daniel2015} to CRTs, a total of $9$ causal pathways from the intervention to outcome exist as represented by Figure \ref{fig:DAG_esimtands_causallyordered}, including (1) $A_i \rightarrow Y_{ij}$, (2) $A_i \rightarrow M_{ij}^{(1)} \rightarrow Y_{ij}$, (3) $A_i \rightarrow \mathbf{M}_{i(-j)}^{(1)} \rightarrow Y_{ij}$, (4) $A_i \rightarrow M_{ij}^{(1)} \rightarrow M_{ij}^{(2)}  \rightarrow Y_{ij}$, (5) $A_i \rightarrow M_{ij}^{(1)}  \rightarrow \mathbf{M}_{i(-j)}^{(2)} \rightarrow Y_{ij}$, (6) $A_i \rightarrow \mathbf{M}_{i(-j)}^{(1)} \rightarrow M_{ij}^{(2)}  \rightarrow Y_{ij}$, (7) $A_i \rightarrow \mathbf{M}_{i(-j)}^{(1)}  \rightarrow \mathbf{M}_{i(-j)}^{(2)} \rightarrow Y_{ij}$, (8) $A_i \rightarrow M_{ij}^{(2)} \rightarrow Y_{ij}$, and (9) $A_i \rightarrow \mathbf{M}_{i(-j)}^{(2)} \rightarrow Y_{ij}$.
The exit indirect effect through $\mathbf{M}_i^{(1)}$, $\mathrm{EIE}_{\mathrm{C}}^{(1)}$, compares the potential outcome that activates all of the $9$ pathways with one that deactivates pathways (2) and (3), i.e., the two blue-hightlighted pathways in Figure \ref{fig:DAG_esimtands_causallyordered} (A). In other words, $\mathrm{EIE}_{\mathrm{C}}^{(1)}$ picks up all interventions pathways that immediately set through $\mathbf{M}_i^{(1)}$ and then immediately move toward the outcome after wards. As illustrated in Figure \ref{fig:DAG_esimtands_causallyordered} (B), the exit indirect effect through $\mathbf{M}_i^{(2)}$, $\mathrm{EIE}_{\mathrm{C}}^{(2)}$, compares the potential outcome that activates all pathways with one that deactivates pathways (4)-(9). That is, $\mathrm{EIE}_{\mathrm{C}}^{(2)}$ combines all causal pathways that set through $\mathbf{M}_i^{(2)}$ regardless of whether or not they previously set through $\mathbf{M}_i^{(1)}$. In this case, because $\mathbf{M}_i^{(2)}$ is affected by $\mathbf{M}_i^{(1)}$ but cannot causally affect $\mathbf{M}_i^{(1)}$, we have that 
\begin{align*}
    Y_{ij}(1,\mathbf{M}_i^{(1)}(1), \mathbf{M}_i^{(2)}(a)) = Y_{ij}(1, \mathbf{M}_i^{(2)}(a))
\end{align*}
by composition of potential values, and $\mathrm{EIE}_{\mathrm{C}}^{(2)}$ reduces to the more familiar natural indirect effect through $\mathbf{M}_i^{(2)}$:
\begin{align*}
    \mathrm{NIE}_C^{(2)} = \E\left[ \frac{1}{N_i} \sum_{j=1}^{N_i} \left\{ Y_{ij}(1, \mathbf{M}_i^{(2)}(1)) - Y_{ij}(1, \mathbf{M}_i^{(2)}(0)) \right\} \right],
\end{align*}
which compares the potential outcomes when $\mathbf{M}_i^{(2)}$ switches from its natural value under the control condition to that under the treated condition, while maintaining the treated condition in all other arguments of the potential outcomes.

\begin{figure}[htbp]
\centering 
 \begin{tikzpicture}
 \node[] (xx) at (2.7,2.5) {(A) $\text{EIE}_{c}^{(1)}$};
 
    \node[state] (a) at (0,0) {$A_i$};
    \node[state] (m1) at (1.5,1.5) {$M_{ij}^{(1)}$};
    \node[state] (m2) at (4, 1.5)  {$\mathbf{M}_{i(-j)}^{(1)}$};
    \node[state] (y) at (6,0) {$Y_{ij}$};
    \node[state] (m3) at (1.5,-1.5) {$M_{ij}^{(2)}$};
    \node[state] (m4) at (4, -1.5) {$\mathbf{M}_{i(-j)}^{(2)}$};

    \path (a) edge (y);
    \path [line width=1pt, blue] (a) edge (m1);
    \path [line width=1pt, blue] (a) edge (m2);
    \path (a) edge (m3);
    \path (a) edge (m4);
    
    \path [line width=1pt, blue] (m1) edge (y);
    \path [line width=1pt, blue] (m2) edge (y);
    \path  (m3) edge (y);
    \path  (m4) edge (y);
    
    \draw [dashed,-] (m1) edge (m2);
    \draw  (m1) edge (m3);
    \draw  (m1) edge (m4);
    \draw  (m2) edge (m3);
    \draw  (m2) edge (m4);
    \draw [dashed,-] (m3) edge (m4);
    
\end{tikzpicture}
\vspace{0.2in}\hspace{0.2in}
 \begin{tikzpicture}
 \node[] (xx) at (2.7,2.5) {(B) $\text{EIE}_{c}^{(2)}$};
 
    \node[state] (a) at (0,0) {$A_i$};
    \node[state] (m1) at (1.5,1.5) {$M_{ij}^{(1)}$};
    \node[state] (m2) at (4, 1.5)  {$\mathbf{M}_{i(-j)}^{(1)}$};
    \node[state] (y) at (6,0) {$Y_{ij}$};
    \node[state] (m3) at (1.5,-1.5) {$M_{ij}^{(2)}$};
    \node[state] (m4) at (4, -1.5) {$\mathbf{M}_{i(-j)}^{(2)}$};

    \path (a) edge (y);
    \path [line width=1pt, blue] (a) edge (m1);
    \path [line width=1pt, blue] (a) edge (m2);
    \path [line width=1pt, blue] (a) edge (m3);
    \path [line width=1pt, blue] (a) edge (m4);
    
    \path  (m1) edge (y);
    \path  (m2) edge (y);
    \path [line width=1pt, blue] (m3) edge (y);
    \path [line width=1pt, blue] (m4) edge (y);
    
    \draw [dashed,-] (m1) edge (m2);
    \draw [line width=1pt, blue] (m1) edge (m3);
    \draw [line width=1pt, blue] (m1) edge (m4);
    \draw [line width=1pt, blue] (m2) edge (m3);
    \draw [line width=1pt, blue] (m2) edge (m4);
    \draw [dashed,-] (m3) edge (m4);
    
\end{tikzpicture}
\vspace{0.2in}\hspace{0.2in}
\begin{tikzpicture}
 \node[] (xx) at (2.7,2.5) {(C) $\text{ESME}_{c}^{(1)}$};
 
    \node[state] (a) at (0,0) {$A_i$};
    \node[state] (m1) at (1.5,1.5) {$M_{ij}^{(1)}$};
    \node[state] (m2) at (4, 1.5)  {$\mathbf{M}_{i(-j)}^{(1)}$};
    \node[state] (y) at (6,0) {$Y_{ij}$};
    \node[state] (m3) at (1.5,-1.5) {$M_{ij}^{(2)}$};
    \node[state] (m4) at (4, -1.5) {$\mathbf{M}_{i(-j)}^{(2)}$};

    \path (a) edge (y);
    \path (a) edge (m1);
    \path [line width=1pt, blue] (a) edge (m2);
    \path (a) edge (m3);
    \path (a) edge (m4);
    
    \path  (m1) edge (y);
    \path [line width=1pt, blue] (m2) edge (y);
    \path  (m3) edge (y);
    \path  (m4) edge (y);
    
    \draw [dashed,-] (m1) edge (m2);
    \draw  (m1) edge (m3);
    \draw  (m1) edge (m4);
    \draw   (m2) edge (m3);
    \draw   (m2) edge (m4);
    \draw [dashed,-] (m3) edge (m4);
    
\end{tikzpicture}
\vspace{0.2in}\hspace{0.2in}
 \begin{tikzpicture}
 \node[] (xx) at (2.7,2.5) {(D) $\text{ESME}_{c}^{(2)}$};
 
    \node[state] (a) at (0,0) {$A_i$};
    \node[state] (m1) at (1.5,1.5) {$M_{ij}^{(1)}$};
    \node[state] (m2) at (4, 1.5)  {$\mathbf{M}_{i(-j)}^{(1)}$};
    \node[state] (y) at (6,0) {$Y_{ij}$};
    \node[state] (m3) at (1.5,-1.5) {$M_{ij}^{(2)}$};
    \node[state] (m4) at (4, -1.5) {$\mathbf{M}_{i(-j)}^{(2)}$};

    \path (a) edge (y);
    \path [line width=1pt, blue] (a) edge (m1);
    \path [line width=1pt, blue] (a) edge (m2);
    \path  (a) edge (m3);
    \path [line width=1pt, blue] (a) edge (m4);
    
    \path  (m1) edge (y);
    \path  (m2) edge (y);
    \path  (m3) edge (y);
    \path [line width=1pt, blue] (m4) edge (y);
    
    \draw [dashed,-] (m1) edge (m2);
    \draw  (m1) edge (m3);
    \draw [line width=1pt, blue] (m1) edge (m4);
    \draw  (m2) edge (m3);
    \draw [line width=1pt, blue] (m2) edge (m4);
    \draw [dashed,-] (m3) edge (m4);
    
\end{tikzpicture}
\vspace{0.2in}\hspace{0.2in}
 \begin{tikzpicture}
 \node[] (xx) at (2.7,2.5) {(E) $\text{EIME}_{c}^{(1)}$};
 
    \node[state] (a) at (0,0) {$A_i$};
    \node[state] (m1) at (1.5,1.5) {$M_{ij}^{(1)}$};
    \node[state] (m2) at (4, 1.5)  {$\mathbf{M}_{i(-j)}^{(1)}$};
    \node[state] (y) at (6,0) {$Y_{ij}$};
    \node[state] (m3) at (1.5,-1.5) {$M_{ij}^{(2)}$};
    \node[state] (m4) at (4, -1.5) {$\mathbf{M}_{i(-j)}^{(2)}$};

    \path (a) edge (y);
    \path [line width=1pt, blue] (a) edge (m1);
    \path  (a) edge (m2);
    \path (a) edge (m3);
    \path (a) edge (m4);
    
    \path [line width=1pt, blue] (m1) edge (y);
    \path  (m2) edge (y);
    \path  (m3) edge (y);
    \path  (m4) edge (y);
    
    \draw [dashed,-] (m1) edge (m2);
    \draw  (m1) edge (m3);
    \draw  (m1) edge (m4);
    \draw  (m2) edge (m3);
    \draw  (m2) edge (m4);
    \draw [dashed,-] (m3) edge (m4);
    
\end{tikzpicture}
\vspace{0.2in}\hspace{0.2in}
 \begin{tikzpicture}
 \node[] (xx) at (2.7,2.5) {(F) $\text{EIME}_{c}^{(2)}$};
 
    \node[state] (a) at (0,0) {$A_i$};
    \node[state] (m1) at (1.5,1.5) {$M_{ij}^{(1)}$};
    \node[state] (m2) at (4, 1.5)  {$\mathbf{M}_{i(-j)}^{(1)}$};
    \node[state] (y) at (6,0) {$Y_{ij}$};
    \node[state] (m3) at (1.5,-1.5) {$M_{ij}^{(2)}$};
    \node[state] (m4) at (4, -1.5) {$\mathbf{M}_{i(-j)}^{(2)}$};

    \path (a) edge (y);
    \path [line width=1pt, blue] (a) edge (m1);
    \path [line width=1pt, blue] (a) edge (m2);
    \path [line width=1pt, blue] (a) edge (m3);
    \path (a) edge (m4);
    
    \path  (m1) edge (y);
    \path  (m2) edge (y);
    \path [line width=1pt, blue] (m3) edge (y);
    \path  (m4) edge (y);
    
    \draw [dashed,-] (m1) edge (m2);
    \draw [line width=1pt, blue] (m1) edge (m3);
    \draw  (m1) edge (m4);
    \draw [line width=1pt, blue] (m2) edge (m3);
    \draw  (m2) edge (m4);
    \draw [dashed,-] (m3) edge (m4);
    
\end{tikzpicture}
\caption{Mediation directed acyclic graph when $\mathbf{M}_i^{(1)}$ causes $\mathbf{M}_i^{(2)}$.}
\label{fig:DAG_esimtands_causallyordered}
\end{figure}

\subsubsection{Causally independent mediators}
If $\mathbf{M}_i^{(1)}$ and $\mathbf{M}_i^{(2)}$ are causally independent, no pathways from one mediator to the other mediator exists (i.e., no pathways (4)-(7)), leaving only 5 causal pathways from intervention to the outcome as represented in Figure \ref{fig:DAG_esimtands_causallyindenpendent} (A) and (B). We denote these 5 causal pathways as (1) $A_i \rightarrow Y_{ij}$, (2) $A_i \rightarrow M_{ij}^{(1)} \rightarrow Y_{ij}$, (3) $A_i \rightarrow \mathbf{M}_{i(-j)}^{(1)} \rightarrow Y_{ij}$, (4) $A_i \rightarrow M_{ij}^{(2)} \rightarrow Y_{ij}$, and (5) $A_i \rightarrow \mathbf{M}_{i(-j)}^{(2)} \rightarrow Y_{ij}$.
In this case, $\mathrm{EIE}_{\mathrm{C}}^{(k)}$ compares the potential outcome that activates all pathways with one that deactivates the pathways setting through $\mathbf{M}_i^{(k)}$; that is, $\mathrm{EIE}_{\mathrm{C}}^{(1)}$ summarizes the causal pathways through (2) and (3) and $\mathrm{EIE}_{\mathrm{C}}^{(2)}$ summarizes the causal pathways through (4) and (5). Also, because $\mathbf{M}_i^{(1)}$ and $\mathbf{M}_i^{(2)}$ are causally independent, we have that 
\begin{align*}
    Y_{ij}(1,\mathbf{M}_i^{(1)}(1), \mathbf{M}_i^{(2)}(a)) &= Y_{ij}(1, \mathbf{M}_i^{(2)}(a)),\\
    Y_{ij}(1,\mathbf{M}_i^{(2)}(1), \mathbf{M}_i^{(1)}(a)) &= Y_{ij}(1, \mathbf{M}_i^{(1)}(a)),
\end{align*}
by composition of potential values. Thus, $\mathrm{EIE}_{\mathrm{C}}^{(k)}$ reduces to natural indirect effect through $\mathbf{M}_i^{(k)}$, that is, for $k=1,2$, we have
\begin{align*}
    \mathrm{NIE}_C^{(k)} = \E\left[ \frac{1}{N_i} \sum_{j=1}^{N_i} \left\{ Y_{ij}(1, \mathbf{M}_i^{(k)}(1)) - Y_{ij}(1, \mathbf{M}_i^{(k)}(0)) \right\} \right].
\end{align*}

\begin{figure}[p]
\centering 
 \begin{tikzpicture}
 \node[] (xx) at (2.7,2.5) {(A) $\text{EIE}_{c}^{(1)}$};
  
    \node[state] (a) at (0,0) {$A_i$};
    \node[state] (m1) at (1.5,1.5) {$M_{ij}^{(1)}$};
    \node[state] (m2) at (4, 1.5)  {$\mathbf{M}_{i(-j)}^{(1)}$};
    \node[state] (y) at (6,0) {$Y_{ij}$};
    \node[state] (m3) at (1.5,-1.5) {$M_{ij}^{(2)}$};
    \node[state] (m4) at (4, -1.5) {$\mathbf{M}_{i(-j)}^{(2)}$};

    \path (a) edge (y);
    \path [line width=1pt, blue] (a) edge (m1);
    \path [line width=1pt, blue] (a) edge (m2);
    \path (a) edge (m3);
    \path (a) edge (m4);
    
    \path [line width=1pt, blue] (m1) edge (y);
    \path [line width=1pt, blue] (m2) edge (y);
    \path (m3) edge (y);
    \path (m4) edge (y);
    
    
    
    \draw [dashed,-] (m1) edge (m2);
    \draw [dashed,-] (m3) edge (m4);
    
\end{tikzpicture}
\vspace{0.2in}\hspace{0.2in}
 \begin{tikzpicture}
 \node[] (xx) at (2.7,2.5) {(B) $\text{EIE}_{c}^{(2)}$};
 
    \node[state] (a) at (0,0) {$A_i$};
    \node[state] (m1) at (1.5,1.5) {$M_{ij}^{(1)}$};
    \node[state] (m2) at (4, 1.5)  {$\mathbf{M}_{i(-j)}^{(1)}$};
    \node[state] (y) at (6,0) {$Y_{ij}$};
    \node[state] (m3) at (1.5,-1.5) {$M_{ij}^{(2)}$};
    \node[state] (m4) at (4, -1.5) {$\mathbf{M}_{i(-j)}^{(2)}$};

    \path (a) edge (y);
    \path (a) edge (m1);
    \path (a) edge (m2);
    \path [line width=1pt, blue] (a) edge (m3);
    \path [line width=1pt, blue] (a) edge (m4);
    
    \path (m1) edge (y);
    \path (m2) edge (y);
    \path [line width=1pt, blue] (m3) edge (y);
    \path [line width=1pt, blue] (m4) edge (y);
    
    \draw [dashed,-] (m1) edge (m2);
    \draw [dashed,-] (m3) edge (m4);
    
\end{tikzpicture}
\vspace{0.2in}\hspace{0.2in}
\begin{tikzpicture}
 \node[] (xx) at (2.7,2.5) {(C) $\text{ESME}_{c}^{(1)}$};
 
    \node[state] (a) at (0,0) {$A_i$};
    \node[state] (m1) at (1.5,1.5) {$M_{ij}^{(1)}$};
    \node[state] (m2) at (4, 1.5)  {$\mathbf{M}_{i(-j)}^{(1)}$};
    \node[state] (y) at (6,0) {$Y_{ij}$};
    \node[state] (m3) at (1.5,-1.5) {$M_{ij}^{(2)}$};
    \node[state] (m4) at (4, -1.5) {$\mathbf{M}_{i(-j)}^{(2)}$};

    \path (a) edge (y);
    \path (a) edge (m1);
    \path [line width=1pt, blue] (a) edge (m2);
    \path (a) edge (m3);
    \path (a) edge (m4);
    
    \path (m1) edge (y);
    \path [line width=1pt, blue] (m2) edge (y);
    \path (m3) edge (y);
    \path (m4) edge (y);
    
    \draw [dashed,-] (m1) edge (m2);
    \draw [dashed,-] (m3) edge (m4);
    
\end{tikzpicture}
\vspace{0.2in}\hspace{0.2in}
 \begin{tikzpicture}
 \node[] (xx) at (2.7,2.5) {(D) $\text{ESME}_{c}^{(2)}$};
 
    \node[state] (a) at (0,0) {$A_i$};
    \node[state] (m1) at (1.5,1.5) {$M_{ij}^{(1)}$};
    \node[state] (m2) at (4, 1.5)  {$\mathbf{M}_{i(-j)}^{(1)}$};
    \node[state] (y) at (6,0) {$Y_{ij}$};
    \node[state] (m3) at (1.5,-1.5) {$M_{ij}^{(2)}$};
    \node[state] (m4) at (4, -1.5) {$\mathbf{M}_{i(-j)}^{(2)}$};

    \path (a) edge (y);
    \path (a) edge (m1);
    \path (a) edge (m2);
    \path (a) edge (m3);
    \path [line width=1pt, blue] (a) edge (m4);
    
    \path (m1) edge (y);
    \path (m2) edge (y);
    \path (m3) edge (y);
    \path [line width=1pt, blue] (m4) edge (y);
    
    \draw [dashed,-] (m1) edge (m2);
    \draw [dashed,-] (m3) edge (m4);
    
\end{tikzpicture}
\vspace{0.2in}\hspace{0.2in}
\begin{tikzpicture}
 \node[] (xx) at (2.7,2.5) {(E) $\text{EIME}_{c}^{(1)}$};
 
    \node[state] (a) at (0,0) {$A_i$};
    \node[state] (m1) at (1.5,1.5) {$M_{ij}^{(1)}$};
    \node[state] (m2) at (4, 1.5)  {$\mathbf{M}_{i(-j)}^{(1)}$};
    \node[state] (y) at (6,0) {$Y_{ij}$};
    \node[state] (m3) at (1.5,-1.5) {$M_{ij}^{(2)}$};
    \node[state] (m4) at (4, -1.5) {$\mathbf{M}_{i(-j)}^{(2)}$};

    \path (a) edge (y);
    \path [line width=1pt, blue] (a) edge (m1);
    \path (a) edge (m2);
    \path (a) edge (m3);
    \path (a) edge (m4);
    
    \path [line width=1pt, blue] (m1) edge (y);
    \path (m2) edge (y);
    \path (m3) edge (y);
    \path (m4) edge (y);
    
    \draw [dashed,-] (m1) edge (m2);
    \draw [dashed,-] (m3) edge (m4);
    
\end{tikzpicture}
\vspace{0.2in}\hspace{0.2in}
 \begin{tikzpicture}
 \node[] (xx) at (2.7,2.5) {(F) $\text{EIME}_{c}^{(2)}$};
 
    \node[state] (a) at (0,0) {$A_i$};
    \node[state] (m1) at (1.5,1.5) {$M_{ij}^{(1)}$};
    \node[state] (m2) at (4, 1.5)  {$\mathbf{M}_{i(-j)}^{(1)}$};
    \node[state] (y) at (6,0) {$Y_{ij}$};
    \node[state] (m3) at (1.5,-1.5) {$M_{ij}^{(2)}$};
    \node[state] (m4) at (4, -1.5) {$\mathbf{M}_{i(-j)}^{(2)}$};

    \path (a) edge (y);
    \path (a) edge (m1);
    \path (a) edge (m2);
    \path [line width=1pt, blue] (a) edge (m3);
    \path (a) edge (m4);
    
    \path (m1) edge (y);
    \path (m2) edge (y);
    \path [line width=1pt, blue] (m3) edge (y);
    \path (m4) edge (y);
    
    \draw [dashed,-] (m1) edge (m2);
    \draw [dashed,-] (m3) edge (m4);
    
\end{tikzpicture}
\caption{Mediation directed acyclic graph when $\mathbf{M}_i^{(1)}$ are $\mathbf{M}_i^{(2)}$ causally independent.}
\label{fig:DAG_esimtands_causallyindenpendent}
\end{figure}

\subsection{Interpretation as interventional mediation effects}
\label{sec:Interpretation_IE}
The causal estimands defined in the main manuscript employ cross-world counterfactuals about which information cannot be obtained even from experimental data. Thus, the researchers are obligated to make strong untestable assumptions like Assumptions \ref{asmp:cond_homogeneity} and \ref{asmp:no_cross_intra_corr}. 
Interventional effects, introduced by \citet{VanderWeele2014_effect_decomposition}, provide a way to define direct and indirect effects without relying on cross-world counterfactuals. They do so by considering interventions that change the distribution of the mediator rather than setting it to specific values. 
For multiple mediators with unknown causal structures with independent data, \citet{Vansteelandt2017} defined interventional effects and a corresponding decomposition using a random draw of mediators. Here we discuss how our identification results have causal interpretations under the interventional causal mediation framework under Assumptions \ref{asmp:sutva}--\ref{asmp:si}. We adapt the existing definition of the interventional effects for independent data (e.g., \citet{Vansteelandt2017,BenkeserRan2021}) and define the interventional effects under CRTs. In particular, the interventional exit indirect effect (IEIE) and the interventional exit spillover mediation effect (IESME) are defined as:
\begin{equation}
\label{eq:interventional_effects}
\begin{split}
    \mathrm{IEIE}_{\mathrm{C}}^{(k)}    &= \E \left[ \frac{1}{N_i}\sum_{j=1}^{N_i} \int_{\mathcal{M}^{(3-k)}} \int_{\mathcal{M}^{(k)}} \E \left[Y_{ij}(1, \mathbf{m}_i^{(k)}, \mathbf{m}_i^{(3-k)}) \mid \mathbf{C}_i, N_i \right] \right. \\
    & \quad\quad\quad \left. \left \{ dF_{\mathbf{G}_{i}^{(k)}(1) \mid \mathbf{C}_i, N_i}(\mathbf{m}_i^{(k)}) - dF_{\mathbf{G}_{i}^{(k)}(0) \mid \mathbf{C}_i, N_i}(\mathbf{m}_i^{(k)}) \right \} dF_{\mathbf{G}_{i}^{(3-k)}(1) \mid \mathbf{C}_i, N_i}(\mathbf{m}_i^{(3-k)}) \right] \\
    \mathrm{IESME}_{\mathrm{C}}^{(k)} &= \E \left[ \frac{1}{N_i}\sum_{j=1}^{N_i} \int_{\mathcal{M}^{(3-k)}} \int_{\mathcal{M}^{(k)}} \E \left[Y_{ij}(1, m_{ij}^{(k)},\mathbf{m}_{i(-j)}^{(k)}, \mathbf{m}_i^{(3-k)}) \mid \mathbf{C}_i, N_i \right] dF_{G_{ij}^{(k)}(1) \mid \mathbf{C}_i, N_i}(m_{ij}^{(k)}) \right. \\
    & \quad\quad\quad \left.  \left \{ dF_{\mathbf{G}_{i(-j)}^{(k)}(1) \mid \mathbf{C}_i, N_i}(\mathbf{m}_{i(-j)}^{(k)}) - dF_{\mathbf{G}_{i(-j)}^{(k)}(0) \mid \mathbf{C}_i, N_i}(\mathbf{m}_{i(-j)}^{(k)}) \right \} dF_{\mathbf{G}_{i}^{(3-k)}(1) \mid \mathbf{C}_i, N_i}(\mathbf{m}_i^{(3-k)}) \right], 
\end{split}
\end{equation}
where $G_{ij}^{(k)}(a)$, $\mathbf{G}_{i}^{(k)}(a)$ and $\mathbf{G}_{i(-j)}^{(k)}(a)$ denotes a randomly generated mediators from the conditional density of $M_{ij}^{(k)}(a)$, $\mathbf{M}_{i}^{(k)}(a)$ and $\mathbf{M}_{i(-j)}^{(k)}(a)$ given covariates. $\mathrm{IEIE}_{\mathrm{C}}^{(k)}$ represents the effect of shifting the distribution of $\mathbf{M}^{(k)}$ from its counterfactual distribution given covariates at intervention level $0$ to that at level $1$, while keeping the intervention fixed at level $1$ and setting the other mediator $\mathbf{M}^{(3-k)}$ to random subject-specific draws from its distribution at level $0$ for all individuals within the same cluster. Similarly, $\mathrm{IESME}_{\mathrm{C}}^{(k)}$  captures the effect of shifting the distribution of $M^{(k)}$ for all peers in the same cluster (i.e., the distribution of $\mathbf{M}_{i(-j)}^{(k)}$), while fixing the distribution of the individual's own mediator $M_{ij}^{(k)}$ and the other mediator for all units $\mathbf{M}^{(3-k)}$.
Possible differences in an individual's potential outcomes for $\mathrm{IESME}_{\mathrm{C}}^{(k)}$ are attributed to the distribution of counterfactual mediators among peers, interpreted as spillover effects. Interventional and exit indirect effects for $M^{(k)}$ coincide if there is a sufficiently rich set of covariates, such that the joint distribution of potential mediators becomes deterministic.

Under Assumptions \ref{asmp:sutva}--\ref{asmp:si}, it is straightforward to show that the counterfactual means $\E \left[Y_{ij}(1, \mathbf{m}_i^{(k)}, \mathbf{m}_i^{(3-k)}) \mid \mathbf{C}_i, N_i \right]$ and $\E \left[Y_{ij}(1, m_{ij}^{(k)},\mathbf{m}_{i(-j)}^{(k)}, \mathbf{m}_i^{(3-k)}) \mid \mathbf{C}_i, N_i \right]$ are  identified by $\mu_{\mathbf{C}, N}(a, \mathbf{m}^{(k)}, \mathbf{m}^{(3-k)})$ and $\kappa_{\mathbf{C}, N}(a, m_{\cdot j}, \mathbf{m}^{(k)}_{\cdot (-j)}, \mathbf{m}^{(3-k)})$, and $F_{\mathbf{G}_{i}^{(k)}(a) \mid \mathbf{C}_i, N_i}(\mathbf{m}_i^{(k)})$ is identified by $F_{\mathbf{M}_{i}^{(k)} \mid A_i=a, \mathbf{C}_i, N_i}(\mathbf{m}_i^{(k)})$ for $a=0,1$ and $k=1,2$, respectively. Plugging in these for \eqref{eq:interventional_effects} leads to the same identification results in Theorems \ref{thm:eie_identification} and \ref{thm:esme_identification}. Therefore, Theorems \ref{thm:eie_identification} and \ref{thm:esme_identification} can be interpreted as the identification formulas for the interventional effects \ref{eq:interventional_effects}, which are valid without the assumptions involving cross-world mediators and outcomes (Assumptions \ref{asmp:cond_homogeneity} and \ref{asmp:no_cross_intra_corr}).

\subsection{Identification for mediation effects with $K$ mediators}
\label{sec:K_mediator_identification}
In this section, we generalize the identification results with $K=2$ in the main manuscript and provide the definitions of the EIE and ESME effects with $K$ mediators and identification formulae for those effects.

For unit $j$ in cluster $i$, we consider $K$ potential mediators $M_{ij}^{(1)}(a), \ldots, M_{ij}^{(K)}(a)$ and potential outcomes $Y_{ij}(a,\mathbf{m}_i^{(1)}, \ldots, \mathbf{m}_i^{(K)})$ for $a=0,1$. We define the EIE and ESME for the mediator $k$ as:
\begin{align*}
    \mathrm{EIE}_{\mathrm{C}}^{(k)} 
    &= \E \left[ \frac{1}{N_i}\sum_{j=1}^{N_i} \left \{ Y_{ij}(1, \mathbf{M}_i^{(k)}(1), \mathbf{M}_i^{(-k)}(1)) - Y_{ij}(1, \mathbf{M}_i^{(k)}(0), \mathbf{M}_i^{(-k)}(1)) \right \} \right]  \\
    \mathrm{ESME}_{\mathrm{C}}^{(k)} &= \E \left[ \frac{1}{N_i} \sum_{j=1}^{N_i} \left\{ Y_{ij}(1, M_{ij}^{(k)}(1), \mathbf{M}_{i(-j)}^{(k)}(1), \mathbf{M}_i^{(-k)}(1)) - Y_{ij}(1, M_{ij}^{(k)}(1), \mathbf{M}_{i(-j)}^{(k)}(0), \mathbf{M}_i^{(-k)}(1)) \right\} \right],
\end{align*}
where we write $\mathbf{M}_i^{(-k)}(a) = [\mathbf{M}_i^{(1)}(a), \ldots, \mathbf{M}_i^{(k-1)}(a), \mathbf{M}_i^{(k+1)}(a), \ldots,\mathbf{M}_i^{(K)}(a)]^\top$ for $a=0,1$. The interpretations of the EIE and ESME remain the same as in the case with two mediators.

Next, we introduce a set of identification assumptions. 
\begin{assumption}[Sequential ignorability for $K$ mediators]
\label{asmp:si_multiple}
    $$ Y_{ij}(a,\mathbf{m}_i^{(1)}, \ldots, \mathbf{m}_i^{(K)}) \indep \{ \mathbf{M}_i^{(1)}(0), \mathbf{M}_i^{(1)}(1), \ldots, \mathbf{M}_i^{(K)}(0), \mathbf{M}_i^{(K)}(0) \} \mid \{ A_i, \mathbf{C}_i, N_i \} $$
    for all $ i, j$, $ a \in \{0,1\}$, and $\mathbf{m}_i^{(1)}, \ldots, \mathbf{m}_i^{(K)} $ over their valid support.
\end{assumption}

\begin{assumption}[Conditional homogeneity]
\label{asmp:cond_homogeneity_multiple}
    \begin{align*}
        &\E \left[ \frac{1}{N_i} \sum_{j=1}^{N_i} \left\{ Y_{ij}(1, M_{ij}^{(k)}(1), \mathbf{M}_{i(-j)}^{(k)}(1), \mathbf{M}_i^{(-k)}(1)) \right.\right.\\
        &   \left. \left. \qquad \qquad - Y_{ij}(1, M_{ij}^{(k)}(a), \mathbf{M}_{i(-j)}^{(k)}(0), \mathbf{M}_i^{(-k)}(1)) \right\} \mid \mathbf{M}_i^{(-k)}(1)=\mathbf{m}_i^{(-k)}, \mathbf{C}_i, N_i \right] \\
        =& \E \left[ \frac{1}{N_i} \sum_{j=1}^{N_i} \left\{ Y_{ij}(1, M_{ij}^{(k)}(1), \mathbf{M}_{i(-j)}^{(k)}(1), \mathbf{m}_i^{(-k)}) - Y_{ij}(1, M_{ij}^{(k)}(a), \mathbf{M}_{i(-j)}^{(k)}(0), \mathbf{m}_i^{(-k)}) \right\} \mid \mathbf{C}_i, N_i \right],
    \end{align*}
    for $a \in \{0,1\}$, and $\mathbf{m}_i^{(1)}, \ldots, \mathbf{m}_i^{(K)} $  over their valid support.
\end{assumption}

\begin{theorem}
\label{thm:eie_identification_multiple}
        Under Assumptions \ref{asmp:sutva}, \ref{asmp:randomization}, \ref{asmp:superpopulation}, \ref{asmp:si_multiple}, and \ref{asmp:cond_homogeneity_multiple}$, \mathrm{EIE}_{\mathrm{C}}^{(k)} $ are nonparametrically identified as follows: 
        \begin{align*}
             \E_{\mathbf{C}, N} \left[ \frac{1}{N} \sum_{j=1}^{N} \left \{\int_{\mathbf{m}^{(-k)}} \int_{\mathcal{M}^{(k)}} \mu_{\mathbf{C}, N}(1, \mathbf{m}^{(k)}, \mathbf{m}^{(-k)}) \right. dF_{\mathbf{M}^{(k)} \mid A=1, \mathbf{C}, N} (\mathbf{m}^{(k)}) dF_{\mathbf{M}^{(-k)} \mid A=1, \mathbf{C}, N} (\mathbf{m}^{(-k)}) \right. \\
             \left. - \int_{\mathbf{m}^{(-k)}} \int_{\mathcal{M}^{(k)}} \mu_{\mathbf{C}, N}(1, \mathbf{m}^{(k)}, \mathbf{m}^{(-k)})  \left.  dF_{\mathbf{M}^{(k)} \mid A=0, \mathbf{C}, N} (\mathbf{m}^{(k)}) dF_{\mathbf{M}^{(-k)} \mid A=1, \mathbf{C}, N} (\mathbf{m}^{(-k)}) \right\}\right],
        \end{align*}
        where $\mu_{\mathbf{C}, N}(a, \mathbf{m}^{(k)}, \mathbf{m}^{(-k)})=\E \left[ Y_{\cdot j} \mid A=a, \mathbf{M}^{(k)}=\mathbf{m}^{(k)}, \mathbf{M}^{(-k)}=\mathbf{m}^{(-k)}, \mathbf{C}, N \right]$.
\end{theorem}
\begin{proof}
    We apply the same proof procedures as in \ref{proof:eie}, but under Assumptions \ref{asmp:si_multiple} and \ref{asmp:cond_homogeneity_multiple} instead of Assumptions \ref{asmp:si} and \ref{asmp:cond_homogeneity}.
\end{proof}

\begin{theorem} 
\label{thm:esme_identification_multiple}
Under Assumptions \ref{asmp:sutva}, \ref{asmp:randomization}, \ref{asmp:superpopulation}, \ref{asmp:no_cross_intra_corr}, \ref{asmp:si_multiple}, and  \ref{asmp:cond_homogeneity_multiple}, $\mathrm{ESME}_{\mathrm{C}}^{(k)}$ are nonparametrically identified as follows:
    \begin{align*}
        &\E\left[ \frac{1}{N}\sum_{j=1}^{N} \left \{  \int_{\mathbf{m}^{(-k)}} \int_{\mathcal{M}^{(k)}} \kappa_{\mathbf{C}, N}(a, m_{\cdot j}, \mathbf{m}^{(k)}_{\cdot (-j)}, \mathbf{m}^{(-k)}) \right.\right. \\
        & \hspace{10.em} dF_{M^{(k)}_{\cdot j} \mid A=1,\mathbf{C}, N} (m_{\cdot j})dF_{\mathbf{M}^{(k)}_{\cdot, (-j)} \mid A=1,\mathbf{C}, N} ( \mathbf{m}^{(k)}_{\cdot (-j)}) dF_{\mathbf{M}^{(-k)} \mid A=1,\mathbf{C}, N} (\mathbf{m}^{(-k)})\\
        & \left.\left. \hspace{3.5em}  -\int_{\mathbf{m}^{(-k)}} \int_{\mathcal{M}^{(k)}} \kappa_{\mathbf{C}, N}(a, m_{\cdot j}, \mathbf{m}^{(k)}_{\cdot (-j)}, \mathbf{m}^{(-k)}) \right.\right. \\
        & \left.\left. \hspace{9.5em} dF_{M^{(k)}_{\cdot j} \mid A=1,\mathbf{C}, N} (m_{\cdot j})dF_{\mathbf{M}^{(k)}_{\cdot, (-j)} \mid A=0,\mathbf{C}, N} ( \mathbf{m}^{(k)}_{\cdot (-j)}) dF_{\mathbf{M}^{(-k)} \mid A=1,\mathbf{C}, N} (\mathbf{m}^{(-k)}) \right\}\right]
    \end{align*}
    where $\kappa_{\mathbf{C}, N}(a, m_{\cdot j}, \mathbf{m}^{(k)}_{\cdot (-j)}, \mathbf{m}^{(-k)})=\E\left[  Y_{\cdot  j} \middle| A=1, M_{\cdot j}^{(k)}=m_{\cdot j}, \mathbf{M}_{\cdot(-j)}^{(k)}=\mathbf{m}^{(k)}_{\cdot (-j)},  \mathbf{M}^{(-k)}=\mathbf{m}^{(-k)}, \mathbf{C}, N  \right] $.
\end{theorem}
\begin{proof}
    We apply the same proof procedures as in \ref{proof:esme}, but under Assumptions \ref{asmp:si_multiple} and \ref{asmp:cond_homogeneity_multiple} instead of Assumptions \ref{asmp:si} and \ref{asmp:cond_homogeneity}.
\end{proof}

\subsubsection{Interaction effects}
The total interaction effect with $ K $ mediators is expressed as 
$\mathrm{INT}^{(K)} = \sum_{k=1}^{K} \mathrm{EIE}_{\mathrm{C}}^{(k)} - \mathrm{NIE}_{\mathrm{C}}$, representing the difference between the NIE and the EIEs, as in the case where $ K = 2 $. This interaction effect also allows for a finer-grained decomposition using multi-way interaction terms.
We specifically consider the case with $K=3$. To simplify the notation, we suppress the indicator $ij$ and write $Y_{a\mathbf{m}_1\mathbf{m}_2\mathbf{m}_3} = Y(a,\mathbf{m}_1,\mathbf{m}_2, \mathbf{m}_3)$ in this section.

The two-way interaction effect between the mediator $k$ and $l$ ($k \neq l $) measures how the effect of one mediator depends on the level of another mediator. It is defined as:
\begin{align*}
    \mathrm{INT}_{\mathrm{C}}^{(1,2)} &= \E\left[ Y_{1\mathbf{M}^{(1)}(1)\mathbf{M}^{(2)}(1)\mathbf{M}^{(3)}(1)} - Y_{1\mathbf{M}^{(1)}(1)\mathbf{M}^{(2)}(0)\mathbf{M}^{(3)}(1)} \right]  \\
    &- \E \left[ Y_{1\mathbf{M}^{(1)}(0)\mathbf{M}^{(2)}(1)\mathbf{M}^{(3)}(1)} - Y_{1\mathbf{M}^{(1)}(0)\mathbf{M}^{(2)}(0)\mathbf{M}^{(3)}(1)} \right], \\
    \mathrm{INT}_{\mathrm{C}}^{(1,3)} &= \E\left[ Y_{1\mathbf{M}^{(1)}(1)\mathbf{M}^{(2)}(1)\mathbf{M}^{(3)}(1)} - Y_{1\mathbf{M}^{(1)}(1)\mathbf{M}^{(2)}(1)\mathbf{M}^{(3)}(0)} \right] \\
    &- \E \left[ Y_{1\mathbf{M}^{(1)}(0)\mathbf{M}^{(2)}(1)\mathbf{M}^{(3)}(1)} - Y_{1\mathbf{M}^{(1)}(0)\mathbf{M}^{(2)}(1)\mathbf{M}^{(3)}(0)} \right],\\
    \mathrm{INT}_{\mathrm{C}}^{(2,3)} &= \E\left[ Y_{1\mathbf{M}^{(1)}(1)\mathbf{M}^{(2)}(1)\mathbf{M}^{(3)}(1)} - Y_{1\mathbf{M}^{(1)}(1)\mathbf{M}^{(2)}(1)\mathbf{M}^{(3)}(0)} \right] \\
    & - \E\left[ Y_{1\mathbf{M}^{(1)}(1)\mathbf{M}^{(2)}(0)\mathbf{M}^{(3)}(1)} - Y_{1\mathbf{M}^{(1)}(1)\mathbf{M}^{(2)}(0)\mathbf{M}^{(3)}(0)} \right].
\end{align*}
The three-way interaction captures how the two-way interactions between any two mediators change when the third mediator changes level. It reflects the complexity of the combined effects of all three mediators on the outcome. It is defined as:
\begin{align*}
\mathrm{INT}_{\mathrm{C}}^{(1,2,3)} &=  \E \left[ Y_{1\mathbf{M}^{(1)}(1)\mathbf{M}^{(2)}(1)\mathbf{M}^{(3)}(1)} - Y_{1\mathbf{M}^{(1)}(1)\mathbf{M}^{(2)}(1)\mathbf{M}^{(3)}(0)} - Y_{1\mathbf{M}^{(1)}(1)\mathbf{M}^{(2)}(0)\mathbf{M}^{(3)}(1)} + Y_{1\mathbf{M}^{(1)}(1)\mathbf{M}^{(2)}(0)\mathbf{M}^{(3)}(0)} \right]  \\
& -  \E \left[ Y_{1\mathbf{M}^{(1)}(0)\mathbf{M}^{(2)}(1)\mathbf{M}^{(3)}(1)} - Y_{1\mathbf{M}^{(1)}(0)\mathbf{M}^{(2)}(1)\mathbf{M}^{(3)}(0)} - Y_{1\mathbf{M}^{(1)}(0)\mathbf{M}^{(2)}(0)\mathbf{M}^{(3)}(1)} + Y_{1\mathbf{M}^{(1)}(0)\mathbf{M}^{(2)}(0)\mathbf{M}^{(3)}(0)} \right].
\end{align*}
Given these interaction effects, the NIE is decomposed into:
\begin{align*}
    \mathrm{NIE}_{\mathrm{C}} = \sum_{k=1}^{3}\mathrm{EIE}_{\mathrm{C}}^{(k)} - \sum_{1\leq k < l \leq 3}\mathrm{INT}_{\mathrm{C}}^{(k,l)} + \mathrm{INT}_{\mathrm{C}}^{(1,2,3)}
\end{align*}

However, these interaction effects typically become less scientifically interesting as $K$ increases because the number of combinatoric interactions explodes, and consequently, they provide less clear interpretations of the mediation. Therefore, further generalization to $K>3$ mediators and the identification of each component are deferred for future research.

\begin{figure*}[htbp]
    \centering
    \includegraphics[width=\textwidth]{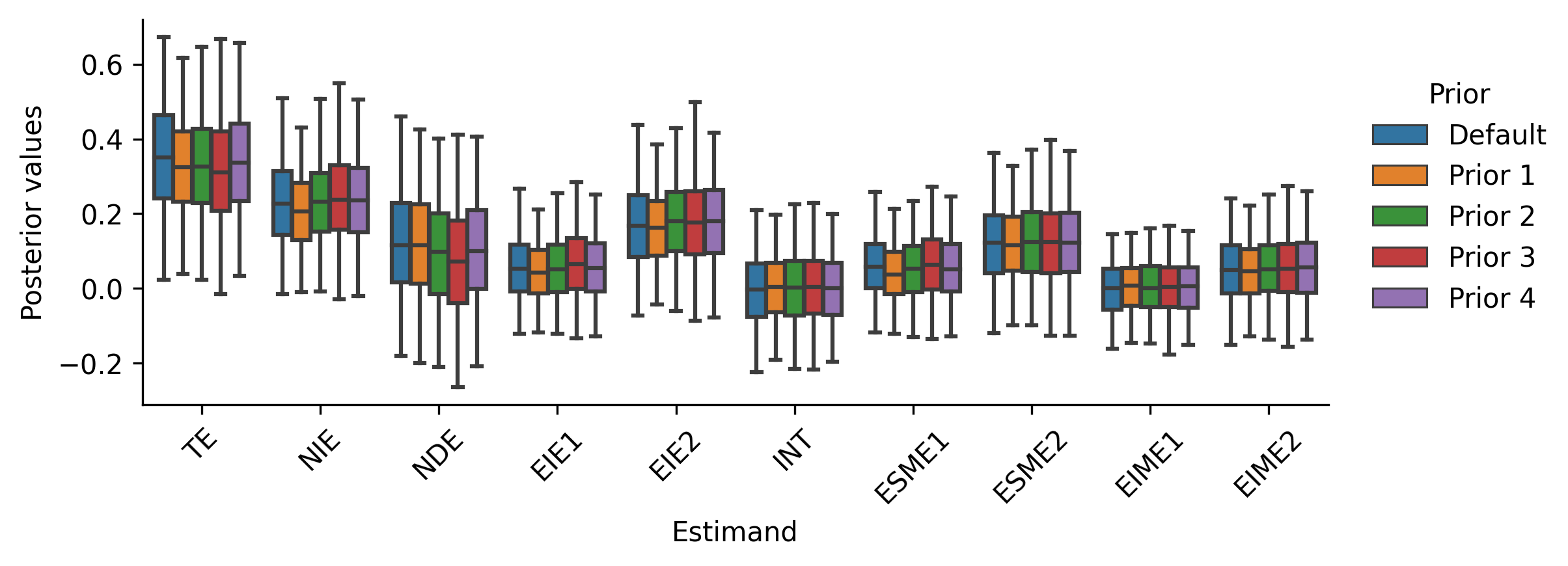}
    \caption{Sensitivity analysis with respect to more or less informative priors. The comparisons of the posterior means of each estimand under different prior specifications.}
    \label{fig:sensitivity}
\end{figure*}

\yo{
\section{Sensitivity analysis}
\label{sec:sensitivity_analysis}

\begin{table*}[htbp]
\centering
\caption{Summary of Prior Specifications}
\begin{adjustbox}{width=\textwidth}
\begin{tabular}{lccccc}
\hline
\textbf{Parameter} & \textbf{Default} & \textbf{Prior 1} & \textbf{Prior 2} & \textbf{Prior 3} & \textbf{Prior 4}\\
\hline
$\boldsymbol{\theta}_{lk}$ 
& $\mathrm{MVN}\bigl(\mathbf{0},10^2 I_{d_y}\bigr)$ 
& $\mathrm{MVN}\bigl(\mathbf{0},3^2 I_{d_y}\bigr)$ 
& $\mathrm{MVN}\bigl(\mathbf{0},30^2 I_{d_y}\bigr)$
& $\mathrm{MVN}\bigl(\mathbf{0},10^2 I_{d_y}\bigr)$
& $\mathrm{MVN}\bigl(\mathbf{0},10^2 I_{d_y}\bigr)$\\

$\sigma^2_{lk}$ 
& $\mathrm{IG}(2.0,\,1.0)$ 
& $\mathrm{IG}(1.0,\,1.0)$ 
& $\mathrm{IG}(1.0,\,1.0)$
& $\mathrm{IG}(0.2,\,0.2)$
& $\mathrm{IG}(5.0,\,5.0)$\\

$\boldsymbol{\gamma}_{1,lk},\,\boldsymbol{\gamma}_{2,lk}$ 
& $\mathrm{MVN}\bigl(\mathbf{0},10^2 I_{d_m}\bigr)$ 
& $\mathrm{MVN}\bigl(\mathbf{0},3^2 I_{d_m}\bigr)$ 
& $\mathrm{MVN}\bigl(\mathbf{0},30^2 I_{d_m}\bigr)$
& $\mathrm{MVN}\bigl(\mathbf{0},10^2 I_{d_m}\bigr)$
& $\mathrm{MVN}\bigl(\mathbf{0},10^2 I_{d_m}\bigr)$\\

$\boldsymbol{\Sigma}$ 
& $\mathrm{IW}\bigl(2.0,\,I_{2}\bigr)$ 
& $\mathrm{IW}\bigl(2.0,\,1.0\,I_{2}\bigr)$ 
& $\mathrm{IW}\bigl(2.0,\,1.0\,I_{2}\bigr)$
& $\mathrm{IW}\bigl(2.0,\,0.1\,I_{2}\bigr)$
& $\mathrm{IW}\bigl(10.0,\,10.0\,I_{2}\bigr)$\\

$\alpha_y,\,\alpha_m$ 
& $\mathrm{Ga}(1.0,\,1.0)$ 
& $\mathrm{Ga}(1.0,\,1.0)$ 
& $\mathrm{Ga}(1.0,\,1.0)$
& $\mathrm{Ga}(0.2,\,0.2)$
& $\mathrm{Ga}(5.0,\,5.0)$\\

$\beta_y,\,\beta_m$ 
& $\mathrm{Ga}(1.0,\,1.0)$ 
& $\mathrm{Ga}(1.0,\,1.0)$ 
& $\mathrm{Ga}(1.0,\,1.0)$
& $\mathrm{Ga}(0.2,\,0.2)$
& $\mathrm{Ga}(5.0,\,5.0)$\\
\hline
\end{tabular}
\end{adjustbox}
\label{tab:prior_speicifications}
\end{table*}

In Bayesian analysis, the choice of priors is a key concern to ensure robust results. In our empirical analysis, we used proper, weakly informative prior distributions for all parameters. We also recommend conducting a sensitivity analysis to assess how variations in prior specifications affect the empirical findings represented by their posteriors.
The default prior choices are as follows:  $\boldsymbol{\theta}_{lk} \sim \mathrm{MVN}(\mathbf{0}, 10^2 \times I_{d_y})$, $\sigma^2_{lk} \sim \mathrm{IG}(2.0, 1.0)$, $\boldsymbol{\gamma}_{1,lk}, \boldsymbol{\gamma}_{2,lk} \sim \mathrm{MVN}(\mathbf{0}, 10^2 \times I_{d_m})$, $\boldsymbol{\Sigma} \sim \mathrm{IW}(2.0, I_2)$, $\alpha_y \sim \mathrm{Ga}(1.0,1.0)$, $\alpha_m \sim \mathrm{Ga}(1.0,1.0)$, $\beta_y \sim \mathrm{Ga}(1.0,1.0)$, and $\beta_m \sim \mathrm{Ga}(1.0,1.0)$, where $I_{d}$ is the identity matrix of dimension $d$. In what follows, we consider several alternative prior scenarios, which are provided in Table \ref{tab:prior_speicifications}. Each parameter's prior is proper and is made more or less informative relative to these default settings. Specifically, 
Priors 1 examine more informative priors on coefficient parameters, while and Priors 2 evaluate the less informative coefficient priors. Priors 3 and 4 evaluate the sensitivity of concentration parameters and variance parameters. Priors 3 can lead to a larger number of cluster components of the nDDP, while Priors 4 tends to generate a conservative number of components.
Figure \ref{fig:sensitivity} compares the posterior distributions of each estimand of alternative priors with those of default priors.
Overall, we observe that the results do not differ significantly, exhibiting the robustness of our methodologies to prior misspecifications. }

\section{Dirichlet process and its related models: A recap}
\label{sec:review_DP}
The Dirichlet process (DP), introduced by \citet{Ferguson1974}, is one of the most widely used nonparametric models for random distributions in Bayesian analysis. 
The most versatile definition of the DP is the stick-breaking representation \citep{Sethuraman1994}: $F(\cdot) = \sum_{h=1}^{\infty}w_h\delta_{\boldsymbol{\theta}_h}(\cdot)$, where $\delta_{\boldsymbol{\theta}}(\cdot)$ is the Dirac measure at $\boldsymbol{\theta}$, $w_h=u_h\prod_{l<h}(1-u_l)$ with $u_h \sim \mathrm{Be}(1,\alpha)$, and $\boldsymbol{\theta}_h \sim G_0$. 
A random probability measure $F$ on a complete and separable metric space $\Theta$ is said to follow a DP prior with a concentration parameter $\alpha>0$ and a base measure $G_0$, $F \sim \mathrm{DP}(\alpha, G_0)$. Because the DP assigns probability one to the space of discrete measures, 
it is more effectively employed as a prior for a mixing distribution, leading to what is known as a Dirichlet Process Mixture (DPM) model \citep{Antoniak1974,Escobar1995}, where a probability density function $f$ is written as $f(\cdot) = \int_{\Theta} p(\cdot \mid \boldsymbol{\theta}) F(d\boldsymbol{\theta}),$ where $p(\cdot \mid \boldsymbol{\theta})$ is a continuous density function parameterized by $\boldsymbol{\theta} \in \Theta$ and $F \sim \mathrm{DP}(\alpha, G_0)$.  Its stick-breaking representation is therefore
$f(\cdot) = \sum_{h=1}^{\infty} w_h p(\cdot \mid \boldsymbol{\theta}_h)$.

Although these applications typically address problems involving exchangeable data from an unknown distribution, incorporating a dependence structure is crucial in many real-world scenarios where the underlying data-generating process is influenced by auxiliary covariates. 
\citet{Maceachern1999,Maceachern2000} proposed the Dependent Dirichlet Process (DDP), 
which introduces a dependency structure whereby the DP is indexed by covariates, enabling the model to capture changes in the distribution as a function of these covariates. Specifically, dependence of covariates $\mathbf{x} \in \mathcal{X}$ is introduced through a modification of the stick-breaking representation as $F_{\mathbf{x}}(\cdot) = \sum_{h=1}^{\infty}w_h(\mathbf{x})\delta_{\boldsymbol{\theta}_h(\mathbf{x})}(\cdot)$, where $w_h$ and $\boldsymbol{\theta}_h$ are replaced with independent stochastic processes $w_h(\mathbf{x})$ and $\boldsymbol{\theta}_h(\mathbf{x})$ with index set $\mathcal{X}$.
This extension makes the DDP particularly powerful in settings where the observed data exhibit heterogeneity across different levels of a covariate. 
Another direction in the advancement of DP was taken by \citet{Rodriguez2008}, who proposed the Nested Dirichlet Process, abbreviated by nDP, or nDPM for its mixture. The nDP extends the DP to nonparametrically model the outcome distributions of multiple groups of data, borrowing information across groups while also allowing groups to be clustered.  
Specifically, a collection of distributions 
$\{F_1, \ldots, F_I\}$ 
is said to follow a nDP if 
$F_i(\cdot) \sim Q \equiv \sum_{k=1}^{\infty} \pi^{*}_k \delta_{F^{*}_k}(\cdot)$ 
for 
$i=1,\ldots,I$ and $F^{*}_k(\cdot)=\sum_{l=1}^{\infty}w^{*}_{lk} \delta_{\boldsymbol{\theta}^{*}_{lk}}(\cdot)$ 
with 
$\boldsymbol{\theta}^{*}_{lk} \sim G_0$, 
$w^{*}_{lk}=u^{*}_{lk}\prod_{m<l}(1-u^{*}_{mk})$, 
$\pi^{*}_{k}=s^{*}_{k}\prod_{m<k}(1-s^{*}_{m})$, 
$s^{*}_{k} \sim \mathrm{Be}(1, \alpha)$ and 
$u^{*}_{lk} \sim \mathrm{Be}(1, \beta)$, 
which is denoted by 
$\{F_1, \ldots, F_I\} \sim \mathrm{nDP}(\alpha, \beta, G_0)$.
There is a similar extension of the DP known as the hierarchical Dirichlet process (HDP) \citep{Teh2006}, and recent research has further extended the HDP by incorporating dependent structures related to covariates \citep{Diana2020,zhang2024}. 
A key difference between the HDP and nDP is that, in the HDP, the random measures share the same atoms but assign them different weights, whereas in the nDP, two distributions either share both atoms and weights or share nothing at all. This distinction enables the nDP to capture distributional heterogeneity across clusters by allowing for clustering at both the outcome and distribution levels, which can be of interest in CRTs. This feature is illustrated by \citet{Ho2013}, who have applied the nDP model to represent the residual distribution in linear model-based analysis of CRTs without intermediate variables. 

\yo{
\subsection{Fully-Dependent Nested Dependent Dirichlet Process Mixture (FD-nDDPM)}
\label{sec:FD_nDDPM}

In the main manuscript, we focused on the development of the Atom-Dependent Nested Dependent Dirichlet Process Mixture (AD-nDDPM) for simplicity and illustration. The AD-nDDPM incorporates covariate dependency into the atoms of the nDPM. In this supplement, we introduce an alternative dependency structure, termed the \textit{Fully-Dependent Nested Dependent Dirichlet Process Mixture (FD-nDDPM)}, where covariate dependency is incorporated into the clustering process through the mixture weights. While several implementations are possible, we employ the Kernel Stick-Breaking Process (KSBP) of \citet{Dunson_2008} to allow the clustering to depend on covariates.

\begin{definition}\label{def:fd-nddpm}
For any $k \in \mathbb{N}$, let $\{F^{*}_{\mathbf{c},k}: \mathbf{c} \in \mathcal{C}\}$ be a $\mathcal{P}(S)$-valued stochastic process on an appropriate probability space $(\Omega, \mathcal{F}, P)$ such that:
\begin{enumerate}[(i)]
    \item $s^{*}_{1}, s^{*}_{2}, \ldots$ are independent random variables of the form $s^{*}_{k}:\Omega \to [0,1]$ for all $k$, with the common Beta distribution with parameter $(1,\alpha)$.
    \item $u^{*}_{1k}, u^{*}_{2k}, \ldots$ are independent random variables of the form $u^{*}_{lk}:\Omega \to [0,1]$ for all $l$, with the common Beta distribution with parameter $(1,\beta)$.
    \item $\boldsymbol{\theta}^{*}_{1k}, \boldsymbol{\theta}^{*}_{2k}, \ldots$ are independent stochastic processes of the form $\boldsymbol{\theta}^{*}_{lk}: \mathcal{C} \times \Omega \to S$ for all $l$, with common finite dimensional distributions determined by the set of copulas $\Psi_{\mathcal{C}}^{\theta}$ and the set of a marginal distribution $G_{\mathcal{C}}^0$.
    \item For every $\mathbf{c} \in \mathcal{C}$, $B \in \mathcal{B}$ and almost every $\omega \in \Omega$,
    \begin{equation*}
        F_k^{*}(\mathbf{c}, \omega)(B) = \sum_{l=1}^{\infty} w^{*}_{lk}(\omega)\delta_{\boldsymbol{\theta}^{*}_{lk}(\mathbf{c}, \omega)(B)},~~~ F(\mathbf{c}, \omega)(B) = \sum_{k=1}^{\infty} \pi^{*}_k(\mathbf{v},\omega) F_k^{*}(\mathbf{c}, \omega)(B),
    \end{equation*}
\end{enumerate}
where $w^{*}_{lk}(\omega) =  u^{*}_{lk}(\omega)\prod_{i=1}^{l-1}(1-u^{*}_{ik}(\omega))$ and $\pi^{*}_k(\mathbf{v},\omega) =  U^{*}_{k}(\mathbf{v},\omega)\prod_{m=1}^{k-1}(1-U^{*}_{m}(\mathbf{v},\omega))$ with 
$U^{*}_k(\mathbf{v}, \omega) = K^{*}(\mathbf{v}; \boldsymbol{\Gamma}_k) s^{*}_{k}(\omega)$, and $K^{*} \to (0,1]$ is a positive bounded function, which is initially assumed to be known.
A process $\mathcal{H} = \{ F(\mathbf{c}, \cdot) : \mathbf{c} \in \mathcal{C} \} $ is referred to as the \emph{Fully-Dependent Nested Dependent Dirichlet Process (FD-nDDP)}. 

\end{definition}
\noindent
Alternatively, we may express that a collection of distributions follows the FD-nDDPM if, for each group $i$ and each value $\mathbf{c} = (\mathbf{v}, \mathbf{x}) \in \mathcal{V} \times \mathcal{X}$,
\begin{equation}
\label{eq:fd-nddpm}
    \begin{split}
        Y_{ij} \mid \mathbf{C}_{ij} = \mathbf{c}, F_{\mathbf{c},i} &\sim \int_{\Theta} p(y \mid \boldsymbol{\theta}) \, dF_{\mathbf{c},i}(\boldsymbol{\theta}), \\
        F_{\mathbf{c},i} &\sim \sum_{k=1}^{\infty} \pi^{*}_k (v)\, \delta_{F^{*}_{\mathbf{c},k}}(\cdot), \\
        F^{*}_{\mathbf{c},k} &= \sum_{l=1}^{\infty} w^{*}_{lk} \, \delta_{\boldsymbol{\theta}^{*}_{lk}(\mathbf{c})}(\cdot),
    \end{split}
\end{equation}
At the cluster level, the weights $\pi^{*}_k(\mathbf{v})$ depend on the cluster-level covariate $\mathbf{v}$, embodying the idea that the prior probability of partitions---that is, the assignment probability of distributions to each cluster---varies with the values of $\mathbf{v}$. Specifically, we define
\begin{equation}
\label{eq:weight_pi_dependent}
    \pi^{*}_k(\mathbf{v}) = K^{*}(\mathbf{v}; \boldsymbol{\Gamma}_k) s^{*}_{k} \prod_{m<k} \left(1 - K^{*}(\mathbf{v}; \boldsymbol{\Gamma}_m) s^{*}_{m}\right),
\end{equation}
where $s^{*}_{k}$ follows the standard stick-breaking representation, and the dependence on $\mathbf{v}$ is expressed through the kernel function $K^{*}(\mathbf{v}; \boldsymbol{\Gamma}_k)$ with a location parameter $\boldsymbol{\Gamma}_k$. This approach is also similar to that of \citet{Reich2007}, who modeled hurricane surface wind fields using a stick-breaking prior that varies spatially according to kernel functions.
In our implementation, we specify the kernel function as
\begin{equation}
    \label{eq:kernel_function}
    K^{*}(\mathbf{v}; \boldsymbol{\Gamma}_k) = \exp\left( - \| \mathbf{v} - \boldsymbol{\Gamma}_k \|^2 / 2 \right),
\end{equation}
where $\|\cdot\|$ denotes the Euclidean norm, and $\boldsymbol{\Gamma}_k$ is an unknown location parameter with a prior $\boldsymbol{\Gamma}_k \sim \mathrm{MVN}(\boldsymbol{\mu}_{\boldsymbol{\Gamma}}, \boldsymbol{\Sigma}_{\boldsymbol{\Gamma}})$. This formulation allows clusters to have weights that vary smoothly over the covariate space, with clusters being more influential near their associated location parameters. 
To ensure the stick-breaking representation at the cluster level is proper, we need to choose priors for $\boldsymbol{\Gamma}_k)$ and $ s^{*}_{k}$ so that $\sum_{k=1}^{\infty}\pi^{*}_k(\mathbf{v})=1$ almost surely for all $\mathbf{v}$, which is proven in \citet{Reich2007}. Specifically, this holds if $\E[s^{*}_{k}]$ and $\E[K^{*}(\mathbf{v})]$ (the expectation is taken over $\boldsymbol{\Gamma}_k$) are both positive, which holds in our setting. For the finite approximation, we set $\pi^{*}_{K_c}(\mathbf{v})=1$ for all $\mathbf{v}$, equivalent to trucating the infinite mixture by attributing all of the mass from the terms with $k \geq K_c$ to $\pi^{*}_{K_c}(\mathbf{v})$.

The FD-nDDP model shares similar statistical properties---expectation, variance, correlation, and weak support properties---with the AD-nDDP model. 
The following proposition formalizes the statistical properties of the FD-nDDP model.
\begin{proposition}
\label{prop:properties_fdnddp}
Let $\{ F_{\mathbf{c},i} : \mathbf{c} \in \mathcal{C}\}$ be an FD-nDDP for each $i$.
The expectation and variance of the FD-nDDP are identical to the AD-nDDP:
\begin{equation*}
    \E[F_{\mathbf{c}, i}(A)] = G_{\mathbf{c}}^0(A) \text{ and } \Var[F_{\mathbf{c}, i}(A)] = \frac{G_{\mathbf{c}}^0(A)(1-G_{\mathbf{c}}^0(A))}{\beta + 1}.
\end{equation*}
Additionally, for any $\mathbf{c}, \mathbf{c}'  \in \mathcal{C}$, cluster $i,j \in \{1,\ldots,I\}$, any measurable sets $A,B \in \mathcal{B}$, with $\rho_{\mathbf{c},\mathbf{c}'}(A,B) = P\left\{ \boldsymbol{\theta}_{lk}^{*}(\mathbf{c}) \in A, \boldsymbol{\theta}_{lk}^{*}(\mathbf{c}') \in B  \right\}$, we have:
    \begin{equation*}
        \Corr(F_{\mathbf{c}, i}(A), F_{\mathbf{c}', j}(B)) = \begin{cases}
            \displaystyle\frac{\rho_{\mathbf{c},\mathbf{c}'}(A,B) - G_{\mathbf{c}}^0(A)G_{\mathbf{c}'}^0(B)}{ \sqrt{G_{\mathbf{c}}^0(A)(1-G_{\mathbf{c}}^0(A))G_{\mathbf{c}'}^0(B)(1-G_{\mathbf{c}'}^0(B))}},  \hfill \text{ if } i=j\\
            \displaystyle h(\alpha, \mathbf{v}, \mathbf{v}') \frac{\rho_{\mathbf{c},\mathbf{c}'}(A,B) - G_{\mathbf{c}}^0(A)G_{\mathbf{c}'}^0(B)}{\sqrt{G_{\mathbf{c}}^0(A)(1-G_{\mathbf{c}}^0(A))G_{\mathbf{c}'}^0(B)(1-G_{\mathbf{c}'}^0(B))}},  \hfill \text{ if } i \neq j,
        \end{cases}
    \end{equation*}
    where $h(\alpha, \mathbf{v}, \mathbf{v}')
    = \frac{2}{(\alpha+1)(\alpha+2)} \sum_{k=1}^{\infty}K^*(\mathbf{v};\Gamma_k)K^*(\mathbf{v}';\Gamma_k)\prod_{j=1}^{k-1} \left\{ 1- \frac{K^*(\mathbf{v};\Gamma_j)+K^*(\mathbf{v}';\Gamma_j)}{\alpha+1} +\frac{2K(\mathbf{v};\Gamma_j)K^*(\mathbf{v}';\Gamma_j)}{(\alpha+1)(\alpha+2)}  \right\}$.
    Finally, if $\Psi_{\mathcal{C}}^{\theta}$ is a collection of copulas with positive density with respect to Lebesgue measure, on the appropriate Euclidean space, and the kernel $K(\cdot)$ is a positive bounded function, then $\mathcal{P}(\Theta)^{\mathcal{C}}$ is the weak support of the process.
\end{proposition}
\noindent
As anticipated, the correlation coincides with that of the AD-nDDP when the kernel function satisfies $K(\cdot) = 1$. The weak support property is also satisfied by the FD-nDDP.
By introducing the FD-nDDPM, we expand the modeling framework for cluster-randomized trials, offering greater flexibility in capturing distributional heterogeneity and complex clustering structures driven by cluster covariates. However, given the limited enhancement observed in our simulation studies, the AD-nDDPM emerges as a practical and effective choice for modeling in CRTs. The individual-level atom-based dependence structure in the AD-nDDPM appears to capture the necessary dependence adequately without the added complexity of covariate-dependent weights. This insight is valuable for practitioners, indicating that simpler models may suffice in certain contexts, thereby reducing computational demands without compromising performance. However, the FD-nDDPM model could offer valuable insights into cluster heterogeneity that varies with cluster-level covariates and might gain efficiency in specific contexts when clusters exhibit extreme heterogeneity. 
Another significant challenge associated with the FD-nDDPM is computational complexity. Unlike the AD-nDDPM, standard closed-form posterior updates are not available for the parameters $\boldsymbol{\Gamma}_k$ and $s^{*}_{k}$ due to the covariate dependence in the mixture weights. To address this, we adopt Metropolis-Hastings steps within the Gibbs sampling algorithm to obtain posterior draws for these parameters. The details of this step are provided in Section \ref{sec:gibbs_FD_nDDPM}. While this approach enables us to estimate the model parameters, it increases computational burden and may affect scalability. Future research could focus on developing more efficient sampling schemes or employing approximate inference methods to enhance computational efficiency.
}

\section{Proofs of the Theorems}
\label{sec:proofs}

\subsection{Proof of Theorem \ref{thm:eie_identification}}
\label{proof:eie}
\begin{proof}
    \begin{align*}
    &\mathrm{EIE}_{\mathrm{C}}^{(k)} \\
    &= \E \left[ \frac{1}{N}\sum_{j=1}^{N} \left\{ Y_{\cdot j}(1, \mathbf{M}^{(k)}(1), \mathbf{M}^{(3-k)}(1)) -  Y_{\cdot j}(1, \mathbf{M}^{(k)}(0), \mathbf{M}^{(3-k)}(1)) \right\} \right] \\
    &= \E\left[\E\left[  \frac{1}{N}\sum_{j=1}^{N} \left\{ Y_{\cdot j}(1, \mathbf{M}^{(k)}(1), \mathbf{M}^{(3-k)}(1)) -  Y_{\cdot j}(1, \mathbf{M}^{(k)}(0), \mathbf{M}^{(3-k)}(1)) \right\} \middle| \mathbf{C}, N  \right]\right] \hfill \tag{$\because$ law of iterated expectations (LIE)} \\
    &= \E\left[ \int_{\mathcal{M}^{(3-k)}} \E\left[  \frac{1}{N}\sum_{j=1}^{N} \left\{ Y_{\cdot j}(1, \mathbf{M}^{(k)}(1), \mathbf{m}^{(3-k)}) -  Y_{\cdot j}(1, \mathbf{M}^{(k)}(0), \mathbf{m}^{(3-k)}) \right\} \middle| \mathbf{M}^{(3-k)}(1)=\mathbf{m}^{(3-k)}, \mathbf{C}, N  \right] \right.\\
    & \left. \hspace{31.5em} dF_{\mathbf{M}^{(3-k)}(1) \mid \mathbf{C}, N} (\mathbf{m}^{(3-k)}) \right] \tag{$\because$ LIE} \\
    &= \E\left[ \int_{\mathcal{M}^{(3-k)}} \E\left[  \frac{1}{N}\sum_{j=1}^{N} \left\{ Y_{\cdot j}(1, \mathbf{M}^{(k)}(1), \mathbf{m}^{(3-k)}) -  Y_{\cdot j}(1, \mathbf{M}^{(k)}(0), \mathbf{m}^{(3-k)}) \right\} \middle| \mathbf{C}, N  \right]  dF_{\mathbf{M}^{(3-k)}(1) \mid \mathbf{C}, N} (\mathbf{m}^{(3-k)}) \right] \tag{$\because$ Assumption \ref{asmp:cond_homogeneity} with $a=0$ and $a'=1$} \\
    &= \E\left[ \frac{1}{N}\sum_{j=1}^{N} \int_{\mathcal{M}^{(3-k)}} \E\left[  Y_{\cdot j}(1, \mathbf{M}^{(k)}(1), \mathbf{m}^{(3-k)}) -  Y_{\cdot j}(1, \mathbf{M}^{(k)}(0), \mathbf{m}^{(3-k)})  \middle| \mathbf{C}, N  \right]  dF_{\mathbf{M}^{(3-k)}(1) \mid \mathbf{C}, N} (\mathbf{m}^{(3-k)}) \right] \\
    &= \E\left[ \frac{1}{N}\sum_{j=1}^{N} \left \{ \underbrace{\int_{\mathcal{M}^{(3-k)}} \E\left[  Y_{\cdot j}(1, \mathbf{M}^{(k)}(1), \mathbf{m}^{(3-k)}) \middle| \mathbf{C}, N  \right]  dF_{\mathbf{M}^{(3-k)}(1) \mid \mathbf{C}, N} (\mathbf{m}^{(3-k)})}_{=\theta_1} \right.\right.\\
    & \left.\left. \hspace{5.em} - \underbrace{\int_{\mathcal{M}^{(3-k)}} \E \left[  Y_{\cdot j}(1, \mathbf{M}^{(k)}(0), \mathbf{m}^{(3-k)})  \middle| \mathbf{C}, N  \right]  dF_{\mathbf{M}^{(3-k)}(1) \mid \mathbf{C}, N} (\mathbf{m}^{(3-k)})}_{=\theta_0} \right\}\right]
\end{align*}
To simplify the exposition, we now consider $\theta_a$ for $a\in\{0,1\}$.
\begin{align*}
    &\theta_a = \int_{\mathcal{M}^{(3-k)}} \E\left[  Y_{\cdot j}(1, \mathbf{M}^{(k)}(a), \mathbf{m}^{(3-k)}) \middle| \mathbf{C}, N  \right]  dF_{\mathbf{M}^{(3-k)}(1) \mid \mathbf{C}, N} (\mathbf{m}^{(3-k)}) \\
    =&\int_{\mathcal{M}^{(3-k)}}\int_{\mathcal{M}^{(k)}} \E\left[  Y_{\cdot j}(1, \mathbf{m}^{(k)}, \mathbf{m}^{(3-k)}) \middle| \mathbf{M}^{(k)}(a)=\mathbf{m}^{(k)},\mathbf{C}, N  \right]  dF_{\mathbf{M}^{(k)}(a) \mid \mathbf{C}, N} (\mathbf{m}^{(k)})dF_{\mathbf{M}^{(3-k)}(1) \mid \mathbf{C}, N} (\mathbf{m}^{(3-k)}) \tag{$\because$ LIE}  \\
    =&\int_{\mathcal{M}^{(3-k)}}\int_{\mathcal{M}^{(k)}} \E\left[  Y_{\cdot j}(1, \mathbf{m}^{(k)}, \mathbf{m}^{(3-k)}) \middle| A=1, \mathbf{M}^{(k)}(a)=\mathbf{m}^{(k)},\mathbf{C}, N  \right]  \\
    & \hspace{15em} dF_{\mathbf{M}^{(k)}(a) \mid A=a,\mathbf{C}, N} (\mathbf{m}^{(k)})dF_{\mathbf{M}^{(3-k)}(1) \mid A=1, \mathbf{C}, N} (\mathbf{m}^{(3-k)}) \tag{$\because$ Assumption \ref{asmp:randomization}}  \\
    =&\int_{\mathcal{M}^{(3-k)}}\int_{\mathcal{M}^{(k)}} \E\left[  Y_{\cdot j}(1, \mathbf{m}^{(k)}, \mathbf{m}^{(3-k)}) \middle| A=1, \mathbf{M}^{(k)}(1)=\mathbf{m}^{(k)}, \mathbf{M}^{(3-k)}(1)=\mathbf{m}^{(3-k)},\mathbf{C}, N  \right]  \\
    & \hspace{15em} dF_{\mathbf{M}^{(k)}(a) \mid A=a,\mathbf{C}, N} (\mathbf{m}^{(k)})dF_{\mathbf{M}^{(3-k)}(1) \mid A=1, \mathbf{C}, N} (\mathbf{m}^{(3-k)}) \tag{$\because$ Assumption \ref{asmp:si}}  \\
    =& \int_{\mathcal{M}^{(3-k)}}\int_{\mathcal{M}^{(k)}} \E\left[  Y_{\cdot j} \middle| A=1, \mathbf{M}^{(k)}=\mathbf{m}^{(k)},\mathbf{M}^{(3-k)}=\mathbf{m}^{(3-k)},\mathbf{C}, N  \right] \\
    & \hspace{15em} dF_{\mathbf{M}^{(k)} \mid A=a,\mathbf{C}, N} (\mathbf{m}^{(k)})dF_{\mathbf{M}^{(3-k)} \mid A=1, \mathbf{C}, N} (\mathbf{m}^{(3-k)}) \tag{$\because$ Assumption \ref{asmp:sutva}}.  
\end{align*}
Reinserting $\theta_a$ completes the proof.
\end{proof}

\subsection{Proof of Theorem \ref{thm:esme_identification}}
\label{proof:esme}
\begin{proof}
    \begin{align*}
        &\mathrm{ESME}_{\mathrm{C}}^{(k)} \\
        &= \E \left[ \frac{1}{N} \sum_{j=1}^{N} \left\{ Y_{ j}(1, M_{\cdot j}^{(k)}(1), \mathbf{M}_{\cdot (-j)}^{(k)}(1), \mathbf{M}^{(3-k)}(1)) - Y_{\cdot j}(1, M_{\cdot j}^{(k)}(1), \mathbf{M}_{\cdot (-j)}^{(k)}(0), \mathbf{M}^{(3-k)}(1)) \right\} \right] \\
        &= \E \left[ \E \left[ \frac{1}{N} \sum_{j=1}^{N} \left\{ Y_{\cdot  j}(1, M_{\cdot j}^{(k)}(1), \mathbf{M}_{\cdot(-j)}^{(k)}(1), \mathbf{M}^{(3-k)}(1)) - Y_{\cdot j}(1, M_{\cdot j}^{(k)}(1), \mathbf{M}_{\cdot (-j)}^{(k)}(0), \mathbf{M}^{(3-k)}(1)) \right\} \right] \middle| \mathbf{C}, N \right] \tag{$\because$ LIE} \\
        &= \E\left[ \int_{\mathcal{M}^{(3-k)}} \E\left[  \frac{1}{N}\sum_{j=1}^{N} \left\{  Y_{\cdot  j}(1, M_{\cdot j}^{(k)}(1), \mathbf{M}_{\cdot(-j)}^{(k)}(1), \mathbf{m}^{(3-k)}) \right.\right.\right.\\
        & \left.\left.\left. \hspace{5.em}  - Y_{\cdot j}(1, M_{\cdot j}^{(k)}(1), \mathbf{M}_{\cdot (-j)}^{(k)}(0), \mathbf{m}^{(3-k)}) \right\} \middle| \mathbf{M}^{(3-k)}(1)=\mathbf{m}^{(3-k)}, \mathbf{C}, N  \right] 
     dF_{\mathbf{M}^{(3-k)}(1) \mid \mathbf{C}, N} (\mathbf{m}^{(3-k)}) \right] \tag{$\because$ LIE} \\
     &= \E\left[ \int_{\mathcal{M}^{(3-k)}} \E\left[  \frac{1}{N}\sum_{j=1}^{N} \left\{  Y_{\cdot  j}(1, M_{\cdot j}^{(k)}(1), \mathbf{M}_{\cdot(-j)}^{(k)}(1), \mathbf{m}^{(3-k)}) \right.\right.\right.\\
        & \left.\left.\left. \hspace{5.em}  - Y_{\cdot j}(1, M_{\cdot j}^{(k)}(1), \mathbf{M}_{\cdot (-j)}^{(k)}(0), \mathbf{m}^{(3-k)}) \right\} \middle|  \mathbf{C}, N  \right] 
     dF_{\mathbf{M}^{(3-k)}(1) \mid \mathbf{C}, N} (\mathbf{m}^{(3-k)}) \right] \tag{$\because$ Assumption \ref{asmp:cond_homogeneity} with $a=1$ and $a'=1$} \\
     &= \E\left[ \frac{1}{N}\sum_{j=1}^{N} \int_{\mathcal{M}^{(3-k)}} \E\left[  \left\{  Y_{\cdot  j}(1, M_{\cdot j}^{(k)}(1), \mathbf{M}_{\cdot(-j)}^{(k)}(1), \mathbf{m}^{(3-k)}) \right.\right.\right.\\
        & \left.\left.\left. \hspace{5.em}  - Y_{\cdot j}(1, M_{\cdot j}^{(k)}(1), \mathbf{M}_{\cdot (-j)}^{(k)}(0), \mathbf{m}^{(3-k)}) \right\} \middle|  \mathbf{C}, N  \right] 
     dF_{\mathbf{M}^{(3-k)}(1) \mid \mathbf{C}, N} (\mathbf{m}^{(3-k)}) \right] \\
     &= \E\left[ \frac{1}{N}\sum_{j=1}^{N} \left \{ \underbrace{\int_{\mathcal{M}^{(3-k)}} \E\left[  Y_{\cdot  j}(1, M_{\cdot j}^{(k)}(1), \mathbf{M}_{\cdot(-j)}^{(k)}(1), \mathbf{m}^{(3-k)}) \middle| \mathbf{C}, N  \right]  dF_{\mathbf{M}^{(3-k)}(1) \mid \mathbf{C}, N} (\mathbf{m}^{(3-k)})}_{=\tau_1} \right.\right.\\
    & \left.\left. \hspace{5.em} - \underbrace{\int_{\mathcal{M}^{(3-k)}} \E \left[  Y_{\cdot j}(1, M_{\cdot j}^{(k)}(1), \mathbf{M}_{\cdot (-j)}^{(k)}(0), \mathbf{m}^{(3-k)})  \middle| \mathbf{C}, N  \right]  dF_{\mathbf{M}^{(3-k)}(1) \mid \mathbf{C}, N} (\mathbf{m}^{(3-k)})}_{=\tau_0} \right\}\right]
    \end{align*}
    Now, we focus on $\tau_a$.
    \begin{align*}
        \tau_a &= \int_{\mathcal{M}^{(3-k)}} \E\left[  Y_{\cdot  j}(1, M_{\cdot j}^{(k)}(1), \mathbf{M}_{\cdot(-j)}^{(k)}(a), \mathbf{m}^{(3-k)}) \middle| \mathbf{C}, N  \right]  dF_{\mathbf{M}^{(3-k)}(1) \mid \mathbf{C}, N} (\mathbf{m}^{(3-k)}) \\
        &= \int_{\mathcal{M}^{(3-k)}} \int_{\mathcal{M}^{(k)}} \E\left[  Y_{\cdot  j}(1, m_{\cdot j}, \mathbf{m}^{(k)}_{\cdot (-j)}, \mathbf{m}^{(3-k)}) \middle| M_{\cdot j}^{(k)}(1)=m_{\cdot j}, \mathbf{M}_{\cdot(-j)}^{(k)}(a)=\mathbf{m}^{(k)}_{\cdot (-j)}, \mathbf{C}, N  \right]  \\
        & \hspace{10.em} dF_{M^{(k)}_{\cdot j}(1), \mathbf{M}^{(k)}_{\cdot, (-j)}(a) \mid \mathbf{C}, N} (m_{\cdot j}, \mathbf{m}^{(k)}_{\cdot (-j)})dF_{\mathbf{M}^{(3-k)}(1) \mid \mathbf{C}, N} (\mathbf{m}^{(3-k)}) \tag{$\because$ LIE}\\
        &= \int_{\mathcal{M}^{(3-k)}} \int_{\mathcal{M}^{(k)}} \E\left[  Y_{\cdot  j}(1, m_{\cdot j}, \mathbf{m}^{(k)}_{\cdot (-j)}, \mathbf{m}^{(3-k)}) \middle| M_{\cdot j}^{(k)}(1)=m_{\cdot j}, \mathbf{M}_{\cdot(-j)}^{(k)}(a)=\mathbf{m}^{(k)}_{\cdot (-j)}, \mathbf{C}, N  \right]  \\
        & \hspace{10.em} dF_{M^{(k)}_{\cdot j}(1) \mid \mathbf{C}, N} (m_{\cdot j})dF_{\mathbf{M}^{(k)}_{\cdot, (-j)}(a) \mid \mathbf{C}, N} ( \mathbf{m}^{(k)}_{\cdot (-j)}) dF_{\mathbf{M}^{(3-k)}(1) \mid \mathbf{C}, N} (\mathbf{m}^{(3-k)}) \tag{$\because$ Assumption \ref{asmp:no_cross_intra_corr}}\\
        &= \int_{\mathcal{M}^{(3-k)}} \int_{\mathcal{M}^{(k)}} \E\left[  Y_{\cdot  j}(1, m_{\cdot j}, \mathbf{m}^{(k)}_{\cdot (-j)}, \mathbf{m}^{(3-k)}) \middle| A=1, M_{\cdot j}^{(k)}(1)=m_{\cdot j}, \mathbf{M}_{\cdot(-j)}^{(k)}(a)=\mathbf{m}^{(k)}_{\cdot (-j)}, \mathbf{C}, N  \right]  \\
        & \hspace{10.em} dF_{M^{(k)}_{\cdot j}(1) \mid A=1,\mathbf{C}, N} (m_{\cdot j})dF_{\mathbf{M}^{(k)}_{\cdot, (-j)}(a) \mid A=a,\mathbf{C}, N} ( \mathbf{m}^{(k)}_{\cdot (-j)}) dF_{\mathbf{M}^{(3-k)}(1) \mid A=1,\mathbf{C}, N} (\mathbf{m}^{(3-k)}) \tag{$\because$ Assumption \ref{asmp:randomization}}\\
        &= \int_{\mathcal{M}^{(3-k)}} \int_{\mathcal{M}^{(k)}} \E\left[  Y_{\cdot  j}(1, m_{\cdot j}, \mathbf{m}^{(k)}_{\cdot (-j)}, \mathbf{m}^{(3-k)}) \middle| A=1, M_{\cdot j}^{(k)}(1)=m_{\cdot j}, \mathbf{M}_{\cdot(-j)}^{(k)}(1)=\mathbf{m}^{(k)}_{\cdot (-j)}, \right.\\
        & \left. \hspace{25.em} \mathbf{M}^{(3-k)}(1)=\mathbf{m}^{(3-k)}, \mathbf{C}, N  \right]  \\
        & \hspace{10.em} dF_{M^{(k)}_{\cdot j}(1) \mid A=1,\mathbf{C}, N} (m_{\cdot j})dF_{\mathbf{M}^{(k)}_{\cdot, (-j)}(a) \mid A=a,\mathbf{C}, N} ( \mathbf{m}^{(k)}_{\cdot (-j)}) dF_{\mathbf{M}^{(3-k)}(1) \mid A=1,\mathbf{C}, N} (\mathbf{m}^{(3-k)}) \tag{$\because$ Assumption \ref{asmp:si}}\\
        &= \int_{\mathcal{M}^{(3-k)}} \int_{\mathcal{M}^{(k)}} \E\left[  Y_{\cdot  j} \middle| A=1, M_{\cdot j}^{(k)}=m_{\cdot j}, \mathbf{M}_{\cdot(-j)}^{(k)}=\mathbf{m}^{(k)}_{\cdot (-j)},  \mathbf{M}^{(3-k)}=\mathbf{m}^{(3-k)}, \mathbf{C}, N  \right]  \\
        & \hspace{10.em} dF_{M^{(k)}_{\cdot j} \mid A=1,\mathbf{C}, N} (m_{\cdot j})dF_{\mathbf{M}^{(k)}_{\cdot, (-j)} \mid A=a,\mathbf{C}, N} ( \mathbf{m}^{(k)}_{\cdot (-j)}) dF_{\mathbf{M}^{(3-k)} \mid A=1,\mathbf{C}, N} (\mathbf{m}^{(3-k)}) \tag{$\because$ Assumption \ref{asmp:sutva}}
    \end{align*}
\end{proof}

\subsection{Identification of interaction effects}
\label{sec:identification_INT}
When we consider two mediators in CRTs, the interaction effect (INT) is expressed as $\mathrm{INT}^{(1,2)}_{\mathrm{C}} = \mathrm{EIE}^{(1)}_{\mathrm{C}} + \mathrm{EIE}^{(2)}_{\mathrm{C}} - \mathrm{NIE}_{\mathrm{C}}$. Therefore, the $\mathrm{INT}^{(1,2)}_{\mathrm{C}}$ effect is identified as the difference between identified $\mathrm{EIE}^{(k)}_{\mathrm{C}}$ and identified $\mathrm{NIE}_{\mathrm{C}}$. 
The following theorem provides the identification result for $\mathrm{NIE}_{\mathrm{C}}$. 

\begin{theorem}
\label{thm:nie_identification}
        Under Assumption \ref{asmp:sutva}--\ref{asmp:cond_homogeneity}, $ \mathrm{NIE}_{\mathrm{C}} $ are nonparametrically identified as follows: 
        \begin{align*}
             \E_{\mathbf{C}, N} \left[ \frac{1}{N} \sum_{j=1}^{N} \left \{\int_{\mathcal{M}^{(3-k)}} \int_{\mathcal{M}^{(k)}} \mu_{\mathbf{C}, N}(1, \mathbf{m}^{(k)}, \mathbf{m}^{(3-k)}) \right. dF_{\mathbf{M}^{(k)} \mid A=1, \mathbf{C}, N} (\mathbf{m}^{(k)}) dF_{\mathbf{M}^{(3-k)} \mid A=1, \mathbf{C}, N} (\mathbf{m}^{(3-k)}) \right. \\
             \left. - \int_{\mathcal{M}^{(3-k)}} \int_{\mathcal{M}^{(k)}} \mu_{\mathbf{C}, N}(1, \mathbf{m}^{(k)}, \mathbf{m}^{(3-k)})  \left.  dF_{\mathbf{M}^{(k)} \mid A=0, \mathbf{C}, N} (\mathbf{m}^{(k)}) dF_{\mathbf{M}^{(3-k)} \mid A=0, \mathbf{C}, N} (\mathbf{m}^{(3-k)}) \right\}\right],
        \end{align*}
        where $\mu_{\mathbf{C}, N}(a, \mathbf{m}^{(k)}, \mathbf{m}^{(3-k)})=\E \left[ Y_{\cdot j} \mid A=a, \mathbf{M}^{(k)}=\mathbf{m}^{(k)}, \mathbf{M}^{(3-k)}=\mathbf{m}^{(3-k)}, \mathbf{C}, N \right]$.
\end{theorem}
\begin{proof}
    \begin{align*}
    &\mathrm{NIE}_{\mathrm{C}} \\
    &= \E \left[ \frac{1}{N}\sum_{j=1}^{N} \left\{ Y_{\cdot j}(1, \mathbf{M}^{(1)}(1), \mathbf{M}^{(2)}(1)) -  Y_{\cdot j}(1, \mathbf{M}^{(1)}(0), \mathbf{M}^{(2)}(0)) \right\} \right] \\
    &= \E\left[\E\left[  \frac{1}{N}\sum_{j=1}^{N} \left\{ Y_{\cdot j}(1, \mathbf{M}^{(1)}(1), \mathbf{M}^{(2)}(0)) -  Y_{\cdot j}(1, \mathbf{M}^{(1)}(0), \mathbf{M}^{(2)}(0)) \right\} \middle| \mathbf{C}, N  \right]\right] \hfill \tag{$\because$ law of iterated expectations (LIE)} \\
    &= \E\left[ \int_{\mathcal{M}^{(2)}} \E\left[  \frac{1}{N}\sum_{j=1}^{N} \left\{ Y_{\cdot j}(1, \mathbf{M}^{(1)}(1), \mathbf{m}^{(2)}) -  Y_{\cdot j}(1, \mathbf{M}^{(1)}(0), \mathbf{m}^{(2)}) \right\} \middle| \mathbf{M}^{(2)}(0)=\mathbf{m}^{(2)}, \mathbf{C}, N  \right] \right.\\
    & \left. \hspace{31.5em} dF_{\mathbf{M}^{(2)}(0) \mid \mathbf{C}, N} (\mathbf{m}^{(2)}) \right] \tag{$\because$ LIE} \\
    &= \E\left[ \int_{\mathcal{M}^{(2)}} \E\left[  \frac{1}{N}\sum_{j=1}^{N} \left\{ Y_{\cdot j}(1, \mathbf{M}^{(1)}(1), \mathbf{m}^{(2)}) -  Y_{\cdot j}(1, \mathbf{M}^{(1)}(0), \mathbf{m}^{(2)}) \right\} \middle| \mathbf{C}, N  \right]  dF_{\mathbf{M}^{(2)}(0) \mid \mathbf{C}, N} (\mathbf{m}^{(2)}) \right] \tag{$\because$ Assumption \ref{asmp:cond_homogeneity} with $a=0$, $a'=0$ and $k=1$} \\
    &= \E\left[ \frac{1}{N}\sum_{j=1}^{N} \int_{\mathcal{M}^{(2)}} \E\left[  Y_{\cdot j}(1, \mathbf{M}^{(1)}(1), \mathbf{m}^{(2)}) -  Y_{\cdot j}(1, \mathbf{M}^{(1)}(0), \mathbf{m}^{(2)})  \middle| \mathbf{C}, N  \right]  dF_{\mathbf{M}^{(2)}(0) \mid \mathbf{C}, N} (\mathbf{m}^{(2)}) \right] \\
    &= \E\left[ \frac{1}{N}\sum_{j=1}^{N} \left \{ \underbrace{\int_{\mathcal{M}^{(2)}} \E\left[  Y_{\cdot j}(1, \mathbf{M}^{(1)}(1), \mathbf{m}^{(2)}) \middle| \mathbf{C}, N  \right]  dF_{\mathbf{M}^{(2)}(0) \mid \mathbf{C}, N} (\mathbf{m}^{(2)})}_{=\theta_1} \right.\right.\\
    & \left.\left. \hspace{5.em} - \underbrace{\int_{\mathcal{M}^{(2)}} \E \left[  Y_{\cdot j}(1, \mathbf{M}^{(1)}(0), \mathbf{m}^{(2)})  \middle| \mathbf{C}, N  \right]  dF_{\mathbf{M}^{(2)}(0) \mid \mathbf{C}, N} (\mathbf{m}^{(2)})}_{=\theta_0} \right\}\right]
\end{align*}
We now consider $\theta_a$ for $a\in\{0,1\}$.
\begin{align*}
    &\theta_a = \int_{\mathcal{M}^{(2)}} \E\left[  Y_{\cdot j}(1, \mathbf{M}^{(1)}(a), \mathbf{m}^{(2)}) \middle| \mathbf{C}, N  \right]  dF_{\mathbf{M}^{(2)}(0) \mid \mathbf{C}, N} (\mathbf{m}^{(2)}) \\
    =&\int_{\mathcal{M}^{(2)}}\int_{\mathcal{M}^{(1)}} \E\left[  Y_{\cdot j}(1, \mathbf{m}^{(1)}, \mathbf{m}^{(2)}) \middle| \mathbf{M}^{(1)}(a)=\mathbf{m}^{(1)},\mathbf{C}, N  \right]  dF_{\mathbf{M}^{(1)}(a) \mid \mathbf{C}, N} (\mathbf{m}^{(1)})dF_{\mathbf{M}^{(2)}(0) \mid \mathbf{C}, N} (\mathbf{m}^{(2)}) \tag{$\because$ LIE}  \\
    =&\int_{\mathcal{M}^{(2)}}\int_{\mathcal{M}^{(1)}} \E\left[  Y_{\cdot j}(1, \mathbf{m}^{(1)}, \mathbf{m}^{(2)}) \middle| A=1, \mathbf{M}^{(1)}(a)=\mathbf{m}^{(1)}, \mathbf{M}^{(2)}(0)=\mathbf{m}^{(2)}, \mathbf{C}, N  \right]  \\
    & \hspace{15em} dF_{\mathbf{M}^{(1)}(a) \mid A=a,\mathbf{C}, N} (\mathbf{m}^{(1)})dF_{\mathbf{M}^{(2)}(0) \mid A=1, \mathbf{C}, N} (\mathbf{m}^{(2)}) \tag{$\because$ Assumption \ref{asmp:randomization} and \ref{asmp:si}}  \\
    =&\int_{\mathcal{M}^{(2)}}\int_{\mathcal{M}^{(1)}} \E\left[  Y_{\cdot j}(1, \mathbf{m}^{(1)}, \mathbf{m}^{(2)}) \middle| A=1, \mathbf{M}^{(1)}(1)=\mathbf{m}^{(1)}, \mathbf{M}^{(2)}(1)=\mathbf{m}^{(2)},\mathbf{C}, N  \right]  \\
    & \hspace{15em} dF_{\mathbf{M}^{(1)}(a) \mid A=a,\mathbf{C}, N} (\mathbf{m}^{(1)})dF_{\mathbf{M}^{(2)}(0) \mid A=1, \mathbf{C}, N} (\mathbf{m}^{(2)}) \tag{$\because$ Assumption \ref{asmp:si}}  \\
    =& \int_{\mathcal{M}^{(2)}}\int_{\mathcal{M}^{(1)}} \E\left[  Y_{\cdot j} \middle| A=1, \mathbf{M}^{(1)}=\mathbf{m}^{(1)},\mathbf{M}^{(2)}=\mathbf{m}^{(2)},\mathbf{C}, N  \right] \\
    & \hspace{15em} dF_{\mathbf{M}^{(1)} \mid A=a,\mathbf{C}, N} (\mathbf{m}^{(1)})dF_{\mathbf{M}^{(2)} \mid A=1, \mathbf{C}, N} (\mathbf{m}^{(2)}) \tag{$\because$ Assumption \ref{asmp:sutva}}. 
\end{align*}
Reinserting $\theta_a$ completes the proof.
\end{proof}

\yo{
\subsection{Proof of Equation \ref{eq:expectation_variance}}
\label{sec:proof_equation1}
\begin{proof}
    Let $Q_{\mathbf{c}} \equiv \sum_{k=1}^{\infty} \pi^{*}_k \, \delta_{F^{*}_{\mathbf{c},k}}$, for a random probability measure $F_{\mathbf{c},i}$ indexed by $\mathbf{c} \in \mathcal{C}$. We have $P(F_{\mathbf{c},i} = F^{*}_{\mathbf{c},k} \mid Q_{\mathbf{c}})=\pi^{*}_k$, and, for two  random probability measures $F_{\mathbf{c},i}$ and $F_{\mathbf{c},j}$,
    \begin{align*}
        P(F_{\mathbf{c},i} = F_{\mathbf{c},j} \mid Q_{\mathbf{c}}) 
        &= \sum_{k=1}^{\infty} P(F_{\mathbf{c},i} = F_{\mathbf{c},j} =F^{*}_{\mathbf{c},k} \mid Q_{\mathbf{c}}) \\
        &= \sum_{k=1}^{\infty} P(F_{\mathbf{c},i}  =F^{*}_{\mathbf{c},k} \mid Q_{\mathbf{c}})P(F_{\mathbf{c},j} =F^{*}_{\mathbf{c},k} \mid Q_{\mathbf{c}}) \\
        &= \sum_{k=1}^{\infty} \pi^{*2}_k.
    \end{align*}
    The second equation follows from the independence of the measures. Therefore, 
    \begin{align*}
        P(F_{\mathbf{c},i} = F_{\mathbf{c},j}) = \E\left[P(F_{\mathbf{c},i} = F_{\mathbf{c},j} \mid Q_{\mathbf{c}}) \right] =\E\left[\sum_{k=1}^{\infty} \pi^{*2}_k \right]=\sum_{k=1}^{\infty} \E[\pi^{*2}_k] = \frac{1}{\alpha + 1}.
    \end{align*}
    For any measurable set $A \in \mathcal{B}$, 
    \begin{align*}
        \E[F_{\mathbf{c},i}(A) \mid Q_{\mathbf{c}} ] = \sum_{k=1}^{\infty} F^{*}_{\mathbf{c},k}(A)P(F_{\mathbf{c},i}=F^{*}_{\mathbf{c},k} \mid Q_{\mathbf{c}}) = \sum_{k=1}^{\infty} \pi^{*}_k F^{*}_{\mathbf{c},k}(A).
    \end{align*}
    Thus, we have 
    \begin{align*}
        \E[F_{\mathbf{c},i}(A) ] = \E\left[ \E[F_{\mathbf{c},i}(A) \mid Q_{\mathbf{c}} ] \right] = \sum_{k=1}^{\infty} \E[\pi^{*}_k] \E[F^{*}_{\mathbf{c},k}(A)] =  \E[F^{*}_{\mathbf{c},k}(A)] = G_{\mathbf{c}}^0(A),
    \end{align*}
    which follows from the independence of weights and atoms.
    Similarily, 
    \begin{align*}
        \E[F_{\mathbf{c},i}^2(A)] = \E\left[ \E[F_{\mathbf{c},i}^2(A) \mid Q_{\mathbf{c}} ] \right] = \E\left[\sum_{k=1}^{\infty} \pi^{*}_k F^{*2}_{\mathbf{c},k}(A) \right] = \E[F^{*2}_{\mathbf{c},k}(A)].
    \end{align*}
    Since $\E[F_{\mathbf{c},i}(A) ] =  \E[F^{*}_{\mathbf{c},k}(A)]$, we also have 
    \begin{align*}
        \Var[F_{\mathbf{c},i}(A)] = \Var[F^{*}_{\mathbf{c},k}(A)] = \frac{G_{\mathbf{c}}^0(A)(1-G_{\mathbf{c}}^0(A))}{\beta+1}.
    \end{align*}
\end{proof}

\subsection{Proof of Proposition \ref{prop:correlation_measures}}
\label{sec:proof_correlation}
\begin{proof}
    For  $\mathbf{c}, \mathbf{c}'  \in \mathcal{C}$, $i \neq i'$ and measurable sets $A, A' \in \mathcal{B}$, 
    \begin{align*}
        \E[F_{\mathbf{c},i}(A)F_{\mathbf{c}',i'}(A')] =  \E \left[ \E[ F_{\mathbf{c},i}(A)F_{\mathbf{c}',i'}(A') \mid Q_{\mathbf{c}}, Q_{\mathbf{c}'}]\right].
    \end{align*}
    Then, the inner expectation can be expressed as:
    \begin{align*}
        &\E[ F_{\mathbf{c},i}(A)F_{\mathbf{c}',i'}(A') \mid Q_{\mathbf{c}}, Q_{\mathbf{c}'}]  \\
        &= \sum_{k \geq 1}\sum_{k' \geq 1} P(F_{\mathbf{c},i}=F^{*}_{\mathbf{c},k}, F_{\mathbf{c}',i'} = F^{*}_{\mathbf{c}',k'} \mid Q_{\mathbf{c}}, Q_{\mathbf{c}'})F^{*}_{\mathbf{c},k}(A)F^{*}_{\mathbf{c}',k'}(A') \\
        &= \sum_{k = k'}P(F_{\mathbf{c},i}=F^{*}_{\mathbf{c},k}, F_{\mathbf{c}',i'} = F^{*}_{\mathbf{c}',k} \mid Q_{\mathbf{c}}, Q_{\mathbf{c}'})F^{*}_{\mathbf{c},k}(A)F^{*}_{\mathbf{c}',k}(A') \\
        &+ \sum_{k \geq 1}\sum_{k' \neq k} P(F_{\mathbf{c},i}=F^{*}_{\mathbf{c},k} \mid Q_{\mathbf{c}} )P(F_{\mathbf{c}',i'} = F^{*}_{\mathbf{c}',k'} \mid Q_{\mathbf{c}'})F^{*}_{\mathbf{c},k}(A)F^{*}_{\mathbf{c}',k'}(A') \\
        &= \sum_{k = k'} \pi^{*2}_k \left \{ \sum_{l\geq 1} w_{lk}^{*}\delta_{\theta_{lk}^{*}(\mathbf{c})}(A) \right\} \left \{ \sum_{m\geq 1} w_{mk}^{*}\delta_{\theta_{mk}^{*}(\mathbf{c}')}(A') \right\} \\
        &+ \sum_{k \geq 1}\sum_{k' \neq k} \pi^{*}_{k}\pi^{*}_{k'} \left \{ \sum_{l\geq 1} w_{lk}^{*}\delta_{\theta_{lk}^{*}(\mathbf{c})}(A) \right\} \left \{ \sum_{m\geq 1} w_{mk'}^{*}\delta_{\theta_{mk'}^{*}(\mathbf{c}')}(A') \right\} \\
        &= \sum_{k = k'} \pi^{*2}_k \left \{ \sum_{l=m} w_{lk}^{*2}\delta_{\theta_{lk}^{*}(\mathbf{c})}(A)\delta_{\theta_{lk}^{*}(\mathbf{c}')}(A') + \sum_{l\geq 1}\sum_{m \neq l} w_{lk}^{*}w_{mk}^{*}\delta_{\theta_{lk}^{*}(\mathbf{c})}(A)\delta_{\theta_{mk}^{*}(\mathbf{c}')}(A') \right\}  \\
        &+ \sum_{k \geq 1}\sum_{k' \neq k} \pi^{*}_{k}\pi^{*}_{k'} \left \{ \sum_{l\geq 1} w_{lk}^{*}\delta_{\theta_{lk}^{*}(\mathbf{c})}(A) \right\} \left \{ \sum_{m\geq 1} w_{mk'}^{*}\delta_{\theta_{mk'}^{*}(\mathbf{c}')}(A') \right\}.
    \end{align*}
    Taking the outer expectation, we obtain
    \begin{align*}
        &\E[F_{\mathbf{c},i}(A)F_{\mathbf{c}',i'}(A')] \\
        &= \sum_{k = k'} \E[\pi^{*2}_k] \left \{ \sum_{l=m} \E[w_{lk}^{*2}]P\left\{ \theta_{lk}^{*}(\mathbf{c}) \in A, \theta_{lk}^{*}(\mathbf{c}') \in A'  \right\} + \sum_{l\geq 1}\sum_{m \neq l} \E[w_{lk}^{*}]\E[w_{mk}^{*}] G_{\mathbf{c}}^0(A)G_{\mathbf{c}'}^0(A') \right\}  \\
        &+ \sum_{k \geq 1}\sum_{k' \neq k} \E[\pi^{*}_{k}\pi^{*}_{k'}] G_{\mathbf{c}}^0(A)G_{\mathbf{c}'}^0(A) \\
        &= \frac{1}{\alpha + 1} \left\{ \frac{P\left\{ \theta_{lk}^{*}(\mathbf{c}) \in A, \theta_{lk}^{*}(\mathbf{c}') \in A'  \right\}}{\beta + 1} + \frac{\beta G_{\mathbf{c}}^0(A)G_{\mathbf{c}'}^0(A')}{\beta + 1} \right\} + \frac{\alpha G_{\mathbf{c}}^0(A)G_{\mathbf{c}'}^0(A')}{\alpha+1}.
    \end{align*}
    Therefore, 
    \begin{align*}
        \Cov(F_{\mathbf{c},i}(A), F_{\mathbf{c}',i'}(A')) &=  \E[F_{\mathbf{c},i}(A)F_{\mathbf{c}',i'}(A')] - \E[F_{\mathbf{c},i}(A)]E[F_{\mathbf{c}',i'}(A')] \\
        &= \frac{1}{\alpha + 1} \left\{ \frac{P\left\{ \theta_{lk}^{*}(\mathbf{c}) \in A, \theta_{lk}^{*}(\mathbf{c}') \in A'  \right\} - G_{\mathbf{c}}^0(A)G_{\mathbf{c}'}^0(A')}{\beta + 1} \right\}.
    \end{align*}
     Plugging in this result for the correlation formula, we obtain the desired result for the case of $i \neq j$.

    When $i = j$, we have
    \begin{align*}
        &\E[ F_{\mathbf{c},i}(A)F_{\mathbf{c}',i}(A') \mid Q_{\mathbf{c}}, Q_{\mathbf{c}'}]  \\
        &= \sum_{k \geq 1}P(F_{\mathbf{c},i}=F^{*}_{\mathbf{c},k}, F_{\mathbf{c}',i} = F^{*}_{\mathbf{c}',k'} \mid Q_{\mathbf{c}}, Q_{\mathbf{c}'})F^{*}_{\mathbf{c},k}(A)F^{*}_{\mathbf{c}',k'}(A') \\
        &= \sum_{k \geq 1} \pi^{*}_k \left \{ \sum_{l\geq 1} w_{lk}^{*}\delta_{\theta_{lk}^{*}(\mathbf{c})}(A) \right\} \left \{ \sum_{m\geq 1} w_{mk}^{*}\delta_{\theta_{mk}^{*}(\mathbf{c}')}(A') \right\} \\
        &= \sum_{k \geq 1} \pi^{*}_k \left \{ \sum_{l=m} w_{lk}^{*2}\delta_{\theta_{lk}^{*}(\mathbf{c})}(A)\delta_{\theta_{lk}^{*}(\mathbf{c}')}(A') + \sum_{l\geq 1}\sum_{m \neq l} w_{lk}^{*}w_{mk}^{*}\delta_{\theta_{lk}^{*}(\mathbf{c})}(A)\delta_{\theta_{mk}^{*}(\mathbf{c}')}(A') \right\}.
    \end{align*}
    Taking the outer expectation, we obtain
    \begin{align*}
        \E[ F_{\mathbf{c},i}(A)F_{\mathbf{c}',i}(A')] =  \frac{P\left\{ \theta_{lk}^{*}(\mathbf{c}) \in A, \theta_{lk}^{*}(\mathbf{c}') \in A'  \right\} - G_{\mathbf{c}}^0(A)G_{\mathbf{c}'}^0(A')}{\beta + 1}.
    \end{align*}
    Plugging in this result for the covariance and correlation formulas, we obtain the desired result.
\end{proof}

\subsection{Proof of Proposition \ref{prop:weak_support}}
\label{sec:proof_weak_support}
\begin{proof}
    We follow the proof technique of \citet{Barrientos2012}, who established a sufficient condition under which dependent Dirichlet processes (DDPs) have full weak support. Specifically, the proofs in \citet{Barrientos2012} consist of two parts. The first part demonstrates that a sufficient condition for achieving full weak support is that the process assigns positive probability mass to a product space of particular simplices \citep[Theorem~1]{Barrientos2012}. This portion of their proof is directly applicable to our setting. Therefore, it suffices to adapt only the second part of their proof to our scenario, similar to Theorem 2 of \citet{Barrientos2012}. In particular, we will show 
    \begin{equation}
    \label{eq:condition1}
        P\Bigl\{\omega \in \Omega: \bigl[F(\mathbf{c}_i,\omega)(A_0), \ldots, F(\mathbf{c}_i,\omega)(A_J)\bigr] \in B\bigl(\mathbf{s}_{\mathbf{c}_i}, \epsilon\bigr),\, i = 1,\ldots,T \Bigr\} > 0,
    \end{equation}
    where 
    $A_j \subset \Theta$ are sets having positive measure with respect to the base measure $G_{\mathbf{c}}^0$, i.e.\ $G_{\mathbf{c}}^0(A_j)>0$ and $T$ is a positive integer. 
    Let $J$ be the number of such sets, $J = \#\{\,j : G_{\mathbf{c}}^0(A_j)>0\}$ (cf.\ \citealt[Theorem~1]{Barrientos2012} for the detailed construction). 
    We set $\mathbf{s}_{\mathbf{c}_i}=(w_{(\mathbf{c}_i,0)},\ldots,w_{(\mathbf{c}_i,J)})=(Q_{\mathbf{c}_i}(A_0), \ldots, Q_{\mathbf{c}_i}(A_J))\in \Delta_J$ for $i=1,\ldots,T$, where $Q_{\mathbf{c}_i}$ is a probability measure absolutely continuous w.r.t.\ $G_{\mathbf{c}_i}^0$, and 
    $\Delta_J=\{(w_0,\ldots,w_J): w_j \geq 0,\, \sum_{j=0}^{J} w_j=1\}$ is the $J$-simplex. For some $\varepsilon>0$, define
    \[
      B(\mathbf{s}_{\mathbf{c}_i}, \varepsilon)=\Bigl\{(w_0,\ldots,w_J)\in \Delta_J : \bigl|w_j - w_{(\mathbf{c}_i,j)}\bigr| < \varepsilon,\; j=0,\ldots,J\Bigr\}.
    \]
    Proving \eqref{eq:condition1} is equivalent to proving the positivity statement in \cite[Equation~(3)]{Barrientos2012}.

    Since the rational numbers are dense in $\mathbb{R}$, there exist $M_i,\,m_{ij}\in\mathbb{N}$ such that, for $i=1,\dots,T$ and $j=0,\dots,J-1$,
    \begin{equation*}
      w_{(\mathbf{c}_i,j)}-\frac{\varepsilon}{4} 
      \;<\; \frac{m_{ij}}{M_i}
      \;<\; w_{(\mathbf{c}_i,j)}+\frac{\varepsilon}{4}.
    \end{equation*}
    Set $N=M_1\times \dots\times M_T$, and define 
    \[
      n_{ij}\;=\; m_{ij}\,\prod_{k\neq i} M_k.
    \]
    It follows that, for $i=1,\dots,T$ and $j=0,\dots,J-1$,
    \begin{equation*}
      w_{(\mathbf{c}_i,j)}-\frac{\varepsilon}{4}
      \;<\; \frac{n_{ij}}{N}
      \;<\; w_{(\mathbf{c}_i,j)}+\frac{\varepsilon}{4}.
    \end{equation*}
    Therefore, for any $\mathbf{p}=(p_1,\dots,p_N)\in\Delta_{N-1}$ satisfying
    \[
      \frac{1}{N}-\frac{\varepsilon}{4N}
      \;<\; p_k
      \;<\; \frac{1}{N}+\frac{\varepsilon}{4N},\quad k=1,\dots,N,
    \]
    we obtain
    \[
      w_{(\mathbf{c}_i,0)}-\frac{\varepsilon}{2}
      \;<\;\sum_{k=1}^{n_{i0}}p_k
      \;<\; w_{(\mathbf{c}_i,0)}+\frac{\varepsilon}{2},
      \quad i=1,\dots,T,
    \]
    and similarly
    \[
      w_{(\mathbf{c}_i,j)}-\frac{\varepsilon}{2}
      \;<\;\sum_{k=n_{i,(j-1)}+1}^{n_{ij}}p_k
      \;<\; w_{(\mathbf{c}_i,j)}+\frac{\varepsilon}{2},
    \]
    for $i=1,\dots,T$ and $j=1,\dots,J-1$.

    Next, define a mapping $a(i,k)$ such that, for $i=1,\dots,T$ and $k=1,\dots,N$,
    \[
      a(i,k)=
      \begin{cases}
        0, & \text{if } k\le n_{i0},\\[4pt]
        1, & \text{if } n_{i0}<k\le n_{i0}+n_{i1},\\[4pt]
        \vdots & \\[4pt]
        J-1, & \text{if }\sum_{j=0}^{J-2}n_{ij}<k\le\sum_{j=0}^{J-1}n_{ij},\\[4pt]
        J, & \text{if }\sum_{j=0}^{J-1}n_{ij}<k\le N.
      \end{cases}
    \]
    Consider a subset $\Omega_0\subseteq \Omega$ such that, for every $\omega\in\Omega_0$, the following conditions hold:

    \begin{enumerate}[(1)]
      \item For $k=1$,
      \[
        \frac{1}{N}-\frac{\varepsilon}{4N}
        \;<\; s_k^{*}(\omega)
        \;<\; \frac{1}{N}+\frac{\varepsilon}{4N}.
      \]

      \item For $k=2,\dots,N-1$,
      \[
        \frac{\tfrac{1}{N}-\tfrac{\varepsilon}{4N}}
             {\prod_{k'<k}[1 - s_{k'}^{*}(\omega)]}
        \;<\; s_k^{*}(\omega)
        \;<\; \frac{\tfrac{1}{N}+\tfrac{\varepsilon}{4N}}
             {\prod_{k'<k}[1 - s_{k'}^{*}(\omega)]}.
      \]

      \item For $k=N$,
      \[
        \frac{\,1 - \sum_{k'=1}^{N-1}\pi^{*}_{k'}(\omega)\;-\;\tfrac{\varepsilon}{2}\,}
             {\prod_{k'<N}(1 - s_{k'}^{*}(\omega))}
        \;<\; s^{*}_N(\omega)
        \;<\; \frac{\,1 - \sum_{k'=1}^{N-1}\pi^{*}_{k'}(\omega)\,}
             {\prod_{k'<N}(1 - s_{k'}^{*}(\omega))},
      \]
      where $\,\pi^{*}_{k-1}(\omega)\;=\;s_k^{*}(\omega)\prod_{k'<k}[\,1-s_{k'}^{*}(\omega)\,]$.

      \item Given any $\delta>0$, $F^{*}_k(\mathbf{c}_i,\omega)\bigl(A_{a(i,k)}\bigr) \;>\; 1 - \delta$,  for $i=1,\dots,T$ and $k=1,\dots,N$. 
    \end{enumerate}

    To prove the proposition, it suffices to show $P(\{\omega:\omega\in\Omega_0\})>0$. 
    Conditions (1)--(3) constrain the outer stick-breaking weights $s_k^{*}(\omega)$ so that the resulting mixture weights approximate the desired rational structure, matching the sums $\sum_{k=1}^{N}p_k$ as above. Condition~(4) ensures that $F^{*}_k(\mathbf{c}_i,\omega)(A_j) \approx 1$ for those $k$ with $a(i,k)=j$, stipulating that each inner measure $F^{*}_k(\mathbf{c}_i,\omega)$ puts arbitrarily large mass into $A_{a(i,k)}$. Since $F^{*}_k(\mathbf{c}_i,\omega)$ is a single weights DDP for each $k$, Theorem~2 of \citet{Barrientos2012} applies under the assumption that $\Psi_{\mathcal{C}^{\theta}}$ (the set of copulas governing the atoms) has a positive density w.r.t.\ Lebesgue measure.  And, due to this single-weights DDP's full weak support property, $F(\mathbf{c}_k,\omega)\bigl(A_{a(i,k)}\bigr)$ can get as close as we wish to a measure that is (nearly) degenerate on $A_{a(i,k)}$. In other words, we can force $F(\mathbf{c}_k,\omega)$ in the block of $k$ mapping to $j$ to be arbitrarily close to any desired distribution $Q_k$, which lumps nearly all of its mass on $A_j$.
    Therefore, if (1)--(4) hold, then it is clear to see that, for each $i=1,\dots,T$,
    \begin{align*}
        \bigl[F(\mathbf{c}_i,\omega)\bigl(A_0\bigr),\dots,F(\mathbf{c}_i,\omega)\bigl(A_J\bigr)\bigr]
      \;\in\; B\bigl(\mathbf{s}_{\mathbf{c}_i},\varepsilon\bigr).
    \end{align*}

    By definition of the AD-nDDP,
    \begin{align*}
      &P\Bigl\{\omega\in\Omega:\bigl[F(\mathbf{c}_i,\omega)(A_0),\dots,F(\mathbf{c}_i,\omega)(A_J)\bigr]\in B\bigl(\mathbf{s}_{\mathbf{c}_i},\varepsilon\bigr),\, i=1,\dots,T\Bigr\}\\
      &\quad \;\ge\; P\Bigl\{\omega\in\Omega: s_k^{*}(\omega)\in Q_k^\omega,\; k=1,\dots,N\Bigr\}
      \;\times\; \prod_{k=1}^N\;\Bigl[\prod_{i=1}^T\,F^{*}_k(\mathbf{c}_i,\omega)\bigl(A_{a(i,k)}\bigr)\Bigr]\\
      &\qquad\qquad \times\;\prod_{k=N+1}^{\infty}P\{\omega\in\Omega : s_k^{*}(\omega)\in[0,1]\}
      \;\times\;\prod_{k=N+1}^{\infty}\Bigl[\prod_{i=1}^T\,F^{*}_k(\mathbf{c}_i,\omega)\bigl(\Theta\bigr)\Bigr],
    \end{align*}
    where the sets $Q_k^\omega$ are the intervals induced by conditions (1)--(3).

    Since $\{s_k^{*}(\omega)\}$ are independent Beta random variables (each having strictly positive density on $(0,1)$), it follows that 
    \[
      P\Bigl\{\omega\in\Omega: s_k^{*}(\omega)\in Q_k^\omega,\; k=1,\dots,N\Bigr\}\;>\;0.
    \]
    Condition~(4) ensures that $\prod_{i=1}^T F^{*}_k(\mathbf{c}_i,\omega)\bigl(A_{a(i,k)}\bigr)$ also has strictly positive probability. Hence the overall product is strictly positive, which completes the proof.
\end{proof}

\subsection{Proof of Proposition \ref{prop:properties_fdnddp}}
Since we have $\sum_{k=1}^{\infty}\pi^{*}_k(\mathbf{v})=1$ \citep{Reich2007}, the expectation and variance are derived in the same way as in Section \ref{sec:proof_equation1}.
For correlation, it is sufficient to consider $\sum_{k = 1}^{\infty} \E[\pi_k(\mathbf{v})\pi_k(\mathbf{v}')]$ for $\mathbf{v} \neq \mathbf{v}'$ in Section \ref{sec:proof_correlation} because other steps follow as they are. By simple calculus, we have
\begin{align*}
    &h(\alpha, \mathbf{v}, \mathbf{v}') \\
    &:= \sum_{k = k'} \E[\pi_k(\mathbf{v})\pi_k(\mathbf{v}')] \\
    &= \frac{2}{(\alpha+1)(\alpha+2)} \sum_{k=1}^{\infty}K^*(\mathbf{v};\Gamma_k)K^*(\mathbf{v}';\Gamma_k)\prod_{j=1}^{i-1} \left\{ 1- \frac{K^*(\mathbf{v};\Gamma_k)+K^*(\mathbf{v}';\Gamma_k)}{\alpha+1} +\frac{2K(\mathbf{v};\Gamma_k)K^*(\mathbf{v}';\Gamma_k)}{(\alpha+1)(\alpha+2)}  \right\}.
\end{align*}
By following the same procedure in Section \ref{sec:proof_correlation}, we obtain the desired result for correlation.

Finally, for the weak support property, using a similar reasoning as in Section \ref{sec:proof_weak_support}, it suffices to prove \ref{eq:condition1} under the FD-nDDP model.
Similar to Section \ref{sec:proof_weak_support}, consider a subset $\Omega_0\subseteq \Omega$ such that, for every $\omega\in\Omega_0$, the following conditions hold:
    \begin{enumerate}[(1)]
      \item For $i=1,\ldots,T$,
      \[
        w_{(\mathbf{c}_i,0)} - \frac{\varepsilon}{2}
        \;<\; U_1^{*}(\mathbf{v}_i, \omega)
        \;<\; w_{(\mathbf{c}_i,0)} + \frac{\varepsilon}{2}
      \]

      \item For $i=1,\ldots,T$ and $j=1,\dots,J-1$,
      \[
        \frac{w_{(\mathbf{c}_i,j)} - \frac{\varepsilon}{2}}{\prod_{l<j+1} (1-U_{l}^{*}(\mathbf{v}_i, \omega))}
        \;<\; U_{j+1}^{*}(\mathbf{v}_i, \omega)
        \;<\; \frac{w_{(\mathbf{c}_i,j)} + \frac{\varepsilon}{2}}{\prod_{l<j+1} (1-U_{l}^{*}(\mathbf{v}_i, \omega))}.
      \]

      \item For $i=1,\ldots,T$,
      \[
        \frac{1 - \sum_{j=1}^{J-1}W^{*}_{j}(\mathbf{v}_i, \omega) - \frac{\varepsilon}{2}}
             {\prod_{l<J+1}(1 - U_{l}^{*}(\mathbf{v}_i, \omega))}
        \;<\; U_{J+1}^{*}(\mathbf{v}_i, \omega)
        \;<\; \frac{1 - \sum_{j=1}^{J-1}W^{*}_{j}(\mathbf{v}_i, \omega)}
             {\prod_{l<J+1}(1 - U_{l}^{*}(\mathbf{v}_i, \omega))}
      \]
      where for $j=1,\ldots,J-1$, $W^{*}_{k-1}(\mathbf{v}_i,\omega)\;=\;U_k^{*}(\mathbf{v}_i,\omega)\prod_{k'<k}[\,1-U_{k'}^{*}(\mathbf{v}_i,\omega)\,]$.

      \item Given any $\delta>0$, $F^{*}_k(\mathbf{c}_i,\omega)\bigl(A_{j}\bigr) \;>\; 1 - \delta$,  for $i=1,\dots,T$ and $j=1,\dots,J+1$. 
    \end{enumerate}
    Then,  it is clear to see that, if (1)--(4) hold, for each $i=1,\dots,T$,
    \begin{align*}
        \bigl[F(\mathbf{c}_i,\omega)\bigl(A_0\bigr),\dots,F(\mathbf{c}_i,\omega)\bigl(A_J\bigr)\bigr]
      \;\in\; B\bigl(\mathbf{s}_{\mathbf{c}_i},\varepsilon\bigr).
    \end{align*}
    It then follows from the FD-nDDP definition that 
    \begin{align*}
      &P\Bigl\{\omega\in\Omega:\bigl[F(\mathbf{c}_i,\omega)(A_0),\dots,F(\mathbf{c}_i,\omega)(A_J)\bigr]\in B\bigl(\mathbf{s}_{\mathbf{c}_i},\varepsilon\bigr),\, i=1,\dots,T\Bigr\}\\
      & \ge P\Bigl\{ \omega\in\Omega: [U_j^{*}(\mathbf{v}_1, \omega), \ldots, U_j^{*}(\mathbf{v}_T, \omega)]\in Q_j^\omega, j=1,\dots,J+1 \Bigr\} 
      \times \prod_{j=1}^{J+1}\prod_{i=1}^T\,F^{*}_k(\mathbf{c}_i,\omega)\bigl(A_{j}\bigr)\\
    & \times \prod_{j=J+2}^{\infty} P\Bigl\{ \omega\in\Omega: [U_j^{*}(\mathbf{v}_1, \omega), \ldots, U_j^{*}(\mathbf{v}_T, \omega)]\in [0,1]^{T} \Bigr\}
    \times \prod_{j=J+2}^{\infty}\prod_{i=1}^T\,F^{*}_k(\mathbf{c}_i,\omega)\bigl(\Theta\bigr),
    \end{align*}
    where the sets $Q_k^\omega$ are the intervals induced by conditions (1)--(3), that is, 
    \begin{align*}
        & Q_1^\omega = \prod_{i=1}^{T}\left[ w_{(\mathbf{c}_i,0)} - \frac{\varepsilon}{2}, w_{(\mathbf{c}_i,0)} + \frac{\varepsilon}{2}\right], \\
        & Q_{j+1}^\omega = \prod_{i=1}^{T}\left[ \frac{w_{(\mathbf{c}_i,j)} - \frac{\varepsilon}{2}}{\prod_{l<j+1} (1-U_{l}^{*}(\mathbf{v}_i, \omega))} , \frac{w_{(\mathbf{c}_i,j)} + \frac{\varepsilon}{2}}{\prod_{l<j+1} (1-U_{l}^{*}(\mathbf{v}_i, \omega))} \right] \text{ for } j=1,\ldots,J-1, \text{ and } \\
        & Q_{J+1}^\omega 
        = \prod_{i=1}^{T}\left[ \frac{1 - \sum_{j=1}^{J-1}W^{*}_{j}(\mathbf{v}_i, \omega) - \frac{\varepsilon}{2}}
             {\prod_{l<J+1}(1 - U_{l}^{*}(\mathbf{v}_i, \omega))}, \frac{1 - \sum_{j=1}^{J-1}W^{*}_{j}(\mathbf{v}_i, \omega)}
             {\prod_{l<J+1}(1 - U_{l}^{*}(\mathbf{v}_i, \omega))} \right].
    \end{align*}
    By the definition of the process, $P\Bigl\{ \omega\in\Omega: [U_j^{*}(\mathbf{v}_1, \omega), \ldots, U_j^{*}(\mathbf{v}_T, \omega)]\in [0,1]^{T} \Bigr\}=1
    $ and $F^{*}_k(\mathbf{c}_i,\omega)\bigl(\Theta\bigr)=1$. Recall that $U^{*}_k(\mathbf{v}, \omega) = K^{*}(\mathbf{v}; \boldsymbol{\Gamma}_k) s^{*}_{k}(\omega)$ with a positive bounded function $K^{*}$, and the non-sigularity of the Beta distribution for $s^{*}_{k}(\omega)$ implies that $P\Bigl\{ \omega\in\Omega: [U_j^{*}(\mathbf{v}_1, \omega), \ldots, U_j^{*}(\mathbf{v}_T, \omega)]\in Q_j^\omega, j=1,\dots,J+1 \Bigr\} > 0$. Condition~(4) ensures that $\prod_{i=1}^T F^{*}_k(\mathbf{c}_i,\omega)\bigl(A_{a(i,k)}\bigr)$ also has strictly positive probability. Hence, the overall product is strictly positive, which completes the proof.
}

\section{Details of Gibbs sampler}
\label{sec:gibbs_details}
\subsection{Gibbs sampler for AD-nDDPM}
The posterior distributions of the model parameters are obtained from the Markov chain Monte Carlo (MCMC) method. We employ an approximated blocked Gibbs sampler \citep{Ishwaran2000} based on a two-level truncation of the stick-breaking representation of the DP proposed by \citet{Rodriguez2008}. As described in the main manuscript, we set conservative upper bounds on the number of latent classes at cluster and individual levels. We set $K_I=5$ and $K_C=10$ by examining the sizes that are large enough for all clusters not to be occupied. This section first details the Gibbs sampling algorithm for the AD-nDDPM model.

\subsubsection{Sample $\zeta_i$}
Given $\mathbf{C}^{m}_{ij}$ and $\mathbf{M}_{ij} = (M^{(1)}_{ij}, M^{(2)}_{ij})^\top$ for $i=1,\ldots,I$ and $j=1,\ldots,N_i$, and $\pi^{*}_{k}$, $w^{*}_{lk}$, $\boldsymbol{\gamma}_{1,lk}$, $\boldsymbol{\gamma}_{2,lk}$, and $\boldsymbol{\Sigma}_{lk}$ for $l=1,\ldots,K_I$ and $k=1,\ldots,K_C$, we sample the cluster-level latent class assignment $\zeta_i$ for each cluster $i$ from the multinomial distribution with probability:

\begin{equation*} 
P(\zeta_i = k \mid \cdot) = \frac{\pi^{*}_k \prod_{j = 1}^{N_i} \left( \sum_{l=1}^{K_I} w^{*}_{lk} \mathrm{MVN}\left( \mathbf{M}_{ij}; (\boldsymbol{\gamma}_{1,lk}^\top \mathbf{C}^{m}_{ij}, \boldsymbol{\gamma}_{2,lk}^\top \mathbf{C}^{m}_{ij})^\top , \boldsymbol{\Sigma}_{lk} \right) \right)}{\sum_{k=1}^{K_C} \pi^{*}_k \prod_{j = 1}^{N_i} \left( \sum_{l=1}^{K_I} w^{*}_{lk} \mathrm{MVN}\left( \mathbf{M}_{ij}; (\boldsymbol{\gamma}_{1,lk}^\top \mathbf{C}^{m}_{ij}, \boldsymbol{\gamma}_{2,lk}^\top \mathbf{C}^{m}_{ij})^\top , \boldsymbol{\Sigma}_{lk} \right) \right)}, 
\end{equation*}
where $\mathrm{MVN}(\mathbf{M}; \boldsymbol{\mu}, \boldsymbol{\Sigma})$ denotes the multivariate Gaussian density with mean $\boldsymbol{\mu}$ and covariance $\boldsymbol{\Sigma}$ evaluated at $\mathbf{M}$.
For each individual $j$ in cluster $i$, set $\zeta_{ij} = \zeta_i$.

\subsubsection{Sample $\xi_{ij}$}

For each individual $j$ within cluster $i$, given the cluster-level class assignment $\zeta_i$, sample the individual-level latent class assignment $\xi_{ij}$ from the multinomial distribution with probability:

\begin{equation*} 
P(\xi_{ij} = l \mid \cdot) = \frac{w^{*}_{l\zeta_i}  \mathrm{MVN}\left( \mathbf{M}_{ij};(\boldsymbol{\gamma}_{1,l\zeta_i}^\top \mathbf{C}^{m}_{ij}, \boldsymbol{\gamma}_{2,l\zeta_i}^\top \mathbf{C}^{m}_{ij})^\top, \boldsymbol{\Sigma}_{l\zeta_i} \right)}{\sum_{l=1}^{K_I}w^{*}_{l\zeta_i}  \mathrm{MVN}\left( \mathbf{M}_{ij};(\boldsymbol{\gamma}_{1,l\zeta_i}^\top \mathbf{C}^{m}_{ij}, \boldsymbol{\gamma}_{2,l\zeta_i}^\top \mathbf{C}^{m}_{ij})^\top, \boldsymbol{\Sigma}_{l\zeta_i} \right)} , 
\end{equation*}

\subsubsection{Sample $\pi^{*}_k$ and $s^{*}_k$}
\label{sec:gibbs_update_s_pi}

Let $s^{*}_{K_C}=1$. Given $\alpha$ and $\zeta_i$, draw $s^{*}_k$ for $k=1,\ldots,K_C-1$ from
\begin{align} 
    s^{*}_k \sim \text{Be}\left(1 + \sum_{i=1}^{I} \mathbbm{1}(\zeta_i = k), \alpha + \sum_{i=1}^{I} \mathbbm{1}(\zeta_i > k)\right).
\end{align}
Then update $\pi^{*}_k = s^{*}_k \prod_{j=1}^{k-1} (1 - s^{*}_j)$.

\subsubsection{Sample $w^{*}_{lk}$ and $u^{*}_{lk}$}

For each class $k$, let $u_{K_{I}k}=1$. Given $\beta_k$ and $\xi_{ij}$, draw  $u_{lk}$ for $l=1,\ldots,K_{I}-1$ from
\begin{align} 
    u_{lk} &\sim \text{Be}\left(1 + \sum_{i=1}^{I}\sum_{j=1}^{N_i} \mathbbm{1}(\xi_{ij} = l, \zeta_{i} = k), \beta_k + \sum_{i=1}^{I}\sum_{j=1}^{N_i} \mathbbm{1}(\xi_{ij} > l, \zeta_{i} = k) \right). 
\end{align}
Then update $w^{*}_{lk} = u^{*}_{lk} \prod_{j=1}^{l-1} (1 - u^{*}_{jk})$ for $k=1,\ldots,K_I$.

\subsubsection{Update $\alpha$ and $\beta_k$}

Assuming the conjugate priors $\alpha \sim \mathrm{Ga}(a_\alpha, b_\alpha)$ and $\beta_k \sim \mathrm{Ga}(a_\beta, b_\beta)$, update the concentration parameters $\alpha$ and $\beta_k$:
\begin{align*} 
    \alpha &\sim \mathrm{Ga}\left( a_\alpha + K_C - 1, b_\alpha - \sum_{k=1}^{K_C - 1} \ln(1 - s_k)  \right), \\
    \beta_k &\sim \mathrm{Ga}\left( a_\beta + K_I - 1, b_\beta - \sum_{l=1}^{K_I - 1} \ln(1 - u_{lk}) \right). 
\end{align*}

\subsubsection{Sample $\boldsymbol{\gamma}_{1,lk}$, $\boldsymbol{\gamma}_{2,lk}$, and $\boldsymbol{\Sigma}_{lk}$}
\label{sec:gibbs_gamma_Sigma}
For each  $l$ and $k$, update the atoms (the regression coefficients and covariance matrix for each component of the mixture). Let $n_{lk} = \sum_{i=1}^{I}\sum_{j=1}^{N_i} \mathbbm{1}(\xi_{ij} = l, \zeta_{i} = k)$.
\begin{itemize}
    \item If $n_{lk} = 0$ (no data assigned to component $(l,k)$), sample from the prior:
    \begin{align*} 
        \boldsymbol{\Sigma}_{lk} &\sim \text{IW}(\nu_0, \boldsymbol{\Psi}_0), \\ 
        \boldsymbol{\gamma}_{1,lk} &\sim \mathrm{MVN}(\mathbf{0}, \mathbf{S}_0), \\ 
        \boldsymbol{\gamma}_{2,lk} &\sim \mathrm{MVN}(\mathbf{0}, \mathbf{S}_0). 
    \end{align*}
    \item If $n_{lk} > 0$, update using the data:
    We assumed the following prior distributions:
    \begin{align*}
        \boldsymbol{\Sigma}_{lk} &\sim \text{IW}(\nu_0, \boldsymbol{\Psi}_0), \\ 
        \boldsymbol{\gamma}_{1,lk} &\sim \mathrm{MVN}(\mathbf{0}, \mathbf{S}_0), \\ 
        \boldsymbol{\gamma}_{2,lk} &\sim \mathrm{MVN}(\mathbf{0}, \mathbf{S}_0). 
    \end{align*}
    \begin{enumerate} 
        \item Collect the data assigned to component $(l,k)$: \begin{itemize} 
            \item Let $\mathbf{M}^{(1)}_{lk}$, $\mathbf{M}^{(2)}_{lk}$ denote $n_{lk}$-dimensional vectors of  $M^{(1)}_{lk}$ and $M^{(2)}_{lk}$ for all $(i,j)$ such that $\xi_{ij} = l$ and $\zeta_{i} = k$.  Let $\mathbf{C}^{m}_{lk}$ denote the $(n_{lk} \times d_m)$-matrix of $\mathbf{C}^{m}_{ij} \in \mathbb{R}^{d_m}$ corresponding to the same indices.
        \end{itemize} 
            \item Update the covariance matrix $\boldsymbol{\Sigma}_{lk}$:
            \begin{align*}
                \boldsymbol{\Sigma}_{lk} &\sim \text{IW}\left( \nu_0 + n_{lk}, \boldsymbol{\Psi}_0 + \mathbf{S} \right), 
            \end{align*}
            where $\mathbf{S} = (\Delta_1, \Delta_2)^\top (\Delta_1, \Delta_2)$ with $\Delta_1 = \mathbf{M}^{(1)}_{lk} - \mathbf{C}^{m}_{lk} \boldsymbol{\gamma}_{1,lk}$ and $\Delta_2 = \mathbf{M}^{(2)}_{lk} - \mathbf{C}^{m}_{lk} \boldsymbol{\gamma}_{2,lk}$.
            
            \item Update the regression coefficients $\boldsymbol{\gamma}_{1,lk}$: 
            \begin{align*} 
                \boldsymbol{\gamma}_{1,lk} \sim \mathrm{MVN}(m_{\boldsymbol{\gamma}_1}, \mathbf{V}_{\boldsymbol{\gamma}_1}),
            \end{align*}
            where $ \mathbf{V}_{\boldsymbol{\gamma}_1} = \left( \boldsymbol{\Sigma}_{lk}^{-1}(1,1) \mathbf{C}^{m\top}_{lk} \mathbf{C}^{m}_{lk} + \mathbf{S}_0^{-1} \right)^{-1}$ 
            
            \noindent
            and $m_{\boldsymbol{\gamma}_1} = \mathbf{V}_{\boldsymbol{\gamma}_1} \left( \boldsymbol{\Sigma}_{lk}^{-1}(1,1) \mathbf{C}^{m\top}_{lk} \mathbf{M}^{(1)}_{lk} + \boldsymbol{\Sigma}_{lk}^{-1}(1,2) \mathbf{C}^{m\top}_{lk} (\mathbf{M}^{(2)}_{lk} - \mathbf{C}^{m}_{lk} \boldsymbol{\gamma}_{2,lk}) \right)$, with $\boldsymbol{\Sigma}_{lk}^{-1}(1,1)$ and $\boldsymbol{\Sigma}_{lk}^{-1}(1,2)$ representing the $(1,1)$ and $(1,2)$ elements of the inverse covariance matrix $\boldsymbol{\Sigma}_{lk}^{-1}$. 
            \item Update the regression coefficients $\boldsymbol{\gamma}_{2,lk}$ in a similar way to the step above, switching the index $1$ with $2$.
    \end{enumerate}
\end{itemize}

\subsubsection{Repeat the same procedure for the Outcome Model}

The same steps are applied to the outcome model. The only difference is the update of $\boldsymbol{\theta}_{lk}$ and $\sigma_{lk}^2$. Assuming prior distributions $\boldsymbol{\theta}_{lk} \sim \mathrm{N}(0, \boldsymbol{\Sigma}_0)$ and $\sigma_{lk}^2 \sim \mathrm{IG}(a_0,b_0)$, sample $\boldsymbol{\theta}_{lk}$ and $\sigma_{lk}^2$ as follows.
\begin{align*} 
            \sigma_{lk}^2 &\sim \text{IG}\left( a_0 + \frac{n_{lk}}{2}, b_0 + \frac{1}{2} \sum_{(i,j)} (Y_{ij} - \mathbf{C}^{y}_{ij} \boldsymbol{\theta}_{lk})^2 \right), \\ 
            \boldsymbol{\theta}_{lk} &\sim \mathrm{MVN}(m_{\boldsymbol{\theta}}, \mathbf{V}_{\boldsymbol{\theta}}),
        \end{align*} 
    where $\mathbf{V}_{\boldsymbol{\theta}} = \left( \frac{1}{\sigma_{lk}^2} \mathbf{C}^{y\top}_{ij} \mathbf{C}^{y}_{ij} +\boldsymbol{\Sigma}^{-1}_0 \right)^{-1},             m_{\boldsymbol{\theta}} = \mathbf{V}_{\boldsymbol{\theta}} \left( \frac{1}{\sigma_{lk}^2} (\mathbf{C}^{y\top}_{ij} \mathbf{C}^{y}_{ij})^{-1}\mathbf{C}^{y \top}_{ij} Y_{lk} \right).$

\subsubsection{G-computation}
The final step is the g-computation step to obtain the draws of causal estimands. 
\begin{enumerate}
    \item Given all parameters at the current iteration, draw $M^{(1)}_{ij}(0)$ and $M^{(2)}_{ij}(0)$ from the posterior predictive distributions of $M^{(1)}_{ij}$ and $M^{(2)}_{ij}$ by letting $A_i=0$. Also draw $M^{(1)}_{ij}(1)$ and $M^{(2)}_{ij}(1)$ from their posterior predictive distributions by letting $A_i=1$. Specifically, for each individual $(ij)$, sample the mediators under different treatments $a = 0$ and $a = 1$: 
    \begin{align*} 
        M^{(1)}_{ij}(a) &\sim \mathrm{N}\left( \mathbf{C}^{m}_{ij}(a)^\top \boldsymbol{\gamma}_{1,\xi_{ij}\zeta_{ij}}, \boldsymbol{\Sigma}_{\xi_{ij}\zeta_{ij}}(1,1) \right), \ M^{(2)}_i(a) &\sim \mathrm{N}\left( \mathbf{C}^{m}_{ij}(a)^\top \boldsymbol{\gamma}_{2,\xi_{ij},\zeta_{ij}}, \boldsymbol{\Sigma}_{\xi_{ij}\zeta_{ij}}(2,2)\right),
    \end{align*}
    where $\mathbf{C}^{m}_{ij}(a)$ is a replication of $\mathbf{C}^{m}_{ij}$ with $A_i$ set to $a$.
    \item For each individual $j$ in each cluster $i$,  construct augmented covariates including the sampled mediators and their cluster means by computing the summary function $g_{ij}^{m}(\mathbf{M}_{i})=\left\{ M_{ij}, \frac{1}{N_i - 1} \sum_{\substack{k=1 \\ k \ne j}}^{N_i} M_{ik} \right\}$ based on the samples of $M^{(1)}_{ij}(a)$ and $M^{(2)}_{ij}(a)$. Note that the value of the summary function varies across individuals depending on the mediator values of other units within the same cluster.
    \item Given the samples of $M^{(1)}_{ij}(a)$ and $M^{(2)}_{ij}(a)$ for $a=0,1$, and the corresponding summary function, sample the outcome under different mediator values: 
    \begin{align*} 
        Y_{ij}(1, M^{(1)}_{ij}(1), M^{(1)}_{i(-j)}(1), &M^{(2)}_{ij}(1), M^{(2)}_{i(-j)}(1)) \\
        &\sim \mathrm{N}\left( \mathbf{C}^{y}_{ij}(1, M^{(1)}_{ij}(1), M^{(1)}_{i(-j)}(1), M^{(2)}_{ij}(1), M^{(2)}_{i(-j)}(1))^\top \boldsymbol{\theta}_{\xi_{ij},\zeta_{ij}}, \sigma_{\xi_{ij},\zeta_{ij}}^2 \right), \\
        Y_{ij}(1, M^{(1)}_{ij}(1), M^{(1)}_{i(-j)}(1), &M^{(2)}_{ij}(0), M^{(2)}_{i(-j)}(0)) \\
        &\sim \mathrm{N}\left( \mathbf{C}^{y}_{ij}(1, M^{(1)}_{ij}(1), M^{(1)}_{i(-j)}(1), M^{(2)}_{ij}(0), M^{(2)}_{i(-j)}(0))^\top \boldsymbol{\theta}_{\xi_{ij},\zeta_{ij}}, \sigma_{\xi_{ij},\zeta_{ij}}^2 \right), \\
        Y_{ij}(1, M^{(1)}_{ij}(1), M^{(1)}_{i(-j)}(1), &M^{(2)}_{ij}(1), M^{(2)}_{i(-j)}(0)) \\
        &\sim \mathrm{N}\left( \mathbf{C}^{y}_{ij}(1, M^{(1)}_{ij}(1), M^{(1)}_{i(-j)}(1), M^{(2)}_{ij}(1), M^{(2)}_{i(-j)}(0))^\top \boldsymbol{\theta}_{\xi_{ij},\zeta_{ij}}, \sigma_{\xi_{ij},\zeta_{ij}}^2 \right), \\
        Y_{ij}(1, M^{(1)}_{ij}(0), M^{(1)}_{i(-j)}(0), &M^{(2)}_{ij}(1), M^{(2)}_{i(-j)}(1)) \\
        &\sim \mathrm{N}\left( \mathbf{C}^{y}_{ij}(1, M^{(1)}_{ij}(0), M^{(1)}_{i(-j)}(0), M^{(2)}_{ij}(1), M^{(2)}_{i(-j)}(1))^\top \boldsymbol{\theta}_{\xi_{ij},\zeta_{ij}}, \sigma_{\xi_{ij},\zeta_{ij}}^2 \right), \\
        Y_{ij}(1, M^{(1)}_{ij}(1), M^{(1)}_{i(-j)}(0), &M^{(2)}_{ij}(1), M^{(2)}_{i(-j)}(1)) \\
        &\sim \mathrm{N}\left( \mathbf{C}^{y}_{ij}(1, M^{(1)}_{ij}(1), M^{(1)}_{i(-j)}(0), M^{(2)}_{ij}(1), M^{(2)}_{i(-j)}(1))^\top \boldsymbol{\theta}_{\xi_{ij},\zeta_{ij}}, \sigma_{\xi_{ij},\zeta_{ij}}^2 \right), \\
        Y_{ij}(1, M^{(1)}_{ij}(0), M^{(1)}_{i(-j)}(0), &M^{(2)}_{ij}(0), M^{(2)}_{i(-j)}(0)) \\
        &\sim \mathrm{N}\left( \mathbf{C}^{y}_{ij}(1, M^{(1)}_{ij}(0), M^{(1)}_{i(-j)}(0), M^{(2)}_{ij}(0), M^{(2)}_{i(-j)}(0))^\top \boldsymbol{\theta}_{\xi_{ij},\zeta_{ij}}, \sigma_{\xi_{ij},\zeta_{ij}}^2 \right), \\
        Y_{ij}(0, M^{(1)}_{ij}(0), M^{(1)}_{i(-j)}(0), &M^{(2)}_{ij}(0), M^{(2)}_{i(-j)}(0)) \\
        &\sim \mathrm{N}\left( \mathbf{C}^{y}_{ij}(0, M^{(1)}_{ij}(0), M^{(1)}_{i(-j)}(0), M^{(2)}_{ij}(0), M^{(2)}_{i(-j)}(0))^\top \boldsymbol{\theta}_{\xi_{ij},\zeta_{ij}}, \sigma_{\xi_{ij},\zeta_{ij}}^2 \right), 
    \end{align*} 
    where $\mathbf{C}^{y}_{ij}(a, M^{(1)}_{ij}, M^{(1)}_{i(-j)}, M^{(2)}_{ij}, M^{(2)}_{i(-j)})$ denotes the augmented covariates with the summary function computed from the baseline covariates and corresponding mediators.
    \item Average the potential outcomes across units to compute the esimands of interest, e.g., $\mathrm{ESME^{(1)}} = \frac{1}{I}\sum_{i=1}^{I}\frac{1}{N_i} \sum_{j=1}^{N_i} \{ Y_{ij}(1, M_{ij}^{(1)}(1), M_{i(-j)}^{(1)}(1), M_i^{(2)}(1)) - Y_{ij}(1, M_{ij}^{(1)}(1), M_{i(-j)}^{(1)}(0), M_i^{(2)}(1)) \}$.
\end{enumerate}

\subsection{Extensions to discrete variables}
\label{sec:extension_discrete_gibbs}
When the outcome or mediators are discrete variables, we adopt the probit data-augmentation approach \citep{Albert1993}. For simplicity, we describe the binary case here.
Let us consider the case where $M^{(1)}_{ij}$ is binary. We introduce a latent variable $Z_{ij}$ and posit the following model:
\begin{align*}
    \begin{pmatrix}
      Z_{ij} \\
      M^{(2)}_{ij}
    \end{pmatrix}
    &\sim \mathrm{MVN}\left( \begin{pmatrix}
      \mathbf{C}^{m}_{ij}\boldsymbol{\gamma}_{1,\xi_{ij}\zeta_{i}} \\
      \mathbf{C}^{m}_{ij}\boldsymbol{\gamma}_{2,\xi_{ij}\zeta_{i}}
    \end{pmatrix}, \boldsymbol{\Sigma}_{\xi_{ij}\zeta_{i}} \right), \\
    p\left(M^{(1)}_{ij}=m \mid Z_{ij} \right) &= p\left(Z_{ij}\leq 0 \right)^{m}\left(1-p\left(Z_{ij} > 0 \right) \right)^{1-m}.
\end{align*}
This modeling approach allows us to effectively capture the underlying correlation between $M^{(1)}_{ij}$ and $M^{(2)}_{ij}$ through the latent variable $Z_{ij}$. As $Z_{ij}$ marginally follows the Gaussian distribution, it also facilitates posterior inference using data-augmentation techniques for probit model. Given all parameters in the current iteration of MCMC, we draw $Z_{ij}$ from
\[
Z_{ij} \sim \begin{cases}
\mathrm{TN}(\mathbf{C}^{m}_{ij} \boldsymbol{\gamma}_{1,\xi_{ij}\zeta_{i}}, \boldsymbol{\Sigma}_{\xi_{ij}\zeta_{i}}(1,1), 0, \infty) & \text{if } M^{(1)}_{ij} = 1, \\
\mathrm{TN}(\mathbf{C}^{m}_{ij} \boldsymbol{\gamma}_{1,\xi_{ij}\zeta_{i}}, \boldsymbol{\Sigma}_{\xi_{ij}\zeta_{i}}(1,1), -\infty, 0) & \text{if } M^{(1)}_{ij} = 0,
\end{cases}
\]
where $\mathrm{TN}(\mu,\sigma^2,l,u)$ denotes the truncated normal distribution with mean, variance, lower bound, and upper bound parameters. Given $Z_{ij}$, the updates for other parameters are straightforward. We simply replace $M^{(1)}_{ij}$ with $Z_{ij}$ in all steps where $M^{(1)}_{ij}$ appears in Section \ref{sec:gibbs_details}.

\subsection{Gibbs sampler for FD-nDDPM}
\label{sec:gibbs_FD_nDDPM}
For the posterior inference of the FD-nDDPm model \eqref{eq:fd-nddpm}, we need to derive a sampling step for an additional parameter $\boldsymbol{\Gamma}_k$ and modify the sampling step of $\pi^{*}_k$ and $s^{*}_k$ in Section \ref{sec:gibbs_update_s_pi}, involving $\boldsymbol{\Gamma}_k$. First, we update $\boldsymbol{\Gamma}_k$ using the Metropolis-Hasting (MH) algorithm as follows. For $k=1,\ldots,K_C$,
\begin{enumerate}
    \item Draw a proposal $\boldsymbol{\Gamma}^{*}_k \sim \mathrm{MVN}(\boldsymbol{\Gamma}^{prev}_k, I_{d_v})$, where $\boldsymbol{\Gamma}^{*}_k$ is the proposal, $\boldsymbol{\Gamma}^{prev}_k$ is the $\boldsymbol{\Gamma}_k$ in the previous iteration, and $d_v$ is the dimension of $\mathbf{v}$.
    \item Accept $\boldsymbol{\Gamma}^{*}_k$ with a probability:
    \begin{align*}
        \frac{\mathrm{MVN}(\boldsymbol{\Gamma}^{*}_k \mid \boldsymbol{\mu}_{\boldsymbol{\Gamma}}, \boldsymbol{\Sigma}_{\boldsymbol{\Gamma}}) \prod_{i:\zeta_{i}=k}K^{*}(\mathbf{v}; \boldsymbol{\Gamma}_k^{*}) \prod_{i:\zeta_{i}>k}  \left\{1-s^{*}_k K^{*}(\mathbf{v}; \boldsymbol{\Gamma}_k^{*}) \right\} }{\mathrm{MVN}(\boldsymbol{\Gamma}^{prev}_k \mid \boldsymbol{\mu}_{\boldsymbol{\Gamma}}, \boldsymbol{\Sigma}_{\boldsymbol{\Gamma}}) \prod_{i:\zeta_{i}=k}K^{*}(\mathbf{v}; \boldsymbol{\Gamma}_k^{prev}) \prod_{i:\zeta_{i}>k}  \left\{1-s^{*}_k K^{*}(\mathbf{v}; \boldsymbol{\Gamma}_k^{prev}) \right\}},
    \end{align*}
    where $K^{*}(\mathbf{v}; \boldsymbol{\Gamma}_k)$ is given in \eqref{eq:kernel_function}.
    If the probability is greater than $1$, accept the sample.
\end{enumerate}

Then, we update $s^{*}_k$ using the MH algorithm. For $k=1,\ldots,K_C$,
\begin{enumerate}
    \item Draw a proposal $s_{k, prop} \sim \mathrm{U}(0,1)$.
    \item Accept $s_{k, prop} $ with a probability:
    \begin{align*}
        \min\left(1, \frac{ s_{k, prop}^{n_k} (1-s_{k, prop})^{\alpha-1}\prod_{i:\zeta_{i}>k} (1-s_{k, prop}K^{*}(\mathbf{v}; \boldsymbol{\Gamma}_k)) }{s_{k, prev}^{n_k} (1-s_{k, prev})^{\alpha-1}\prod_{i:\zeta_{i}>k} (1-s_{k, prev}K^{*}(\mathbf{v}; \boldsymbol{\Gamma}_k)) } \right),
    \end{align*}
    where $n_k =\sum_{i=1}^{I}\mathbbm{1}(\zeta_{i}=k)$, $s_{k, prev}$ is the $s^{*}_{k}$ in the previous iteration and $K^{*}(\mathbf{v}; \boldsymbol{\Gamma}_k)$ is given in \eqref{eq:kernel_function}.
    \item Obtain $\pi_k^{*}(\mathbf{v}_i)$ from Equation \eqref{eq:weight_pi_dependent}.
\end{enumerate}

\section{Baseline simulation details}
\label{sec:simulation_details}

This section details the data-generating process for our simulation study, which involves hierarchical data with clusters and individuals, covariates, treatments, mediators, and outcomes. 
We explain 3 scenarios for the baseline simulation (Section \ref{sec:baseline_simulation}) in detail. Section \ref{sec:DGP_cluster_var} and \ref{sec:DGP_mediator} are common across all scenarios.

\subsection{Cluster-level and individual-level variables}
\label{sec:DGP_cluster_var}

We consider a total of $K=100$ clusters (or groups), indexed by $i=1,2,\dots,K$. For each cluster $i$, we draw the cluster size $N_i \sim \text{DiscreteUniform}(20, 60)$, the cluster-level covariate $V_i \sim \mathrm{N}\left( \dfrac{3N_i}{50},\, 1 \right)$, and the cluster-level treatment $A_i \sim \text{Bernoulli}(0.5).$
Within each cluster $i$, there are $N_i$ individuals, indexed by $j=1,2,\dots,N_i$. For each individual $(i,j)$, we draw $X_{1_{ij}} \sim \mathrm{N}\left( -V_i,\, 2.0^2 \right), X_{2_{ij}} \sim \mathrm{N}\left( 0,\, 1.0^2 \right).$

\subsection{Mediators}
\label{sec:DGP_mediator}

We consider two mediators, $M^{(1)}$ and $M^{(2)}$, for each individual. We consider a scenario where $M^{(1)}$ and $M^{(2)}$ are correlated within the same units, and the same type of mediators are correlated between units within the same cluster as well.

For each individual $(i,j)$, we calculate the mediator mean parameters based on cluster-level and individual-level variables:
\begin{align*}
    \theta_{M^{(1)}_{ij}}(A_i) &= 1.5 \left\{ -2 + 2A_i + \left(0.5 + 0.5A_i\right) \dfrac{N_i}{50} + 0.5X_{1_{ij}} - 0.5X_{2_{ij}} + 0.5V_i \right\}, \\
    \theta_{M^{(2)}_{ij}}(A_i) &= -\theta_{M^{(1)}_{ij}}(A_i).
\end{align*}

Additionally, as discussed in Section \ref{sec:implication_assumption5}, we consider the following correlation structure between mediators for units $j,k$.
\begin{align*}
    \begin{pmatrix}
      M_{ij}^{(1)}(1) \\
      M_{ij}^{(1)}(0) \\
      M_{ij}^{(2)}(1) \\
      M_{ij}^{(2)}(0) \\
      M_{ik}^{(1)}(1) \\
      M_{ik}^{(1)}(0) \\
      M_{ik}^{(2)}(1) \\
      M_{ik}^{(2)}(0) \\
    \end{pmatrix}
    \sim \mathrm{MVN}\!\left( 
    \begin{pmatrix}
      \theta_{M^{(1)}_{ij}}(1) \\
      \theta_{M^{(1)}_{ij}}(0) \\
      \theta_{M^{(2)}_{ij}}(1) \\
      \theta_{M^{(2)}_{ij}}(0) \\
      \theta_{M^{(1)}_{ik}}(1) \\
      \theta_{M^{(1)}_{ik}}(0) \\
      \theta_{M^{(2)}_{ik}}(1) \\
      \theta_{M^{(2)}_{ik}}(0) \\
    \end{pmatrix},  
    \sigma^2 \begin{pmatrix}
      R & M  \\
      M & R
    \end{pmatrix} \right),
\end{align*}
where the correlation matrices are defined as
\[
    R \;=\; \begin{pmatrix}
      1 & \alpha_1 & \alpha_0 & \alpha_2 \\
      \alpha_1 & 1 & \alpha_2 & \alpha_0 \\
      \alpha_0 & \alpha_2 & 1 & \alpha_1 \\
      \alpha_2 & \alpha_0 & \alpha_1 & 1
    \end{pmatrix}, 
    \quad
    M \;=\; \begin{pmatrix}
      \rho_0 & 0 & \rho_1 & 0 \\
      0 & \rho_0 & 0 & \rho_1 \\
      \rho_1 & 0 & \rho_0 & 0 \\
      0 & \rho_1 & 0 & \rho_0 \\
    \end{pmatrix}.
\]
We let $\sigma^2=1.0$, $\alpha_0=\alpha_2=0.05$, $\alpha_1=0.08$,  $\rho_0=0.1$ and $\rho_1=0.0$. These parameters are selected to satisfy Assumption \ref{asmp:cond_homogeneity} (see Section \ref{sec:implication_assumption5} for details).

\subsection{Outcome variable}
\label{sec:GDP_outcome}
For each individual $(i,j)$, the outcome $Y_{ij}$ is generated based on a function of treatments, mediators, covariates, and random effects.

\subsubsection{Latent class of cluster}
\begin{itemize}
    \item \textbf{Scenario 1}: No latent class of cluster.
    \item \textbf{Scenario 2}: Each cluster is assigned to one of three groups: $G_{ij} \sim \text{Categorical}(0.2,\, 0.3,\, 0.5),$
    where the probabilities correspond to groups 1, 2, and 3, respectively.
    \item \textbf{Scenario 3}: $G_{ij} \sim \text{Categorical}(0.2 + 0.01  N_i,\, 0.3 - 0.005 N_i,\, 0.5 - 0.005 N_i),$ where the assignment probability depends on the cluster size $N_i$.
\end{itemize}
\subsubsection{Outcome parameters}

We first compute the following parameters:
\begin{align*}
    \theta_{1_{ij}} &= 1.0 + A_i + \left(0.5 + 0.5A_i\right) \dfrac{N_i}{50} + 0.5\overline{M^{(1)}_i} - 0.5\overline{M^{(2)}_i} + M^{(1)}_{ij} - M^{(2)}_{ij} + 0.5X_{1_{ij}} - 0.5X_{2_{ij}} + 0.5V_i, \\
    \theta_{2_{ij}} &= -1.0 - A_i - \left(0.5 + 0.5A_i\right) \dfrac{N_i}{50} - 0.5\overline{M^{(1)}_i} + 0.5\overline{M^{(2)}_i} - M^{(1)}_{ij} + M^{(2)}_{ij} - 0.5X_{1_{ij}} - 0.5X_{2_{ij}} + 0.5V_i, \\
    \theta_{3_{ij}} &= 1.0 + A_i + \left(0.3 + 0.3A_i\right) \dfrac{N_i}{50} + 0.3\overline{M^{(1)}_i} - 0.3\overline{M^{(2)}_i} + M^{(1)}_{ij} - M^{(2)}_{ij} + 0.3X_{1_{ij}} - 0.3X_{2_{ij}} + 0.3V_i, \\
    \theta_{4_{ij}} &= -1.0 - A_i - \left(0.3 + 0.3A_i\right) \dfrac{N_i}{50} - 0.3\overline{M^{(1)}_i} + 0.3\overline{M^{(2)}_i} - M^{(1)}_{ij} + M^{(2)}_{ij} - 0.3X_{1_{ij}} - 0.3X_{2_{ij}} + 0.3V_i, \\
    \theta_{5_{ij}} &= -1.5 \theta_{1_{ij}}, 
    \theta_{6_{ij}} = -1.5\theta_{2_{ij}}, 
    \theta_{7_{ij}} = -1.5 \theta_{3_{ij}}, 
    \theta_{8_{ij}} = -1.5 \theta_{4_{ij}},
\end{align*}
where $\overline{M^{(1)}_i}$ and $\overline{M^{(2)}_i}$ are the cluster-level means of the mediators:
\[
\overline{M^{(1)}_i} = \dfrac{1}{N_i} \sum_{j=1}^{N_i} M^{(1)}_{ij}, \quad \overline{M^{(2)}_i} = \dfrac{1}{N_i} \sum_{j=1}^{N_i} M^{(2)}_{ij}.
\]

\subsubsection{Outcome generation}
\begin{itemize}
    \item \textbf{Scenario 1}: $Y_{ij} \sim \mathrm{N}(\theta_{1_{ij}}+b_i, 1.0)$, where $b_i \sim \mathrm{N}(0,1.0)$.
    \item \textbf{Scenario 2, 3}: The outcome $Y_{ij}$ is then generated from a mixture distribution assigned to the group assignment variable $G_i$.
    \begin{itemize}
        \item If $G_i=1$, $Y_{ij} \sim 0.5\mathrm{N}( \theta_{1_{ij}},\, 2.0^2 ) + 0.5  \mathrm{N}( \theta_{2_{ij}},\, 1.0^2 ).$
        \item If $G_i=2$, $Y_{ij} \sim 0.5\mathrm{N}( \theta_{3_{ij}},\, 0.5^2 ) + 0.25  \mathrm{N}( \theta_{4_{ij}},\, 2.0^2 ) + 0.25  \mathrm{N}( \theta_{5_{ij}},\, 1.5^2 ).$
        \item If $G_i=3$, $Y_{ij} \sim 0.5\mathrm{N}( \theta_{6_{ij}},\, 1.5^2 ) + 0.25  \mathrm{N}( \theta_{7_{ij}},\, 1.0^2 ) + 0.25  \mathrm{N}( \theta_{8_{ij}},\, 2.0^2 ).$
    \end{itemize}
\end{itemize}

\subsection{Estimands}
Under these simulation setups, computing the true causal estimands in closed form is not straightforward. Therefore, we approximate the true values of the causal estimands using a Monte Carlo simulation approach by generating and averaging the potential outcomes for a vast number of individuals, increasing the number of clusters to $3,000,000$. This number of clusters is chosen because it yields consistent values for all estimands across multiple runs of the approximation. The potential outcomes for an individual in a cluster are generated by changing the values of $A_i$, generating two mediators for all individuals within the cluster based on the mediator generation process detailed in Section \ref{sec:DGP_mediator}, and generating the outcome based on the outcome generation process detailed in Section \ref{sec:GDP_outcome}.

\section{Addtional simulations}
\label{sec:additonal_simulation}

\subsection{Simulation 2: Non-Gaussian errors}
\label{sec:simulationn_error_term}

\begin{figure*}
    \centering
    \includegraphics[width=\textwidth]{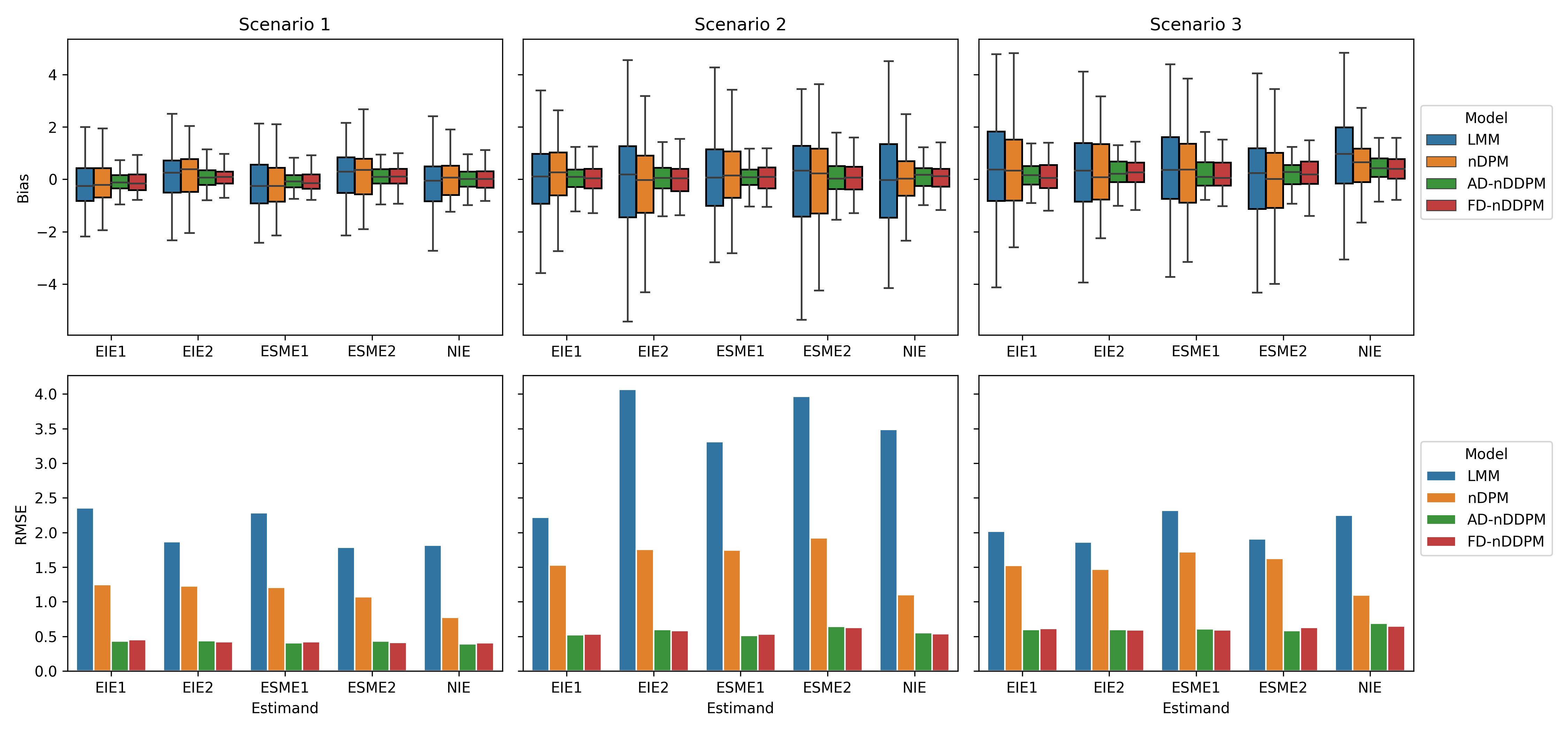}
    \caption{Boxplots of bias and bar charts of root mean squared error (RMSE) for point estimates of the key estimands across three scenarios in Simulation 2. The bias boxplots summarize the distribution over 100 simulation replicates.}
    \label{fig:simulation2_bias_mse_boxplot}
\end{figure*}

We explore the robustness of our models in more challenging yet relevant settings where standard assumptions in CRTs do not hold. First, we consider scenarios where the error terms of the data-generating process are heavy-tailed. Specifically, we replace the Gaussian error terms of the data-generating process with the Student's $t$-distribution with degrees of freedom $\nu=1.5$ in all three scenarios considered in Section \ref{sec:baseline_simulation}.  In Scenarios 2 and 3, the error terms of each component of the mixture distribution are replaced with the $t$-distribution. We fit the same models 
as in Section \ref{sec:baseline_simulation}.

Table \ref{tab:misspecification_error_term} presents the results for all scenarios. The LMM exhibits the worst performance in all metrics across all scenarios, due to its stringent parametric assumption. The nDP model improves upon the performance of the LMM, by successfully capturing the non-Gaussian errors via the use of the nDP for the random effects. Notably, however, our methods demonstrate markedly better performance in all metrics across all scenarios compared to the other methods, highlighting its robustness to heavy-tailed error distributions. 

\subsection{Simulation 3: Non-linear fixed effects}
\label{sec:simulation_nonlinear}

\begin{figure*}
    \centering
    \includegraphics[width=\textwidth]{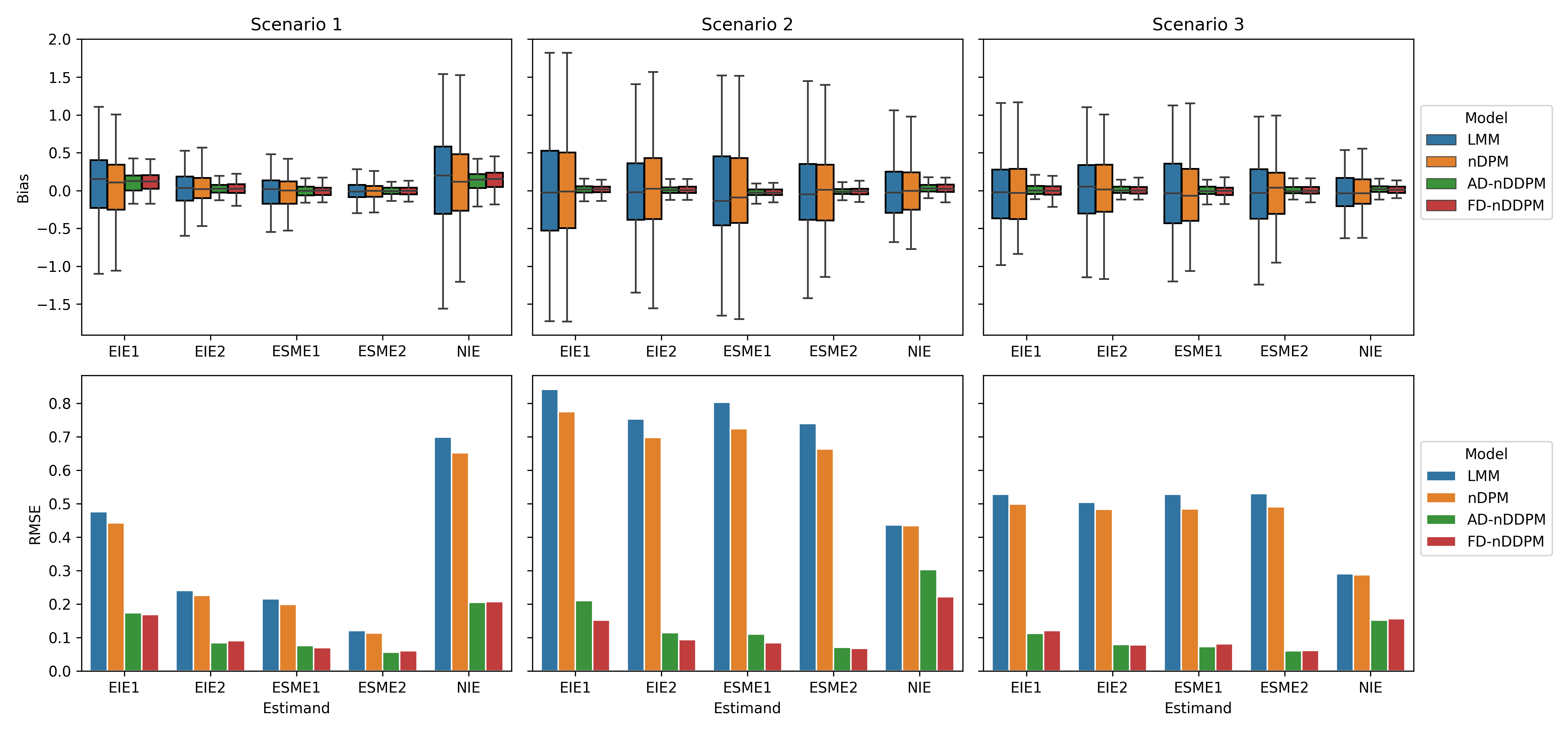}
    \caption{Boxplots of bias and bar charts of root mean squared error (RMSE) for point estimates of the key estimands across three scenarios in Simulation 3. The bias boxplots summarize the distribution over 100 simulation replicates.}
    \label{fig:simulation3_bias_mse_boxplot}
\end{figure*}

Next, we investigate the robustness of our models when the functional form of the fixed effects is misspecified. Specifically, we consider a modified version of the original three scenarios, where non-linear fixed effects---such as second-order terms of covariates and interaction terms---are built into the data-generating processes. These non-linear terms are not directly accounted for in the model specifications; that is, the specifications of the mean function do not involve these non-linear terms.

We consider the same data-generating process detailed in Section \ref{sec:baseline_simulation}, except that the location parameters for the mediator and outcome models are replaced with parameters that include nonlinear higher-order terms and interaction terms. 
The location parameters for mediators are specified as follows:
\begin{align*}
\theta_{M^{(1)}_{ij}} &= -1.0 + A_i + \left(0.5 + 0.5 A_i\right) \dfrac{N_i}{50} + X_{1_{ij}} - X_{2_{ij}} + X_{1_{ij}}^2 + X_{2_{ij}}^2 + X_{1_{ij}} X_{2_{ij}} + 0.5 V_i, \\[1ex]
\theta_{M^{(2)}_{ij}} &= -0.5\, \theta_{M^{(1)}_{ij}}.
\end{align*}

\noindent
The location parameters for outcomes are specified as follows: 
\begin{align*}
\theta_{1_{ij}} &= 1.0 + A_i + \left(0.5 + 0.5 A_i\right) \dfrac{N_i}{50} + \dfrac{0.5}{N_i} \sum_{k=1}^{N_i} M^{(1)}_{ik} - \dfrac{0.5}{N_i} \sum_{k=1}^{N_i} M^{(2)}_{ik} \\
&\quad + 0.5 M^{(1)}_{ij} - 0.5 M^{(2)}_{ij} + 0.3 X_{1_{ij}} A_i - 0.3 X_{2_{ij}} A_i + 0.1 X_{1_{ij}}^2 + 0.1 X_{2_{ij}}^2 + 0.1 X_{1_{ij}} X_{2_{ij}} + 0.5 V_i, \\[1ex]
\theta_{2_{ij}} &= -1.0 - A_i - \left(0.5 + 0.5 A_i\right) \dfrac{N_i}{50} - \dfrac{0.5}{N_i} \sum_{k=1}^{N_i} M^{(1)}_{ik} + \dfrac{0.5}{N_i} \sum_{k=1}^{N_i} M^{(2)}_{ik} \\
&\quad - 0.5 M^{(1)}_{ij} + 0.5 M^{(2)}_{ij} - 0.3 X_{1_{ij}} A_i + 0.3 X_{2_{ij}} A_i - 0.1 X_{1_{ij}}^2 - 0.1 X_{2_{ij}}^2 - 0.1 X_{1_{ij}} X_{2_{ij}} + 0.5 V_i, \\[1ex]
\theta_{3_{ij}} &= 1.0 + A_i + \left(0.3 + 0.3 A_i\right) \dfrac{N_i}{50} + \dfrac{0.3}{N_i} \sum_{k=1}^{N_i} M^{(1)}_{ik} - \dfrac{0.3}{N_i} \sum_{k=1}^{N_i} M^{(2)}_{ik} \\
&\quad + 0.3 M^{(1)}_{ij} - 0.3 M^{(2)}_{ij} + 0.1 X_{1_{ij}} A_i - 0.1 X_{2_{ij}} A_i + 0.1 X_{1_{ij}}^2 + 0.1 X_{2_{ij}}^2 + 0.1 X_{1_{ij}} X_{2_{ij}} + 0.3 V_i, \\[1ex]
\theta_{4_{ij}} &= -1.0 - A_i - \left(0.3 + 0.3 A_i\right) \dfrac{N_i}{50} - \dfrac{0.3}{N_i} \sum_{k=1}^{N_i} M^{(1)}_{ik} + \dfrac{0.3}{N_i} \sum_{k=1}^{N_i} M^{(2)}_{ik} \\
&\quad - 0.3 M^{(1)}_{ij} + 0.3 M^{(2)}_{ij} - 0.1 X_{1_{ij}} A_i + 0.1 X_{2_{ij}} A_i - 0.1 X_{1_{ij}}^2 - 0.1 X_{2_{ij}}^2 - 0.1 X_{1_{ij}} X_{2_{ij}} + 0.3 V_i, \\[1ex]
\theta_{5_{ij}} &= -1.5 \theta_{1_{ij}}, 
\theta_{6_{ij}} = -1.5 \theta_{2_{ij}}, 
\theta_{7_{ij}} = -1.5 \theta_{3_{ij}}, 
\theta_{8_{ij}} = -1.5 \theta_{4_{ij}}.
\end{align*}

Table \ref{tab:simulation3_functional_form} presents the results for all scenarios. The LMM exhibits poor performance across all metrics and scenarios, with significant biases and lower coverage probabilities. Unlike previous simulations, the nDP model does not improve upon the performance of the LMM in this setting. This is because the flexible specification of the random effects via the nDP does not address the misspecification of the functional form of the fixed effects. In contrast, our proposed methods demonstrate substantially better performance across all metrics and scenarios. Despite the inclusion of nonlinear fixed effects, our methods maintain the lowest bias and MSE, smallest interval lengths, and achieve close to nominal frequentist coverage probabilities.

\subsection{Simulation 4: Dichotomous mediators}
\label{sec:simulation_binary}

\begin{figure*}
    \centering
    \includegraphics[width=\textwidth]{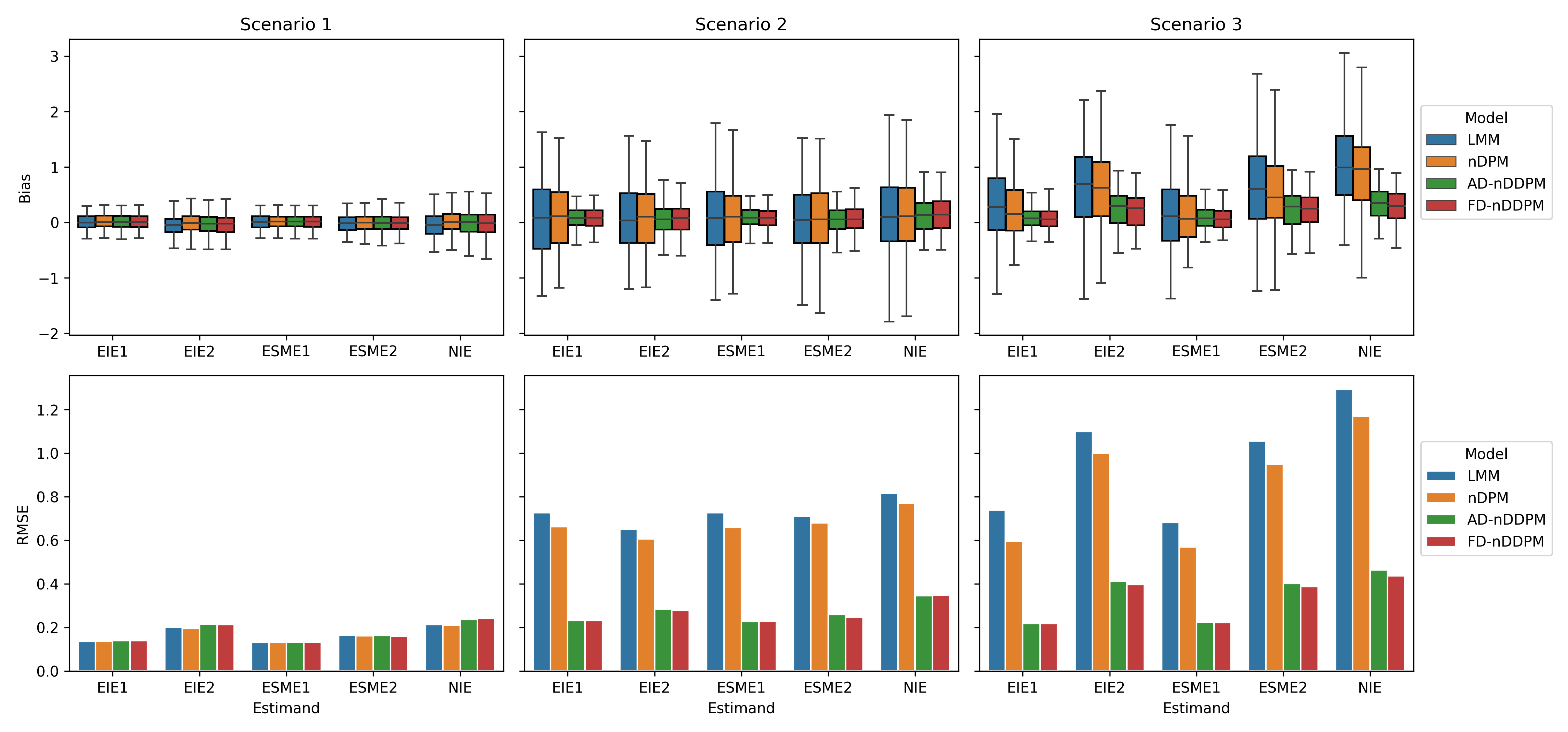}
    \caption{Boxplots of bias and bar charts of root mean squared error (RMSE) for point estimates of the key estimands across three scenarios in Simulation 4. The bias boxplots summarize the distribution over 100 simulation replicates.}
    \label{fig:simulation4_bias_mse_boxplot}
\end{figure*}

This simulation scenario builds on the baseline simulation, except that the first mediator $M^{(1)}_{ij}$  is transformed into a binary mediator using the logit transformation. For inference, we adopt the probit data augmentation as described in \ref{sec:extension_discrete_gibbs}. Figure \ref{fig:simulation4_bias_mse_boxplot} shows the bias and RMSE for the simulation with dichotomous mediators, while Table \ref{tab:simulation4_binary_mediator} provides the complete simulation results. Under this scenario, we also observe significant improvements in bias and RMSE for our methodologies compared to the LMM and nDPM.

\subsection{Simulation 5: Fewer clusters}
\label{sec:simulation_I30}
\begin{figure*}
    \centering
    \includegraphics[width=\textwidth]{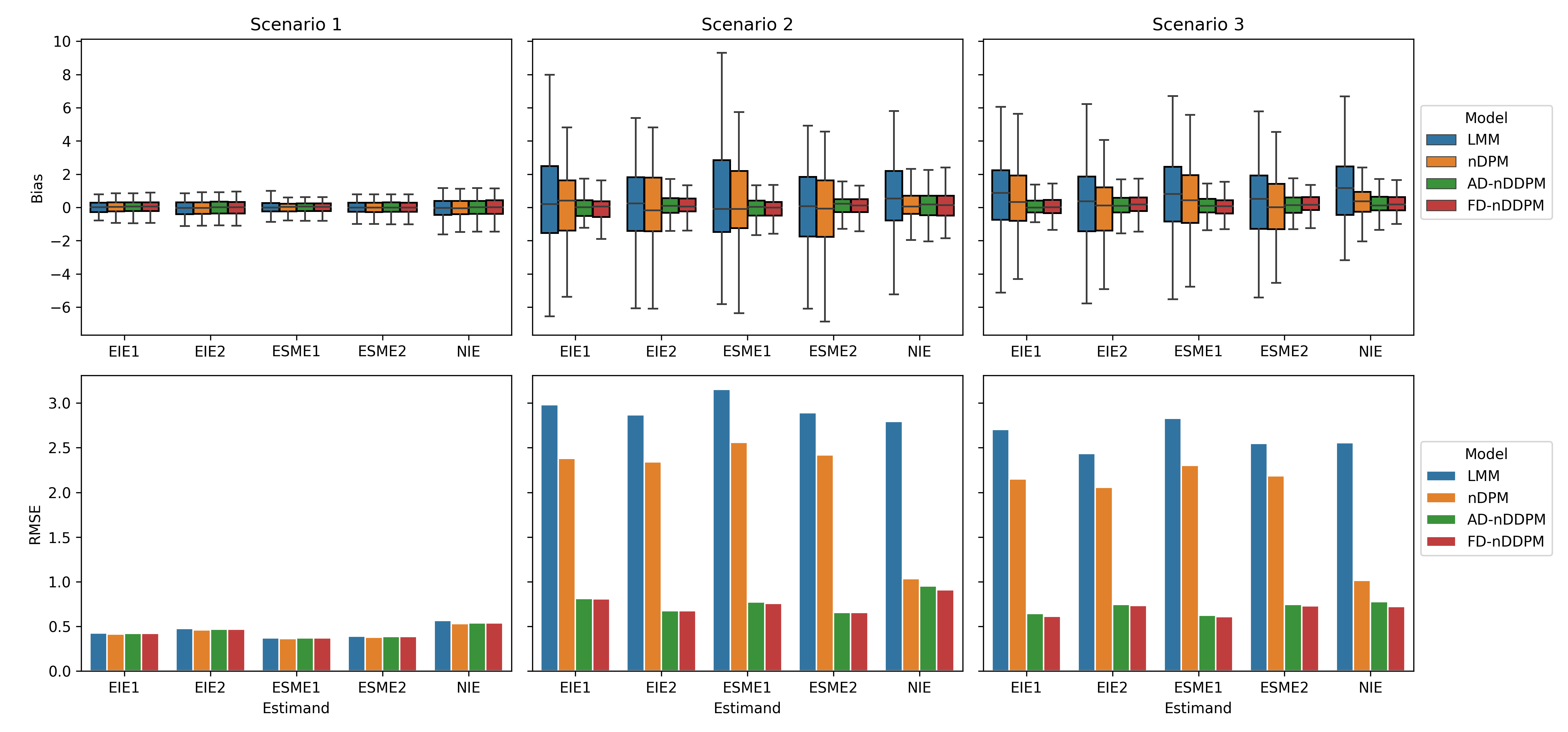}
    \caption{Boxplots of bias and bar charts of root mean squared error (RMSE) for point estimates of the key estimands across three scenarios in Simulation 5. The bias boxplots summarize the distribution over 100 simulation replicates.}
    \label{fig:simulation5_bias_mse_boxplot}
\end{figure*}

Figure \ref{fig:simulation5_bias_mse_boxplot} and Table \ref{tab:baseline_I30} provide additional simulation results with a smaller number of clusters $I=30$. Overall, we observe the same trends as those seen with $I=100$ in the main manuscript. These results reinforce the superiority of our methodology even under a smaller number of clusters.

\subsection{Numerical results}
\label{sec:results_tables}
Tables \ref{tab:baseline}-\ref{tab:baseline_I30} present the numerical results of the simulation studies. We report several performance metrics for both point and interval estimators. Specifically, for point estimators, we provide the bias and root mean squared error (RMSE), while for interval estimators, we report the coverage probability (CP) and the average relative interval length (RAL), where the interval length is scaled by the true value of the causal estimand:
$
{ N_{\mathrm{sim}}}^{-1} \sum_{n=1}^{N_{\mathrm{sim}}} {(\hat{\tau}_n^{u} - \hat{\tau}_n^{l})}/{\tau},
$
where \(N_{\mathrm{sim}}\) is the number of simulations, \(\tau\) denotes the true causal estimand, and \(\hat{\tau}_n^{l}\) and \(\hat{\tau}_n^{u}\) represent the estimates of the lower and upper bounds of the 95\% credible interval in the $n$-th simulation, respectively. By scaling the interval widths relative to the true parameter, this metric likely provides a clearer sense of uncertainty in relation to the actual magnitude of the causal estimand.

\begin{table*}
    \centering
    \caption{Bias and root mean squared error (RMSE) of point estimates and relative average length (RAL) and coverage probability (CP) of 95\% confidence/credible intervals of the key estimands under three scenarios in Simulation 1 (baseline scenarios).}
    \begin{adjustbox}{width=\textwidth}
        \begin{tabular}{l l rrrr rrrr rrrr}
            \toprule
            & & \multicolumn{4}{c}{Scenario 1} & \multicolumn{4}{c}{Scenario 2} & \multicolumn{4}{c}{Scenario 3} \\
            \cmidrule(lr){3-6} \cmidrule(lr){7-10} \cmidrule(lr){11-14}
            \textbf{Model} & \textbf{Estimand} & Bias & RMSE & RAL & CP & Bias & RMSE & RAL & CP & Bias & RMSE & RAL & CP \\
            \midrule
            \multirow{5}{*}{LMM}
             & $\mathrm{EIE}^{(1)}$   & -0.0608  & 0.26429  & 0.1472  & 89.0\% & -0.0865  & 1.40766  & 2.8605  & 94.0\% & 0.8778  & 1.58233  & 6.2339  & 87.0\% \\
             & $\mathrm{EIE}^{(2)}$   & -0.0389  & 0.25607  & 0.1481  & 87.0\% & 0.1685   & 1.44619  & 2.8737  & 93.0\% & 0.2831  & 1.29529  & 6.2694  & 93.0\% \\
             & $\mathrm{ESME}^{(1)}$  & -0.0274  & 0.21970  & 0.4286  & 91.0\% & -0.0441  & 1.37925  & 9.5559  & 94.0\% & 0.8185  & 1.48850  & 17.9291 & 92.0\% \\
             & $\mathrm{ESME}^{(2)}$  &  0.0005  & 0.21029  & 0.4260  & 93.0\% & 0.1475   & 1.50406  & 9.6431  & 91.0\% & 0.2361  & 1.34131  & 18.1468 & 92.0\% \\
             & $\mathrm{NIE}$         & -0.0996  & 0.34419  & 0.0838  & 86.0\% & 0.0804   & 1.32871  & 1.4678  & 95.0\% & 1.1612  & 1.80045  & 3.2582  & 84.0\% \\
            \midrule
            \multirow{5}{*}{nDPM}
             & $\mathrm{EIE}^{(1)}$   & -0.0088  & 0.25681  & 0.1564  & 90.0\% & -0.0782  & 1.30500  & 2.7877  & 95.0\% & 0.6899  & 1.36767  & 6.0022  & 91.0\% \\
             & $\mathrm{EIE}^{(2)}$   &  0.0076  & 0.25611  & 0.1560  & 91.0\% & 0.1998   & 1.33387  & 2.7825  & 93.0\% & 0.0935  & 1.17368  & 6.0057  & 95.0\% \\
             & $\mathrm{ESME}^{(1)}$  & -0.0091  & 0.21822  & 0.4383  & 90.0\% & -0.0543  & 1.29467  & 9.5025  & 97.0\% & 0.6177  & 1.28232  & 17.7058 & 94.0\% \\
             & $\mathrm{ESME}^{(2)}$  &  0.0153  & 0.21062  & 0.4354  & 94.0\% & 0.1610   & 1.41802  & 9.5045  & 93.0\% & 0.0251  & 1.26383  & 17.7633 & 96.0\% \\
             & $\mathrm{NIE}$         & -0.0012  & 0.32962  & 0.0826  & 86.0\% & 0.1194   & 0.91557  & 1.2066  & 99.0\% & 0.7831  & 1.20934  & 2.6819  & 92.0\% \\
            \midrule
            \multirow{5}{*}{AD-nDDPM}
             & $\mathrm{EIE}^{(1)}$   & -0.0054  & 0.25828  & 0.1571  & 90.0\% & 0.1190   & 0.37531  & 0.6752  & 90.0\% & 0.0943  & 0.33950  & 1.4551  & 90.0\% \\
             & $\mathrm{EIE}^{(2)}$   &  0.0112  & 0.25768  & 0.1569  & 91.0\% & 0.1173   & 0.35429  & 0.6739  & 91.0\% & 0.1795  & 0.36775  & 1.4471  & 89.0\% \\
             & $\mathrm{ESME}^{(1)}$  & -0.0076  & 0.21888  & 0.4403  & 92.0\% & 0.1055   & 0.35595  & 2.2477  & 95.0\% & 0.1055  & 0.31892  & 4.1596  & 94.0\% \\
             & $\mathrm{ESME}^{(2)}$  &  0.0167  & 0.21145  & 0.4394  & 94.0\% & 0.0835   & 0.31126  & 2.2353  & 98.0\% & 0.1837  & 0.34180  & 4.1496  & 93.0\% \\
             & $\mathrm{NIE}$         &  0.0060  & 0.33336  & 0.0750  & 81.0\% & 0.2346   & 0.51075  & 0.3535  & 78.0\% & 0.2743  & 0.50414  & 0.7714  & 75.0\% \\
            \midrule
            \multirow{5}{*}{FD-nDDPM}
             & $\mathrm{EIE}^{(1)}$   & -0.0033  & 0.25811  & 0.1570  & 89.0\% & 0.1109   & 0.37117  & 0.6754  & 90.0\% & 0.0980  & 0.33966  & 1.4934  & 91.0\% \\
             & $\mathrm{EIE}^{(2)}$   &  0.0107  & 0.25760  & 0.1559  & 91.0\% & 0.1011   & 0.36746  & 0.6695  & 88.0\% & 0.1721  & 0.37212  & 1.4778  & 87.0\% \\
             & $\mathrm{ESME}^{(1)}$  & -0.0068  & 0.21881  & 0.4387  & 92.0\% & 0.1118   & 0.35060  & 2.2370  & 93.0\% & 0.1183  & 0.32929  & 4.2680  & 95.0\% \\
             & $\mathrm{ESME}^{(2)}$  &  0.0163  & 0.21128  & 0.4386  & 94.0\% & 0.0728   & 0.31477  & 2.2314  & 97.0\% & 0.1798  & 0.34774  & 4.2566  & 91.0\% \\
             & $\mathrm{NIE}$         &  0.0075  & 0.33011  & 0.0746  & 82.0\% & 0.2101   & 0.49654  & 0.3497  & 82.0\% & 0.2703  & 0.49241  & 0.7848  & 79.0\% \\
            \bottomrule
        \end{tabular}
    \end{adjustbox}
    \label{tab:baseline}
\end{table*}

\begin{table*}
    \centering
    \caption{Evaluation metrics under three scenarios in Simulation 2 with error terms following the Student's t-distribution with degrees of freedom $\nu =1.5$}
    \begin{adjustbox}{width=\textwidth}
        \begin{tabular}{l l rrrr rrrr rrrr}
            \toprule
            & & \multicolumn{4}{c}{Scenario 1} & \multicolumn{4}{c}{Scenario 2} & \multicolumn{4}{c}{Scenario 3} \\
            \cmidrule(lr){3-6} \cmidrule(lr){7-10} \cmidrule(lr){11-14}
            \textbf{Model} & \textbf{Estimand} & Bias & RMSE & RAL & CP & Bias & RMSE & RAL & CP & Bias & RMSE & RAL & CP \\
            \midrule
            \multirow{5}{*}{LMM}
             & $\mathrm{EIE}^{(1)}$ & 0.1886 & 2.35722 & 0.8745 & 89.0\% & -0.0599 & 2.21769 & 4.2663 & 97.0\% & 0.4767 & 2.01633 & 8.6900 & 93.0\% \\
             & $\mathrm{EIE}^{(2)}$ & -0.0195 & 1.86593 & 0.8380 & 93.0\% & -0.0146 & 4.06578 & 4.6938 & 93.0\% & 0.3278 & 1.86085 & 8.5045 & 96.0\% \\
             & $\mathrm{ESME}^{(1)}$ & 0.1604 & 2.28409 & 2.7372 & 89.0\% & -0.0274 & 3.31175 & 15.1520 & 95.0\% & 0.4515 & 2.31940 & 26.0464 & 93.0\% \\
             & $\mathrm{ESME}^{(2)}$ & 0.0054  & 1.78773 & 2.5527 & 95.0\% & 0.1259  & 3.96711 & 15.3254 & 94.0\% & 0.1196 & 1.90574 & 24.6032 & 94.0\% \\
             & $\mathrm{NIE}$        & 0.1690   & 1.81588 & 0.3903 & 93.0\% & -0.0761 & 3.48543 & 2.3863 & 95.0\% & 0.8049 & 2.24888 & 4.4871  & 90.0\% \\
            \midrule
            \multirow{5}{*}{nDPM}
             & $\mathrm{EIE}^{(1)}$ & -0.0426 & 1.24678 & 0.8588 & 98.0\% & 0.1430   & 1.52875 & 4.1103 & 97.0\% & 0.3456 & 1.52405 & 8.4596 & 97.0\% \\
             & $\mathrm{EIE}^{(2)}$ & 0.1085  & 1.23009 & 0.8528 & 98.0\% & -0.0437 & 1.75427 & 4.1489 & 96.0\% & 0.1945 & 1.46967 & 8.4792 & 98.0\% \\
             & $\mathrm{ESME}^{(1)}$ & -0.0971 & 1.21042 & 2.6855 & 94.0\% & 0.2021  & 1.74537 & 14.1045 & 96.0\% & 0.2898 & 1.72117 & 25.1012 & 94.0\% \\
             & $\mathrm{ESME}^{(2)}$ & 0.1194  & 1.07367 & 2.6707 & 98.0\% & 0.0786  & 1.92257 & 14.1585 & 97.0\% & -0.0377& 1.62715 & 25.1770 & 98.0\% \\
             & $\mathrm{NIE}$        & 0.0658  & 0.77464 & 0.3327 & 100.0\%& 0.0987  & 1.10116 & 1.5315 & 98.0\% & 0.5376 & 1.09587 & 3.2852 & 99.0\% \\
            \midrule
            \multirow{5}{*}{AD-nDDPM}
             & $\mathrm{EIE}^{(1)}$ & -0.0940  & 0.43482 & 0.4405 & 96.0\% & 0.0662  & 0.52277 & 1.4558 & 97.0\% & 0.1921 & 0.60044 & 2.9951 & 96.0\% \\
             & $\mathrm{EIE}^{(2)}$ & 0.0947  & 0.43710 & 0.4389 & 93.0\% & 0.0709  & 0.59963 & 1.4834 & 96.0\% & 0.2553 & 0.59799 & 2.9970 & 94.0\% \\
             & $\mathrm{ESME}^{(1)}$ & -0.0721 & 0.40687 & 1.3283 & 98.0\% & 0.0856  & 0.51245 & 4.8219 & 98.0\% & 0.1900   & 0.61025 & 8.4806 & 95.0\% \\
             & $\mathrm{ESME}^{(2)}$ & 0.0939  & 0.43215 & 1.3133 & 95.0\% & 0.0838  & 0.64361 & 4.8461 & 96.0\% & 0.2042 & 0.58462 & 8.4467 & 98.0\% \\
             & $\mathrm{NIE}$        & -0.0001 & 0.39570 & 0.1946 & 96.0\% & 0.1357  & 0.55603 & 0.6792 & 96.0\% & 0.4481 & 0.69120 & 1.3706 & 83.0\% \\
            \midrule
            \multirow{5}{*}{FD-nDDPM}
             & $\mathrm{EIE}^{(1)}$ & -0.1123 & 0.45436 & 0.4618 & 97.0\% & 0.0514  & 0.53207 & 1.4717 & 99.0\% & 0.1223 & 0.61534 & 3.0924 & 95.0\% \\
             & $\mathrm{EIE}^{(2)}$ & 0.0977  & 0.42429 & 0.4592 & 95.0\% & 0.0291  & 0.58676 & 1.4740 & 99.0\% & 0.2862 & 0.59680 & 3.1441 & 96.0\% \\
             & $\mathrm{ESME}^{(1)}$ & -0.0974 & 0.42282 & 1.3777 & 98.0\% & 0.0803  & 0.53569 & 4.8561 & 100.0\%& 0.1283 & 0.59462 & 8.7798 & 94.0\% \\
             & $\mathrm{ESME}^{(2)}$ & 0.0994  & 0.41536 & 1.3683 & 96.0\% & 0.0933  & 0.63210 & 4.8829 & 97.0\% & 0.2546 & 0.62983 & 8.9172 & 94.0\% \\
             & $\mathrm{NIE}$        & -0.0138 & 0.40966 & 0.2075 & 96.0\% & 0.0775  & 0.54136 & 0.6900 & 97.0\% & 0.4092 & 0.65167 & 1.4383 & 89.0\% \\
            \bottomrule
        \end{tabular}
    \end{adjustbox}
    \label{tab:misspecification_error_term}
\end{table*}

\begin{table*}
    \centering
    \caption{Evaluation metrics under three scenarios in Simulation 3 with non-linear fixed effects.}
    \begin{adjustbox}{width=\textwidth}
        \begin{tabular}{l l rrrr rrrr rrrr}
            \toprule
            & & \multicolumn{4}{c}{Scenario 1} & \multicolumn{4}{c}{Scenario 2} & \multicolumn{4}{c}{Scenario 3} \\
            \cmidrule(lr){3-6} \cmidrule(lr){7-10} \cmidrule(lr){11-14}
            \textbf{Model} & \textbf{Estimand} & Bias & RMSE & RAL & CP & Bias & RMSE & RAL & CP & Bias & RMSE & RAL & CP \\
            \midrule
            \multirow{5}{*}{LMM} 
             & $\mathrm{EIE}^{(1)}$ & 0.0925 & 0.47538 & 0.7426 & 74.0\% & -0.0173 & 0.84118 & 6.6728 & 92.0\% & -0.0067 & 0.52764 & 12.0304 & 98.0\% \\
             & $\mathrm{EIE}^{(2)}$ & 0.0080  & 0.24064 & 0.7492 & 74.0\% & 0.0541  & 0.75300 & 12.6297 & 94.0\% & -0.0001 & 0.50350 & 22.9643 & 97.0\% \\
             & $\mathrm{ESME}^{(1)}$ & -0.0117 & 0.21557 & 0.7105 & 73.0\% & -0.0665 & 0.80316 & 13.8072 & 94.0\% & -0.0322 & 0.52845 & 25.6587 & 99.0\% \\
             & $\mathrm{ESME}^{(2)}$ & -0.0110 & 0.12042 & 0.7908 & 76.0\% & 0.0446  & 0.73904 & 24.9734 & 95.0\% & -0.0339 & 0.53048 & 49.6564 & 96.0\% \\
             & $\mathrm{NIE}$        & 0.1004  & 0.69826 & 0.7243 & 72.0\% & 0.0348  & 0.43675 & 1.9709 & 83.0\% & -0.0077 & 0.29014 & 3.4207  & 96.0\% \\
            \midrule
            \multirow{5}{*}{nDPM} 
             & $\mathrm{EIE}^{(1)}$ & 0.0949  & 0.44210 & 0.8950 & 86.0\% & -0.0054 & 0.77477 & 8.7528  & 99.0\% & -0.0015 & 0.49899 & 15.6754 & 100.0\% \\
             & $\mathrm{EIE}^{(2)}$ & 0.0229  & 0.22647 & 0.9067 & 83.0\% & 0.0521  & 0.69793 & 17.1982 & 100.0\% & 0.0116  & 0.48292 & 30.8610 & 100.0\% \\
             & $\mathrm{ESME}^{(1)}$ & -0.0117 & 0.19857 & 0.8534 & 85.0\% & -0.0614 & 0.72400 & 17.5372 & 99.0\% & -0.0503 & 0.48382 & 32.6994 & 100.0\% \\
             & $\mathrm{ESME}^{(2)}$ & -0.0043 & 0.11345 & 0.9662 & 84.0\% & 0.0386  & 0.66294 & 33.1052 & 100.0\% & -0.0308 & 0.49048 & 64.6144 & 100.0\% \\
             & $\mathrm{NIE}$        & 0.1177  & 0.65170 & 0.6633 & 71.0\% & 0.0442  & 0.43385 & 4.2520  & 100.0\% & 0.0078  & 0.28700 & 7.5871  & 100.0\% \\
            \midrule
            \multirow{5}{*}{AD-nDDPM} 
             & $\mathrm{EIE}^{(1)}$ & 0.1153  & 0.17367 & 0.2469 & 61.0\% & -0.0166 & 0.21012 & 0.8022  & 94.0\% & 0.0015  & 0.11221 & 1.2569  & 95.0\% \\
             & $\mathrm{EIE}^{(2)}$ & 0.0199  & 0.08462 & 0.3770 & 88.0\% & -0.0022 & 0.11415 & 1.5232  & 97.0\% & 0.0050  & 0.07912 & 2.3809  & 97.0\% \\
             & $\mathrm{ESME}^{(1)}$ & -0.0033 & 0.07596 & 0.3097 & 80.0\% & -0.0361 & 0.11072 & 1.3323  & 91.0\% & -0.0028 & 0.07266 & 2.3023  & 95.0\% \\
             & $\mathrm{ESME}^{(2)}$ & -0.0063 & 0.05657 & 0.5498 & 90.0\% & -0.0106 & 0.07113 & 2.5117  & 97.0\% & 0.0040  & 0.06099 & 4.5103  & 96.0\% \\
             & $\mathrm{NIE}$        & 0.1350  & 0.20560 & 0.1986 & 65.0\% & -0.0208 & 0.30343 & 0.4837  & 84.0\% & 0.0054  & 0.15202 & 0.7116  & 84.0\% \\
            \midrule
            \multirow{5}{*}{FD-nDDPM} 
             & $\mathrm{EIE}^{(1)}$ & 0.1207  & 0.16855 & 0.2774 & 65.0\% & -0.0021 & 0.15241 & 0.7967  & 97.0\% & -0.0121 & 0.12145 & 1.3379  & 91.0\% \\
             & $\mathrm{EIE}^{(2)}$ & 0.0231  & 0.09121 & 0.4286 & 84.0\% & 0.0514  & 0.09391 & 1.5482  & 97.0\% & 0.0007  & 0.07829 & 2.6307  & 95.0\% \\
             & $\mathrm{ESME}^{(1)}$ & -0.0004 & 0.06943 & 0.3379 & 86.0\% & -0.0264 & 0.08485 & 1.3054  & 96.0\% & -0.0113 & 0.08118 & 2.3790  & 93.0\% \\
             & $\mathrm{ESME}^{(2)}$ & -0.0048 & 0.06017 & 0.5875 & 87.0\% & -0.0117 &  0.06833 & 2.6283  & 98.0\% & 0.0023  & 0.06124 & 4.8812  & 97.0\% \\
             & $\mathrm{NIE}$        & 0.1437  & 0.20763 & 0.2276 & 62.0\% & -0.0021 & 0.22188 & 0.4938  & 83.0\% & -0.0123 & 0.15582 & 0.7992  & 87.0\% \\
            \bottomrule
        \end{tabular}
    \end{adjustbox}
    \label{tab:simulation3_functional_form}
\end{table*}

\begin{table*}
    \centering
    \caption{Evaluation metrics for Simulation 4 with dichotomous mediators.}
    \begin{adjustbox}{width=\textwidth}
        \begin{tabular}{l l rrrr rrrr rrrr}
            \toprule
            & & \multicolumn{4}{c}{Scenario 1} & \multicolumn{4}{c}{Scenario 2} & \multicolumn{4}{c}{Scenario 3} \\
            \cmidrule(lr){3-6} \cmidrule(lr){7-10} \cmidrule(lr){11-14}
            \textbf{Model} & \textbf{Estimand} & Bias & RMSE & RAL & CP & Bias & RMSE & RAL & CP & Bias & RMSE & RAL & CP \\
            \midrule
            \multirow{5}{*}{LMM} 
            & $\mathrm{EIE}^{(1)}$ & 0.0123 & 0.13616 & 0.8035 & 96.0\% & 0.0585 & 0.72675 & 10.4081 & 91.0\% & 0.3247 & 0.74007 & 25.3501 & 95.0\% \\
            & $\mathrm{EIE}^{(2)}$ & -0.0467 & 0.20204 & 0.1374 & 92.0\% & 0.0652 & 0.65185 & 1.7470  & 99.0\% & 0.6611 & 1.09925 & 3.9918  & 87.0\% \\
            & $\mathrm{ESME}^{(1)}$ & 0.0155 & 0.13164 & 2.4235 & 94.0\% & 0.0607 & 0.72646 & 34.2359 & 92.0\% & 0.1449 & 0.68209 & 75.4673 & 98.0\% \\
            & $\mathrm{ESME}^{(2)}$ & -0.0098 & 0.16459 & 0.3796 & 95.0\% & 0.0502 & 0.71051 & 5.7989  & 97.0\% & 0.0727 & 0.22197 & 21.3140 & 94.0\% \\
            & $\mathrm{NIE}$        & -0.0342 & 0.21232 & 0.1213 & 89.0\% & 0.1238 & 0.81633 & 1.6029  & 92.0\% & 0.9863 & 1.29292 & 3.7526  & 83.0\% \\
            \midrule
            \multirow{5}{*}{nDPM} 
            & $\mathrm{EIE}^{(1)}$ & 0.0208 & 0.13568 & 0.8339 & 98.0\% & 0.0615 & 0.66272 & 10.3416 & 98.0\% & 0.2469 & 0.59637 & 21.0769 & 94.0\% \\
            & $\mathrm{EIE}^{(2)}$ & -0.0017 & 0.19540 & 0.1455 & 94.0\% & 0.0796 & 0.60677 & 1.7221  & 99.0\% & 0.6192 & 0.99981 & 3.7955  & 89.0\% \\
            & $\mathrm{ESME}^{(1)}$ & 0.0213 & 0.13069 & 2.5092 & 97.0\% & 0.0623 & 0.65960 & 33.5848 & 97.0\% & 0.1627 & 0.57004 & 61.9968 & 95.0\% \\
            & $\mathrm{ESME}^{(2)}$ & 0.0040 & 0.16245 & 0.3905 & 95.0\% & 0.0482 & 0.68057 & 5.8308  & 99.0\% & 0.5028 & 0.94884 & 11.0469 & 91.0\% \\
            & $\mathrm{NIE}$        & 0.0193 & 0.21109 & 0.1277 & 92.0\% & 0.1413 & 0.76982 & 1.5754  & 95.0\% & 0.8668 & 1.16977 & 3.4707  & 83.0\% \\
            \midrule
            \multirow{5}{*}{AD-nDDPM} 
            & $\mathrm{EIE}^{(1)}$ & 0.0214 & 0.13900 & 0.8336 & 96.0\% & 0.0799 & 0.23240 & 3.5034  & 95.0\% & 0.0863 & 0.21739 & 7.2648  & 94.0\% \\
            & $\mathrm{EIE}^{(2)}$ & -0.0286 & 0.21403 & 0.1572 & 95.0\% & 0.0630 & 0.28534 & 0.6201  & 95.0\% & 0.2474 & 0.41261 & 1.3530  & 79.0\% \\
            & $\mathrm{ESME}^{(1)}$ & 0.0202 & 0.13285 & 2.5220 & 96.0\% & 0.0774 & 0.22742 & 11.3365 & 95.0\% & 0.0833 & 0.22358 & 21.3438 & 91.0\% \\
            & $\mathrm{ESME}^{(2)}$ & -0.0032 & 0.16398 & 0.3971 & 95.0\% & 0.0668 & 0.25911 & 1.9878  & 95.0\% & 0.2541 & 0.40117 & 3.8162  & 81.0\% \\
            & $\mathrm{NIE}$        & -0.0071 & 0.23666 & 0.1376 & 93.0\% & 0.1432 & 0.34627 & 0.5705  & 89.0\% & 0.3342 & 0.46460 & 1.2569  & 76.0\% \\
            \midrule
            \multirow{5}{*}{FD-nDDPM} 
            & $\mathrm{EIE}^{(1)}$ & 0.0204 & 0.14018 & 0.8314 & 97.0\% & 0.0806 & 0.23268 & 3.4965  & 97.0\% & 0.0747 & 0.21774 & 7.2794  & 93.0\% \\
            & $\mathrm{EIE}^{(2)}$ & -0.0327 & 0.21342 & 0.1630 & 93.0\% & 0.0700 & 0.27882 & 0.6126  & 95.0\% & 0.2198 & 0.39747 & 1.4768  & 84.0\% \\
            & $\mathrm{ESME}^{(1)}$ & 0.0205 & 0.13251 & 2.5141 & 96.0\% & 0.0777 & 0.22843 & 11.3327 & 93.0\% & 0.0727 & 0.22197 & 21.3140 & 94.0\% \\
            & $\mathrm{ESME}^{(2)}$ & -0.0053 & 0.15962 & 0.4050 & 96.0\% & 0.0690 & 0.24868 & 1.9894  & 96.0\% & 0.2379 & 0.38814 & 3.9986  & 85.0\% \\
            & $\mathrm{NIE}$        & -0.0121 & 0.24172 & 0.1426 & 92.0\% & 0.1507 & 0.34847 & 0.5632  & 89.0\% & 0.2950 & 0.43650 & 1.3649  & 77.0\% \\
            \bottomrule
        \end{tabular}
    \end{adjustbox}
    \label{tab:simulation4_binary_mediator}
\end{table*}

\begin{table*}
  \centering
  \caption{Evaluation metrics for Simulation 5 with a smaller number of clusters $I=30$.}
  \begin{adjustbox}{width=\textwidth}
  \begin{tabular}{l l rrrr rrrr rrrr}
    \toprule
    & & \multicolumn{4}{c}{Scenario 1} & \multicolumn{4}{c}{Scenario 2} & \multicolumn{4}{c}{Scenario 3} \\
    \cmidrule(lr){3-6} \cmidrule(lr){7-10} \cmidrule(lr){11-14}
    \textbf{Model} & \textbf{Estimand} & Bias & RMSE & RAL & CP & Bias & RMSE & RAL & CP & Bias & RMSE & RAL & CP \\
    \midrule
      & $\mathrm{EIE}^{(1)}$ & -0.0058 & 0.42757 & 0.2120 & 91.0\% & 0.0316 & 3.25613 & 6.5765  & 91.0\% & 0.8743 & 3.10120 & 14.4358 & 95.0\% \\
         & $\mathrm{EIE}^{(2)}$ & -0.0868 & 0.49059 & 0.2081 & 73.0\% & -0.0883 & 2.66963 & 6.5604  & 96.0\% & 0.6362 & 3.10251 & 14.2972 & 98.0\% \\
     LMM    & $\mathrm{ESME}^{(1)}$ &  0.0233 & 0.29275 & 0.6071 & 93.0\% & 0.0494 & 3.24522 & 21.3826 & 93.0\% & 0.7476 & 3.26733 & 40.6103 & 90.0\% \\
         & $\mathrm{ESME}^{(2)}$ & -0.0294 & 0.30617 & 0.5926 & 91.0\% & -0.1761 & 2.69842 & 21.4506 & 97.0\% & 0.4595 & 2.90014 & 40.5416 & 95.0\% \\
         & $\mathrm{NIE}$       & -0.0926 & 0.68128 & 0.1453 & 70.0\% & -0.0584 & 3.05335 & 2.9856 & 93.0\% & 1.5108 & 2.97078 & 6.6374 & 93.0\% \\
    \midrule
     & $\mathrm{EIE}^{(1)}$ &  0.0371 & 0.41469 & 0.2996 & 94.0\% & 0.1352 & 2.43880 & 5.7831 & 96.0\% & 0.3595 & 2.59098 & 12.6871 & 94.0\% \\
         & $\mathrm{EIE}^{(2)}$ & -0.0354 & 0.46263 & 0.2948 & 90.0\% & -0.0041 & 2.18977 & 5.7796 & 97.0\% & 0.1013 & 2.55343 & 12.7579 & 95.0\% \\
     nDPM   & $\mathrm{ESME}^{(1)}$ &  0.0434 & 0.36413 & 0.8424 & 95.0\% & 0.1016 & 2.61532 & 19.4215 & 98.0\% & 0.2798 & 2.77734 & 36.6592 & 91.0\% \\
         & $\mathrm{ESME}^{(2)}$ & -0.0099 & 0.38095 & 0.8266 & 94.0\% & 0.1293 & 2.25666 & 19.3928 & 98.0\% & -0.0342 & 2.43065 & 36.8818 & 97.0\% \\
         & $\mathrm{NIE}$       &  0.0017 & 0.53110 & 0.1555 & 88.0\% & 0.1306 & 1.17421 & 1.7356 & 98.0\% & 0.4615 & 1.11609 & 3.8924 & 98.0\% \\
    \midrule
     & $\mathrm{EIE}^{(1)}$ &  0.0489 & 0.42019 & 0.3020 & 95.0\% & -0.0409 & 0.74801 & 1.6522 & 95.0\% & 0.1023 & 0.65331 & 3.6859 & 98.0\% \\
             & $\mathrm{EIE}^{(2)}$ & -0.0190 & 0.46761 & 0.2985 & 92.0\% & 0.1056 & 0.79684 & 1.6394 & 94.0\% & 0.1551 & 0.69205 & 3.6834 & 98.0\% \\
         AD-nDDPM & $\mathrm{ESME}^{(1)}$ &  0.0464 & 0.37051 & 0.8464 & 95.0\% & -0.0343 & 0.67004 & 5.3261 & 98.0\% & 0.1312 & 0.67201 & 10.2851 & 98.0\% \\
             & $\mathrm{ESME}^{(2)}$ & -0.0045 & 0.38875 & 0.8350 & 94.0\% & 0.0954 & 0.74983 & 5.3129 & 94.0\% & 0.1436 & 0.67335 & 10.3048 & 98.0\% \\
             & $\mathrm{NIE}$       &  0.0299 & 0.53839 & 0.1425 & 81.0\% & 0.0631 & 0.82754 & 0.7754 & 94.0\% & 0.2578 & 0.70333 & 1.7399  & 95.0\% \\
    \midrule
     & $\mathrm{EIE}^{(1)}$ &  0.0522 & 0.42250 & 0.3045 & 95.0\% & 0.0811 & 0.74517 & 1.6227 & 97.0\% & 0.0723 & 0.69495 & 3.6635 & 96.0\% \\
             & $\mathrm{EIE}^{(2)}$ & -0.0217 & 0.46864 & 0.2990 & 89.0\% & 0.0385 & 0.78673 & 1.6220 & 95.0\% & 0.2032 & 0.68792 & 3.7713 & 96.0\% \\
         FD-nDDPM  & $\mathrm{ESME}^{(1)}$ &  0.0485 & 0.37099 & 0.8527 & 95.0\% & 0.0528 & 0.69425 & 5.2569 & 98.0\% & 0.0860 & 0.69651 & 10.2399 & 97.0\% \\
             & $\mathrm{ESME}^{(2)}$ & -0.0065 & 0.38747 & 0.8346 & 95.0\% & -0.0190 & 0.74279 & 5.2414 & 96.0\% & 0.1791 & 0.64770 & 10.6053 & 97.0\% \\
             & $\mathrm{NIE}$       &  0.0304 & 0.53982 & 0.1425 & 85.0\% & 0.1181 & 0.76852 & 0.7800 & 94.0\% & 0.2764 & 0.85913 & 1.7510  & 89.0\% \\
    \bottomrule
  \end{tabular}
  \end{adjustbox}
  \label{tab:baseline_I30}
\end{table*}

\section{Additional remarks on the empirical analysis}
\label{sec:remarks_analysis}
We evaluate the predictive performance of the models using the log pseudo marginal likelihood (LPML; \citealp{Geisser1979}). The LPML is a Bayesian model-fit criterion derived from leave-one-out (LOO) predictive assessments of the data. Since our Bayesian simulation-based approach relies on imputing missing potential outcomes at its core, predictive accuracy plays a critical role in overall estimation quality. Consequently, the LPML, derived from LOO predictive densities, serves as a suitable criterion for model comparison. The LPML calculation relies on the conditional predictive ordinate (CPO) \citep{Gelfand_Dey1994}. For each individual $j$ in cluster $i$, and for the $t$-th iteration of the MCMC, we obtain $\text{CPO}_{ij} \approx \left( \frac{1}{T} \sum_{t=1}^{T} \frac{1}{\ell_{ij}^{(t)}}  \right)^{-1},$ where $\ell_{ij}^{(t)}$ is the likelihood of the observed data $\bigl(M_{ij}^{(1)}, M_{ij}^{(2)}, Y_{ij}\bigr)$ given the model parameters at the $t$-th MCMC iteration. We then compute the LPML as $\text{LPML} = \sum_{i=1}^{I} \sum_{j=1}^{N_i} \log\left(\text{CPO}_{ij}\right).$ For the three BNP models under consideration---nDPM, AD-nDDPM, and FD-nDDPM---the LPML values are $30.72$, $33.81$, and $31.81$, respectively. Since a higher LPML indicates a better predictive fit in terms of LOO predictive densities, the AD-nDDPM emerges as the superior model among the three. The numerical results of the empirical analysis in Section \ref{sec:analysis} are reported in Table \ref{tab:posterior_estimates}.

\begin{table}[ht]
    \centering
    \caption{Posterior estimates of causal estimands. ``Est'', ``$95\%$ CI'', and ``PP'' represent the posterior mean, $95\%$ central credible interval, and the posterior probability that the estimand is greater than zero, respectively. The superscript $^{(1)}$ represents the mediator effects for child health check-ups, while the superscript $^{(2)}$ represents the mediator effects for the household dietary diversity $z$-score.}
    \label{tab:posterior_estimates}
    \begin{adjustbox}{width=12cm}
    \begin{tabular}{lcccccc}
        \toprule
        & \multicolumn{3}{c}{AD-nDDPM} & \multicolumn{3}{c}{FD-nDDPM} \\
        \cmidrule(lr){2-4}\cmidrule(lr){5-7}
        Estimand & Est & $95\%$ CI & PP (\%) & Est & $95\%$ CI & PP (\%) \\
        \midrule
        TE          & $0.353$  & $(0.023,\ 0.673)$   & $98.1$ & $0.315$  & $(0.018,\ 0.615)$   & $97.8$ \\
        NIE         & $0.232$  & $(-0.015,\ 0.509)$  & $96.3$ & $0.213$  & $(-0.023,\ 0.478)$  & $95.5$ \\
        NDE         & $0.121$  & $(-0.180,\ 0.460)$  & $78.3$ & $0.102$  & $(-0.207,\ 0.403)$  & $73.5$ \\
        $\mathrm{EIE}^{(1)}$   & $0.058$  & $(-0.122,\ 0.268)$  & $72.4$ & $0.039$  & $(-0.124,\ 0.221)$  & $66.9$ \\
        $\mathrm{EIE}^{(2)}$   & $0.171$  & $(-0.072,\ 0.437)$  & $91.1$ & $0.180$  & $(-0.047,\ 0.412)$  & $93.4$ \\
        INT         & $-0.004$ & $(-0.224,\ 0.209)$  & $48.9$ & $0.005$  & $(-0.218,\ 0.218)$  & $52.1$ \\
        $\mathrm{ESME}^{(1)}$  & $0.060$  & $(-0.117,\ 0.257)$  & $75.4$ & $0.042$  & $(-0.127,\ 0.222)$  & $67.5$ \\
        $\mathrm{ESME}^{(2)}$  & $0.119$  & $(-0.119,\ 0.363)$  & $83.8$ & $0.134$  & $(-0.098,\ 0.371)$  & $87.2$ \\
        $\mathrm{EIME}^{(1)}$  & $-0.002$ & $(-0.161,\ 0.145)$  & $50.8$ & $-0.003$ & $(-0.155,\ 0.143)$  & $50.3$ \\
        $\mathrm{EIME}^{(2)}$  & $0.052$  & $(-0.152,\ 0.242)$  & $70.9$ & $0.045$  & $(-0.128,\ 0.233)$  & $67.8$ \\
        \bottomrule
    \end{tabular}
    \end{adjustbox}
\end{table}

\end{document}